\documentclass[a4paper,11pt]{article}

\usepackage{jheppub} 
\usepackage{graphicx,color}
\usepackage{bm}
\usepackage{amsmath}
\usepackage{amssymb}    
\usepackage{pifont}
\usepackage{stmaryrd}

\usepackage{standalone}
\usepackage{slashed}
\usepackage{multirow}
\usepackage{makecell}
\usepackage{tabu}
\usepackage{hhline}
\usepackage{colortbl}

\usepackage[explicit]{titlesec}
\usepackage{centernot}
\usepackage[english]{babel}
\usepackage{amsmath}
\usepackage{braket}
\usepackage{graphicx}
\usepackage{lipsum} 
\usepackage{bbold}
\usepackage{enumitem}
\usepackage{mathtools}
\usepackage{fancyhdr}
\usepackage{natbib}
\usepackage{hyperref}
\usepackage{feynmp-auto}
\usepackage{import}
\usepackage{tocloft}
\usepackage{mathrsfs}
\usepackage[dvipsnames]{xcolor}

\newcommand{\nbar}{{\bar{n}}}
\newcommand{\cbar}{{\bar{c}}}
\newcommand{\nn}{\nonumber}
\newcommand{\img}{{\rm{i}}}
\newcommand{\tr}{\mathrm{tr}}

\newcommand{\df}{{\rm{d}}}

\newcommand{\fsl}[1]{{\centernot{#1}}}
\newcommand{\colorOne}[1]{{\color[rgb]{0.98,0.1,0.1} #1}}
\newcommand{\colorTwo}[1]{{\color[rgb]{0.7,0.05,0.98} #1}}
\newcommand{\colorThree}[1]{{\color[rgb]{0.9,0.8,0.2} #1}}

\renewcommand{\[}{\left[}

\newcommand{\qbar}{\bar{q}}

\definecolor{green}{rgb}{0.133,0.56,0}

\newcommand{\grn}[1]{\textcolor{ForestGreen}{#1}}
\newcommand{\blu}[1]{\textcolor{RoyalBlue}{#1}}
\newcommand{\orn}[1]{\textcolor{YellowOrange}{#1}}

\title{
Soft background fields at next-to-leading power in transverse momentum dependent SIDIS with jets
}

\newcommand*{\UCM}{Departamento de F\'isica Te\'orica \& IPARCOS, Universidad Complutense de Madrid, Plaza de Ciencias 1, E-28040 Madrid, Spain}
\newcommand*{\UvA}{Institute for Theoretical Physics Amsterdam and Delta Institute for Theoretical Physics, University of
Amsterdam, Science Park 904, 1098 XH Amsterdam, The Netherlands}
\newcommand*{\Nikhef}{Nikhef, Theory Group, Science Park 105, 1098 XG, Amsterdam, The Netherlands}
\author{Max Jaarsma$^{a,b}$,}
\emailAdd{m.jaarsma@uva.nl}
\author{Oscar~del~Rio$^c$,}
\emailAdd{oscgar03@ucm.es}
\author{Ignazio~Scimemi$^c$,}
\emailAdd{ignazios@ucm.es}
\author{Wouter Waalewijn$^{a,b}$}
\emailAdd{w.j.waalewijn@uva.nl}
\affiliation{$^a$\UvA}
\affiliation{$^b$\Nikhef}
\affiliation{$^c$\UCM}

\preprint{IPARCOS-UCM-25-036}

\abstract{We present a factorization formula for the $e + h \rightarrow e + \text{jet} + X$ cross section at small transverse momenta up to next-to-leading power (NLP), derived using the background field method (BFM) with explicit inclusion of soft modes. 
We discuss the relation between soft modes and overlap subtractions at NLP, showing that the inclusion of soft modes enables a definition of twist-3 operators with properly subtracted rapidity and endpoint divergences.
We also derive the effective current operator at NLP and identify additional structures compared to previous approaches based on soft-collinear effective theory and the BFM without soft modes. Nevertheless, for the hadronic tensor, we find agreement at the perturbative order that we are working at, and identify a condition that needs to be satisfied for this agreement to extend to higher orders.
Furthermore, we construct the most general form factors for this process, taking into account the polarization of the initial states. These involve perturbatively-calculable jet functions, opening a path to precise determinations of twist-3 hadronic distributions. 
Finally, our formalism is illustrated with phenomenological results for a specific azimuthal asymmetry, whose leading-logarithmic contribution depends solely on a twist-3 hadronic distribution and a twist-2 jet function.}

\begin{document} 
\allowdisplaybreaks
\maketitle 
\newpage
\section{Introduction}

Factorization of the cross section is crucial to separate the perturbative process-dependent scattering from universal nonperturbative QCD structures.
The case of transverse momentum dependent (TMD) distributions factorization~\cite{Collins:2011zzd, Becher:2010tm, Echevarria:2011epo,Chiu:2012ir,Vladimirov:2021hdn,Ebert:2021jhy}, 
is worthwhile for several reasons: First, the universality of TMD factorization allows us to compare theory and experiment among a range of different processes at different colliders. Thus, unknown theory parameters can be fit to data and compared with results from different experiments. Second, TMD distributions 
allow us to probe the three-dimensional structure of hadrons, extending our knowledge beyond the one-dimensional collinear parton distribution functions (PDFs). 
Finally, TMD factorization can also be applied beyond the realm of classic TMD observables. One example is provided by energy correlators, where the back-to-back limit of the two-point energy correlator is described by TMD factorization~\cite{ Moult:2018jzp,Li:2021txc,Kang:2023big,Kang:2024otf,Bris2025}.

The history of factorization of the differential cross section for TMD observables started a long time ago with the seminal works of \cite{Altarelli:1977kt,Dokshitzer:1978hw,Collins:1981uk,Collins:1984kg}. There, it was observed that perturbative calculations involve a new type of divergences, nowadays called rapidity divergences, that are not regulated by dimensional regularization. This pointed to a more involved form of factorization than for the pure collinear case involving PDFs, a nontrivial soft factor, and overlap (or zero-bin) subtraction.
TMD factorization has been achieved in more recent times, using the method of regions, soft-collinear effective theory (SCET), and the background field method (BFM), with special regulators for rapidity divergences~\cite{Collins:2011zzd, Becher:2010tm, Echevarria:2011epo,Chiu:2012ir,Vladimirov:2021hdn,Ebert:2021jhy}. TMD factorization results from expanding the cross section to leading power in $\lambda$, which is the ratio of a small transverse momentum $q_T$ and the scale $Q$ of the hard scattering process. Key processes in which TMDs play a role are Drell-Yan (DY), semi-inclusive deep-inelastic scattering (SIDIS), and semi-inclusive $e^+ e^-$ annihilation to 2 hadrons or jets (SIA). Here $q_T$ is the small transverse momentum of the di-lepton pair (in an appropriate frame) and $Q$ their invariant mass.

The combination of TMD distributions with jet physics provides a powerful framework for probing hadronization and hadronic substructure. While jets have been proposed as probes of TMDs a long time ago, see e.g.~\cite{Boer:2003tx,Bacchetta:2005rm,Bacchetta:2007sz}, a jet was often synonymous with a final-state parton. A more detailed description of jets in the context of TMD measurements was put forward in ref.~\cite{Gutierrez-Reyes:2018qez}. This is needed to achieve high perturbative accuracy and has been applied more broadly~\cite{Gutierrez-Reyes:2019vbx,Gutierrez-Reyes:2019msa,Chien:2020hzh,Lai:2022aly,Chien:2022wiq}.
Developments of jet substructure have also led to many studies of TMDs inside jets~\cite{DAlesio:2010sag,DAlesio:2011kkm,Bain:2016rrv,Neill:2016vbi,Kang:2017glf,Makris:2018npl,Neill:2018wtk,Kang:2019ahe,Kang:2020xyq,DAlesio:2025jmr}.
This line of research is further driven by the prospect of precise measurements in SIDIS at the upcoming Electron-Ion Collider (EIC)~\cite{Arratia:2019vju,Arratia:2020azl,Arratia:2020nxw,Arratia:2020ssx,Liu:2020dct,Kang:2021ffh,Yang:2022xwy,Caucal:2024vbv}. 

TMD factorization at leading power has been understood for some time now. The resulting factorization formula allows one to resum the large Sudakov double logarithms, and with the existing perturbative ingredients, this can be done to up next-to-next-to-next-to-next-to-leading logarithmic order (N$^4$LL) \cite{Billis:2019evv,Scimemi:2019cmh,Bacchetta:2022awv,Moos:2023yfa,Camarda:2023dqn,Bacchetta:2024qre,Billis:2024dqq,Bacchetta:2025ara,Moos:2025sal}. However, while the resummation of these logarithms is important, at this high level of theoretical precision, sub-leading power effects can no longer be ignored. Therefore, a key factor in achieving high-precision measurements is the systematic treatment of factorization beyond leading-power (LP)~\cite{Ebert:2018gsn,Vladimirov:2021hdn,Gamberg:2022lju,Ebert:2021jhy,Rodini:2022wki,Ferrera:2023vsw,Balitsky:2024ozy,Liu:2024yee}. In broad brush strokes, the derivation of TMD factorization at next-to-leading power proceeds as follows: First, one integrates out off-shell modes corresponding to large momentum transfer (or so-called Glauber exchanges). Then, one successively derives an expression for the effective electromagnetic current, hadronic tensor, and differential cross section. For the cross section to be manifestly free of unphysical divergences, the intermediate result for the hadronic tensor must be reorganized in such a way that all divergences cancel within the individual ingredients of the factorization. These steps have been explored for Drell–Yan~\cite{Gamberg:2022lju,Piloneta:2024aac,Balitsky:2024ozy}, SIDIS~\cite{Ebert:2021jhy,Rodini:2023plb,Gamberg:2022lju}, and SIA~\cite{delCastillo:2023rng}. 

There are, however, different ways one can approach the process of integrating out the off-shell modes from the theory. In particular, there are two approaches for TMD factorization in SIDIS at next-to-leading power with seemingly different results. One using the background field method (BFM) in ref.~\cite{Vladimirov:2021hdn}, and another in ref.~\cite{Ebert:2021jhy} is based on earlier foundations of Soft Collinear Effective Theory (SCET)~\cite{Bauer:2001ct,Bauer:2001yt,Bauer:2002nz,Beneke:2002ph,Becher:2014oda}. These approaches differ mainly in the treatment of the soft sector. In the BFM, soft modes are not introduced as separate degrees of freedom, whereas in SCET, they form an independent sector. This difference in approach results in an apparent disagreement at the level of the cross section. In order for these two approaches to agree, the sub-leading soft contributions found in ref.~\cite{Ebert:2021jhy} have to appear in such a form that they can be absorbed into the collinear and anti-collinear ingredients, which is a non-trivial statement. This makes a detailed comparison between the two frameworks particularly relevant. 

In this work, we introduce a third approach aimed at bridging the gap between the two existing methods and resolving this apparent disagreement. Specifically, we adopt a modified version of the BFM, which includes an additional soft background field. We start by deriving the effective current to NLP, extending beyond leading order in perturbation theory by matching to gauge invariant SCET operators and using symmetry arguments. In the hadronic tensor, many of these sub-leading operators do not contribute.
We subtract the overlap between soft and collinear sectors by introducing a background field for the overlap sector, leading to definitions of physical TMD distributions with a manifestly finite cross section. While our approach parallels SCET in many ways, we differ in the operator basis. This is because we work in position space instead of the label formalism, we work top-down instead of bottom-up and we do not separate the off-shell modes into hard and hard-collinear modes. Furthermore, the background field method does not employ a strict expansion in the power counting parameter (though these effects are beyond the order in the power counting that we consider).

We then apply our result for the hadronic tensor to jet production in SIDIS, deriving the factorized cross section for $e+H\rightarrow e+\text{jet}+X$ at NLP and presenting the explicit expressions for the form factors of the cross section at small transverse momenta. These form factors receive contributions from both twist-2 and twist-3 components of the jet function and from nonperturbative hadronic TMD distributions. We also provide phenomenological predictions for the $\sin\phi_J$ form factor to leading-logarithmic accuracy. In our predictions, we treat both the twist-2 and the twist-3 (dimension minus spin) jet functions as perturbative objects~\cite{delCastillo:2023rng}, and we do not consider possible nonperturbative T-odd jet effects~\cite{Lai:2022aly}. This structure enables an extraction of higher-twist hadronic distributions from azimuthal and spin asymmetry observables.

This work is organized as follows. We start in sec.~\ref{sec:SoftBFM} by introducing our approach to computing NLP corrections to the effective current, incorporating background fields with collinear, anti-collinear, and soft momentum scaling. In sec.~\ref{sec:NLP}, we show the calculation of the effective current operator at NLP, present the final expression in terms of SCET fields, and compare it to previous results. We continue in sec.~\ref{sec:HadronicTensor} by deriving the NLP hadronic tensor with distributions that have open spinor indices. Here, we also perform the subtraction of the overlap between the different modes, which leads to definitions of twist-3 distributions free of rapidity and endpoint divergences. Section \ref{sec:JetSIDISFactorizationNLP} contains the full factorization formula of the cross section for SIDIS with a jet measurement at NLP, along with expressions for all relevant form factors. Additionally, we provide a phenomenological application, analyzing a specific spin asymmetry for unpolarized-electrons scattering off longitudinally polarized protons, using a model for the twist-3 distribution involved. We present our conclusions in sec.~\ref{sec:Conclusion}. Technical details are reported in appendices.

\section{Soft mode and background field method}\label{sec:SoftBFM}

In this section, we present our method for deriving TMD factorization formulas. Lets first discuss some notation and conventions. In this paper, we employ lightcone coordinates, where four-vectors are decomposed as
\begin{align}\label{eq:VectorComponents}
    v^\mu=v^+ \nbar^\mu + v^- n^\mu + v_T^\mu\,,
\end{align}
and use the notation $(v^+,v^-,v_T)$ to represent the components of a vector. Here, $n$ and $\nbar$ are light-like reference vectors which satisfy $n^2=\bar{n}^2=0$ and $n\cdot\bar{n}=1$. The definition of the reference vectors $n$ and $\nbar$ in terms of physical quantities, as well as the definition of the transverse component, will be clarified when we discuss the kinematics of SIDIS in section \ref{sec:JetSIDISFactorizationNLP}. We introduce a parameter $\lambda$, which is of the order of $q_T/Q$, where $q_T$ is the transverse momentum of the intermediate photon and $Q$ is the center-of-mass (COM) energy. For small transverse momentum, the external states are dominated by the following modes of momentum,
\begin{align}\label{eq:sectors}
    &\blu{n\text{-collinear}\sim(\lambda^2,1,\lambda)}\,,&
    &\grn{\nbar\text{-collinear}\sim(1,\lambda^2,\lambda)}\,,&
    &\orn{\text{soft}\sim(\lambda,\lambda,\lambda)}\,.
\end{align}
In the remainder of this paper, we will use the above color scheme to distinguish between the three sectors.

In the background field approach to factorization \cite{Vladimirov:2021hdn,Rodini:2023plb,delCastillo:2023rng, Rodini:2023mnh,Vladimirov:2023aot}, a background field is introduced for each mode of momentum, and the power counting parameter $\lambda$ is used to establish a power counting for the fields and their derivatives. The QCD quark and gluon fields are decomposed into background fields and a remainder dynamical field describing the hard scattering, according to
\begin{align}
    &\psi=\blu{\psi}+\grn{\psi}+\orn{\psi}+\varphi\,,&
    &A^\mu=\blu{A^\mu}+\grn{A^\nu}+\orn{A^\mu}+B^\mu\,.
\end{align}
The fields in black are the dynamical fields, and the others are background fields for the modes according to the color coding. The power counting manifests itself in the scaling of the derivatives of the fields as,
\begin{align}
   \partial^\mu \blu{\phi}\sim(\lambda^2,1,\lambda)\blu{\phi}\,,\qquad
   \partial^\mu \grn{\phi}\sim(1,\lambda^2,\lambda)\grn{\phi}\,,\qquad
  \partial^\mu  \orn{\phi}\sim(\lambda,\lambda,\lambda)\orn{\phi}\,,
\end{align}
where $\phi$ is a placeholder for either fermion or gauge fields.

The background fields themselves also obey a power counting. This power counting can be derived from evaluating the two-point correlators of the fields, and is for gauge fields given by
\begin{align}
    \blu{A^\mu}\sim(\lambda^2,1,\lambda)\,,\quad
    \grn{A^\mu}\sim(1,\lambda^2,\lambda)\,,\quad
    \orn{A^\mu}\sim(\lambda,\lambda,\lambda)\,.
\end{align}
For the collinear and anti-collinear fermion fields, the power counting is also non-uniform. To keep the power counting explicit, we decompose these fields as
\begin{align}
    \blu{\psi}=\tfrac{1}{2}\gamma^+\gamma^-\blu{\psi}
    +\tfrac{1}{2}\gamma^-\gamma^+\blu{\psi}=\blu{\xi}+\blu{\eta}\,,
    \qquad
    \grn{\psi}=\tfrac{1}{2}\gamma^-\gamma^+\grn{\psi}
    +\tfrac{1}{2}\gamma^+\gamma^-\grn{\psi}=\grn{\xi}+\grn{\eta}\,.
\end{align}
The power counting for the fermion background fields then reads
\begin{align}
    \blu{\xi}\sim\grn{\xi}\sim\lambda\,,\qquad
    \orn{\psi}&\sim\lambda^{3/2}\,,\qquad
    \blu{\eta}\sim\grn{\eta}\sim\lambda^2\,.
\end{align}

Following the mode decomposition of the fields, the dynamical fields are integrated out, which correspond to the off-shell modes of momentum that are not part of the collinear, soft, or anti-collinear sectors. To do this, we first apply the mode decomposition to the full QCD action to write,
\begin{align}
    S_{\text{QCD}}[\bar\psi,\psi,A]
    &=\blu{S_{\text{QCD}}[\bar\psi,\psi,A]}
     +\grn{S_{\text{QCD}}[\bar\psi,\psi,A]}
     +\orn{S_{\text{QCD}}[\bar\psi,\psi,A]}
    +S_{\text{QCD}}[\bar\varphi,\varphi,B]
    \nn\\
    &\quad
    +S_{\text{int}}[\bar\varphi,\varphi,B;\blu{\bar\psi,\psi,A};
        \grn{\bar\psi,\psi,A};\orn{\bar\psi,\psi,A}]\,.
\end{align}
Here, the first line contains four copies of the QCD action for the three sectors and the dynamical fields, while the second line contains the interactions between the different sectors and dynamical fields. Since external states involve only collinear, soft, and anti-collinear modes of momentum, we can integrate out the dynamical fields $\varphi$, $\bar\varphi$, and $B$ without affecting matrix elements of the full theory. This gives rise to an effective action for the background fields,
\begin{align}
    e^{\img S_{\text{eff}}[\blu{\bar\psi,\psi,A};
        \grn{\bar\psi,\psi,A};\orn{\bar\psi,\psi,A}]}
    &=
    e^{\img\blu{S_{\text{QCD}}[\bar\psi,\psi,A]}}
    e^{\img\grn{S_{\text{QCD}}[\bar\psi,\psi,A]}}
    e^{\img\orn{S_{\text{QCD}}[\bar\psi,\psi,A]}}
    \nn\\
    &\quad\times
    \int\mathcal{D}\bar\varphi\,\mathcal{D}\varphi\,\mathcal{D}B\,
    e^{\img S_{\text{int}}[\bar\varphi,\varphi,B;\blu{\bar\psi,\psi,A};
        \grn{\bar\psi,\psi,A};\orn{\bar\psi,\psi,A}]}
    e^{\img S_{\text{QCD}}[\bar\varphi,\varphi,B]}\,.
\end{align}

When integrating out the dynamical fields of the theory, two types of off-shell modes play a key role: hard modes and Glauber modes. Hard modes have an off-shellness of order $Q^2$ and are responsible for mediating the hard scattering in high-energy processes. In addition to these, the power counting defined in eq.~\eqref{eq:sectors} also leads to hard modes with an off-shellness of order $\lambda Q^2$. These are referred to as hard-collinear modes~\cite{Bauer:2001yt}, which emerge, for instance, when a soft emission pushes a collinear or anti-collinear particle off-shell by an amount $\lambda Q^2$. In this work, we do not distinguish between hard and hard-collinear modes, and we collectively refer to both as hard modes.

Glauber modes, on the other hand, are low-energy off-shell modes that can mediate interactions across the different sectors. The momenta of these Glauber modes scale as
\begin{align}
    \text{Glauber}\sim(\lambda^a,\lambda^b,\lambda)\,,
\end{align}
with $a+b>2$. To keep track of which contributions originate from integrating out hard modes and which from Glauber modes, we decompose the effective action as
\begin{align}
    S_{\text{eff}}[\blu{\bar\psi,\psi,A};\grn{\bar\psi,\psi,A};\orn{\bar\psi,\psi,A}]
    &=
     \blu{S_\text{QCD}[\bar\psi,\psi,A]}
    +\grn{S_\text{QCD}[\bar\psi,\psi,A]}
    +\orn{S_\text{QCD}[\bar\psi,\psi,A]}
    \\\nonumber
    &\quad
    +S_\text{hard}[\blu{\bar\psi,\psi,A};\grn{\bar\psi,\psi,A};\orn{\bar\psi,\psi,A}]
    \\\nonumber
    &\quad
    +S_\text{Glauber}[\blu{\bar\psi,\psi,A};\grn{\bar\psi,\psi,A};\orn{\bar\psi,\psi,A}]\,.
    \nonumber
\end{align}
We refer the reader to ref.~\cite{Rothstein:2016bsq} for an extensive discussion of the Glauber modes and the corresponding effective action.

Factorization of the cross section depends crucially on the absence of leading-power interactions between the different sectors. For hard interactions encoded in $S_\text{hard}$, this is immediate, as all terms are power-suppressed. While hard-collinear modes can, in principle, lead to leading-power interactions, these can be resummed into Wilson lines and do not spoil factorization. In contrast, leading-power interactions in $S_\text{Glauber}$ require more care. Indeed, Glauber interactions can contribute at leading power. However, for TMD observables, it was shown in refs.~\cite{Collins:1984kg,Collins:1988ig} that these leading-power Glauber contributions cancel in the cross section. Therefore, in what follows, we neglect Glauber modes altogether.

Since all leading-power interactions between the different sectors are either power-suppressed or canceled, all remaining interactions can be treated as power-suppressed perturbations. This allows us to factorize the Hilbert spaces of the different sectors and, in turn, separate external states according to
\begin{align}
    \ket{\Psi}&=
    \Psi\bigl[\blu{\phi};\orn{\phi};\grn{\phi}\bigr]\ket{0}
    =\blu{\ket{\Psi_c}}\orn{\ket{\Psi_s}}\grn{\ket{\Psi_\cbar}}\,.
\end{align}

To derive the factorization formula for TMD factorization, we start from the hadronic tensor, which for SIDIS is defined as
\begin{align}
    W^{\mu\nu}
    &=
    \int\frac{\df^4 b}{(2\pi)^4}\,e^{+\img b \cdot q}\,
    \bra{P} J^\mu(b) \ket{p,X} \bra{p,X} J^\nu(0) \ket{P}\,,
\end{align}
where the state $P$ denotes the incoming proton, $p$ the outgoing hadron or jet, $X$ the outgoing unidentified radiation, and $J^\mu$ is the electromagnetic current operator. Here, it is not necessary to distinguish different flavors of quarks or their precise electric charge, and therefore, we write the electromagnetic current as
\begin{align}
    J^\mu(x)&=\bar\psi(x)\gamma^\mu\psi(x)\,.
\end{align}
Taking the proton to be $\nbar$-collinear and taking the outgoing hadron or jet to be $n$-collinear, we can apply the above formalism to factorize the hadronic tensor. To do this, we use that matrix elements of the electromagnetic current operator can be calculated in the effective theory by means of the effective current operator,
\begin{align}\label{eq:EffectiveCurrent}
    J_\text{eff}^\mu
    &=
    \int\mathcal{D}\bar\varphi\,\mathcal{D}\varphi\,\mathcal{D}B\,
    (\bar\varphi+\blu{\bar\psi}+\orn{\bar\psi}+\grn{\bar\psi})
    \gamma^\mu(\varphi+\blu{\psi}+\orn{\psi}+\grn{\psi})\,
    \nn\\
    &\quad\times
    e^{\img S_{\text{int}}[\bar\varphi,\varphi,B;\blu{\bar\psi,\psi,A};
        \grn{\bar\psi,\psi,A};\orn{\bar\psi,\psi,A}]}
    e^{\img S_{\text{QCD}}[\bar\varphi,\varphi,B]}\,\,,
\end{align}
where it is understood that only \emph{connected graphs} contribute to the effective operator. In the next section, we will derive an expression for the effective current operator to next-to-leading power, and in section~\ref{sec:HadronicTensor} we will use this result to derive a factorized expression for the SIDIS hadronic tensor.

\section{The current operator at to next-to-leading power}\label{sec:NLP}

In this section, we construct the effective current operator at next-to-leading power to all orders in perturbation theory, using the background field method and according to the following steps:
\begin{enumerate}
    \item Integrate out the hard modes by first expanding the exponent in eq.~\eqref{eq:EffectiveCurrent} in the strong coupling. To do this, one needs to multipole expand the background fields in the presence of a dynamical field describing a hard mode. Additionally, terms that do not correspond to a current with a hard photon, terms that are suppressed in the power counting, and terms that can be eliminated via the field equations of motion need to be discarded. This results in an expression for the current operator in terms of the background field operators.
    \item Match onto the gauge invariant building block operators of SCET. This allows one to obtain an expression for the current operator that is manifestly gauge invariant.
    \item Generalize the result to all orders in perturbation theory. This is can be done by introducing a Wilson coefficient for each of the operator that appears and then using  symmetries to constrain said coefficients. 
\end{enumerate}
In the subsequent sections, we follow this recipe step-by-step to construct the SCET-II effective current operator to next-to-leading power. We then present the final result for the effective current operator, and provide a tree-level cross-check with the full-QCD result, in sec.~\ref{sec:NLPresult}. Lastly, in sec.~\ref{sec:ComparisonCurrentNLP}, we compare our result to the existing literature. 

\subsection{Background field calculation}\label{sec:NLPwithSoftBckg}

\begin{figure}[t]
    \centering
    \includegraphics[width=0.6\textwidth]{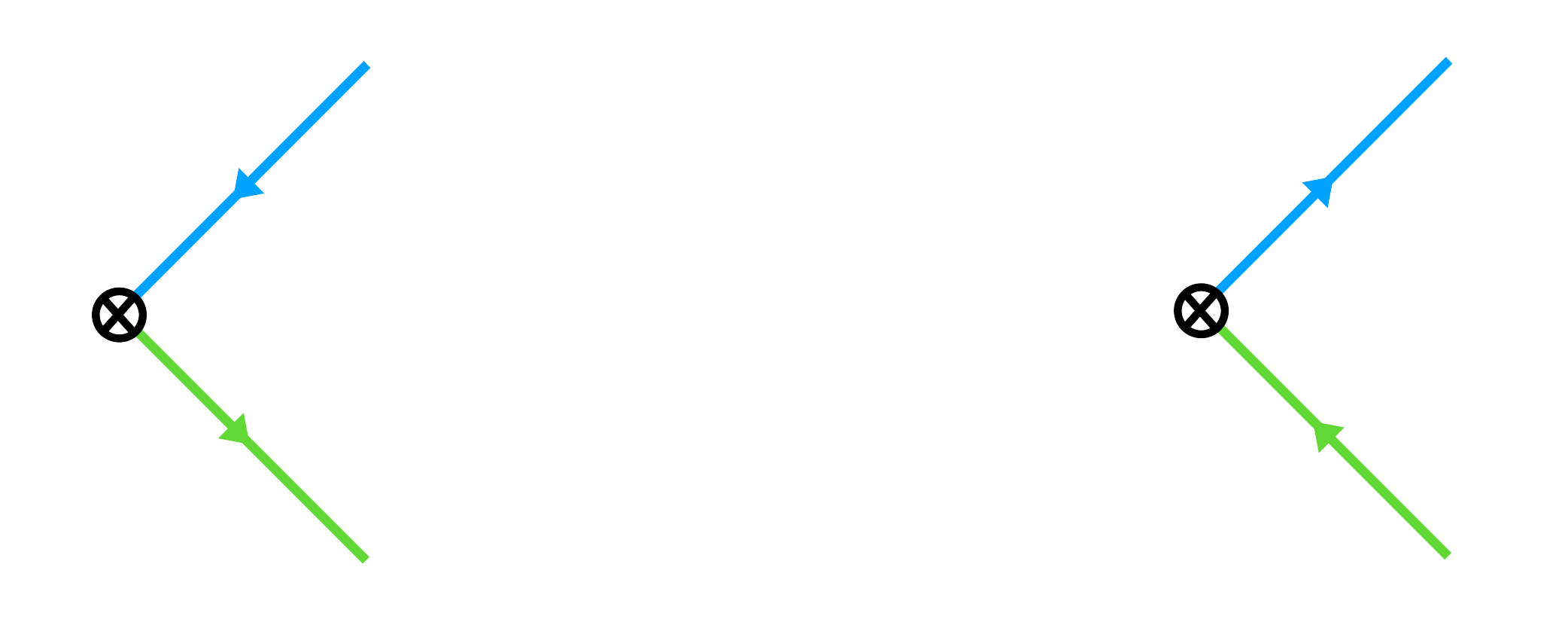}
    \includegraphics[width=0.8\textwidth]{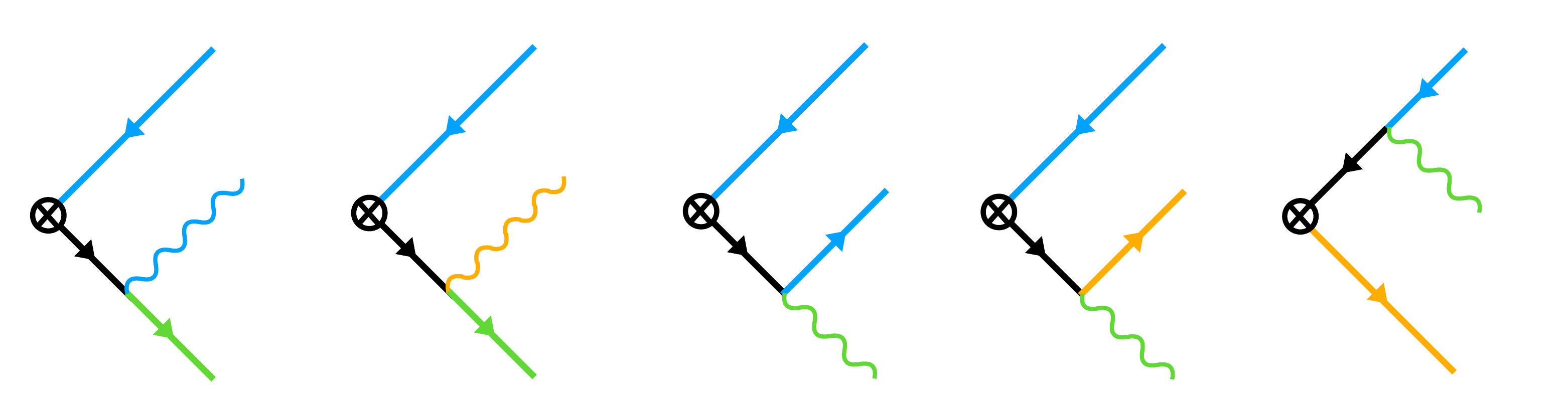}    
    \caption{\label{fig:matching} The current at leading power (top row) and next-to-leading power (bottom row) and leading perturbative order. In the bottom row mirrored diagrams are not shown.}
\end{figure}

Let us now derive the SCET-II effective current operator at leading-power using the method outlined above. We start with the functional integral expression for the effective current evaluated at position $0$, as given in eq.~\eqref{eq:EffectiveCurrent}. For the order we are working at, it suffices to expand the exponential containing the interactions with the hard modes to first order in the coupling $g$. After this expansion, we can perform the functional integration over the hard modes. At leading order in the coupling, this simply gives us the position space propagator,
\begin{align}
    \int\mathcal{D}\bar\varphi\,\mathcal{D}\varphi\,\mathcal{D}B\,
    \varphi_\alpha(x)\,\bar\varphi_\beta(y)\,
    e^{\img S_\text{QCD}[\varphi]}
    &=
    \frac{\Gamma(2-\epsilon)}{2\pi^{2-\epsilon}}
    \frac{\img(\fsl{x}-\fsl{y})_{\alpha\beta}}{[-(x-y)^2+\img0]^{2-\epsilon}}+\mathcal{O}(g^2)\,,
\end{align}
resulting in
\begin{align}
    J^\mu_{\text{eff}}&=
    \blu{\bar\psi}\gamma^\mu\grn{\psi}+
    \bigl[\blu{\bar\psi}+\orn{\bar\psi}+\grn{\bar\psi}\bigr]
    \gamma^\mu
    \biggl[
    \img g\int\df^d w\,
    \frac{\img\Gamma(2-\epsilon)}{2\pi^{d/2}}
    \frac{-\fsl{w}}{(-w^2+\img0)^{2-\epsilon}}
    \\
    &\quad\qquad\times
    \Bigl[
     \blu{\fsl{A}}\orn{\psi}
    +\blu{\fsl{A}}\grn{\psi}
    +\orn{\fsl{A}}\blu{\psi}
    +\orn{\fsl{A}}\grn{\psi}
    +\grn{\fsl{A}}\blu{\psi}
    +\grn{\fsl{A}}\orn{\psi}
    \Bigr](w)
    \biggr]+\text{h.c.}+\mathcal{O}(g^2)\,.
    \nonumber
\end{align}
A subset of the diagrams that contribute to the effective current operator are shown in fig.~\ref{fig:matching}, and their mirrored counterparts are included in the hermitian conjugate.

To perform the integral over the vertex position, we multipole expand the background fields in the presence of a hard propagator. In performing the multipole expansion, one compares the scaling of the different components of the  momenta for each field involved at the vertex. Note that in some cases we have hard internal lines, which scale as $(1,1,1)$, while in other cases we have hard-collinear  or hard-anti-collinear internal lines, which scale as $(1,\lambda,\lambda^{\frac{1}{2}})$ and $(\lambda,1,\lambda^{\frac{1}{2}})$ respectively. This leads to the following all-order multipole expansion, \cite{Vladimirov:2021hdn,Vladimirov:2023aot}
\begin{align}
    \varphi(w)\blu{\mathcal{O}(w)}\grn{\mathcal{O}(w)}&=
    \sum_{k,l,m=0}^\infty
    \frac{(w_T\cdot\partial_T)^k}{k!} 
    \frac{(w^+)^l}{l!}\frac{(w^-)^m}{m!}
    \biggl[\varphi(w)
    \blu{(\partial^+)^m \mathcal{O}(w^+ \nbar)}
    \grn{(\partial^-)^l \mathcal{O}(w^- n)}
    \biggr],
    \nonumber\\
    \varphi(w)\blu{\mathcal{O}(w)}\orn{\mathcal{O}(w)}&=
    \sum_{k,l,m=0}^\infty
    \frac{(w_T\cdot\partial_T)^k}{k!} 
    \frac{(w^+)^l}{l!}\frac{(w^-)^m}{m!}
    \biggl[\varphi(w)
    \blu{(\partial^+)^m \mathcal{O}(w^+ \nbar)}
    \orn{(\partial^-)^l \mathcal{O}(w^- n)}\biggr],
    \nonumber\\
    \varphi(w)\orn{\mathcal{O}(w)}\grn{\mathcal{O}(w)}&=
    \sum_{k,l,m=0}^\infty
    \frac{(w_T\cdot\partial_T)^k}{k!} 
    \frac{(w^+)^l}{l!}\frac{(w^-)^m}{m!}
    \biggl[\varphi(w)
    \orn{(\partial^+)^m \mathcal{O}(w^+ \nbar)}
    \grn{(\partial^-)^l \mathcal{O}(w^- n)}
    \biggr],
\end{align}
where $\varphi$ is a place-holder for a dynamical field associated to hard mode in the first line, to a hard-collinear mode in the second line, and to a hard-anti-collinear mode in the third line (the hard-collinear nature of these modes can be seen from momentum conservation).

From here, we perform the integration over the vertex position $w$ using the integrals from section 3 and appendix B of ref.~\cite{Vladimirov:2021hdn}, which can be combined to write
\begin{align}
    &-\img\int\df^d w\,\frac{\img\Gamma(2-\epsilon)}{2\pi^{d/2}}
    \frac{w_T^{\nu_1}\dotsm w_T^{\nu_{2n}}}{(-w^2 + i0)^{2-\epsilon}}\,
    (w^+)^{\alpha_+}\,(w^-)^{\alpha_-}\,f(w^+)\,g(w^-)\,
    \\
    &=
    2^{n-1} s_{T}^{\nu_1\dotsm\nu_{2n}}
    (-\img)^{1 + \alpha_+ + \alpha_-}
    \frac{\Gamma(1+n-\epsilon)}{\Gamma(1-\epsilon)}
    \frac{\Gamma(n+\alpha_+)\Gamma(n+\alpha_-)}{\Gamma(n)}
    \biggl[\frac{1}{(\img\partial^-)^{n+\alpha_+}}f\biggr]
    \biggl[\frac{1}{(\img\partial^+)^{n+\alpha_-}}g\biggr]\,,
    \nonumber
\end{align}
where $s_T^{\nu_1\dotsm\nu_{2n}}$ is a completely symmetric tensor with a total trace equal to $1$. Additionally, we have defined the inverse derivative,
\begin{align}
    \label{eq:MinusInverseDerivative}\frac{1}{(\partial^-)^k} f(y)&=
    \frac{(-1)^{k-1}}{\Gamma(k)}\int_{\bar{L}}^0\df w^+\,(w^+)^{k-1}\,f(y+w^+\nbar)\,,
    \\
    \label{eq:PlusInverseDerivative}\frac{1}{(\partial^+)^k} g(y)&=
    \frac{(-1)^{k-1}}{\Gamma(k)}\int_L^0\df w^-\,(w^-)^{k-1}\,f(y+w^-n)\,.
\end{align}
The derivation of this identity is sensitive to the analytic structure of the background fields, which in turn depends on the process. This process dependence is encoded in the definition of the inverse derivative through the boundary $L$ and $\bar{L}$ of the integral ($L, \overline L=+\infty$ for final states in SIDIS and $e^+e^-$ and $L,\overline L=-\infty$ for initial states for SIDIS and DY).

Finally, we discard all terms that do not result in a hard photon momentum, terms that are power suppressed, and terms that can be eliminated via the equations of motion. The final result for the effective current operator in terms of the background field operators can then be organized as
\begin{align}
    J^\mu_{\text{eff}}(0)
    &=
    \blu{\bar\psi}\gamma^\mu\biggl[1
    -g\blu{\frac{1}{\img\partial^-}A^-}
    +g\orn{\frac{1}{\img\partial^+}A^+}
    -g\orn{\frac{1}{\img\partial^-}A^-}
    +g\grn{\frac{1}{\img\partial^+}A^+}
    \biggr]\grn{\psi}
    +(\blu{n}\leftrightarrow\grn{\nbar})
    \\ \nn
    &\quad+g\biggl\{
    -n^\mu\blu{\bar\psi}
    \blu{\biggl(\fsl{A}_T-\frac{\img\fsl{\partial}_T}{\img\partial^-}A^-\biggr)}
        \grn{\frac{1}{\img\partial^+}\psi}
    -n^\mu\blu{\bar\psi}
    \orn{\biggl(\fsl{A}_T-\frac{\img\fsl{\partial}_T}{\img\partial^-}A^-\biggr)}
        \grn{\frac{1}{\img\partial^+}\psi}
    \\ \nn
    &\quad\qquad
    -\blu{\bar\psi}\gamma_T^\mu
    \orn{\biggl(\frac{1}{\img\partial^-}A_T^\rho
            -\frac{\img\partial_T^\rho}{(\img\partial^-)^2}A^-\biggr)}
        \grn{\frac{\img\partial_T^{\rho}}{\img\partial^+}\psi}
    -\frac{1}{2}\blu{\bar\psi}\gamma_T^\mu
        \orn{\biggl(\frac{\img\fsl{\partial}_T}{\img\partial^-}\fsl{A}_T
            -\frac{(\img\fsl{\partial}_T)^2}{(\img\partial^-)^2}A^-\biggr)}
        \grn{\frac{1}{\img\partial^+}\psi}
    \\ \nn
    &\quad\qquad
    -\frac{1}{2}\blu{\bar\psi}\gamma_T^\mu\gamma^-
    \grn{\biggl(\fsl{A}_T-\frac{\img\fsl{\partial}_T}{\img\partial^+}A^+\biggr)}
        \orn{\frac{1}{\img\partial^-}\psi}
    -\frac{1}{2}\blu{\bar\psi}\gamma_T^\mu\gamma^-
    \grn{\biggl(\fsl{A}_T-\frac{\img\fsl{\partial}_T}{\img\partial^+}A^+\biggr)}
        \blu{\frac{1}{\img\partial^-}\psi}
    \\ \nn
    &\quad\qquad
    +(\blu{n}\leftrightarrow\grn{\nbar})+\text{h.c.}\biggr\}
    +\mathcal{O}\bigl(g^2,\lambda^{7/2}\bigr)\,.
\end{align}
Here, the notation $\blu{n}\leftrightarrow\grn{\nbar}$ implies swapping $n$ and $\nbar$, which simultaneously interchanges \blu{collinear} operators by \grn{anti-collinear} ones. Note that some terms that appeared at intermediate steps have disappeared simply because they were anti-hermitian. One example is 
\begin{align}
    \blu{\bar\psi} \gamma^- \grn{\frac{1}{\img\partial^+}A^+} \blu{\psi}\,.
\end{align}

\subsection{Current operator in SCET basis}\label{sec:NLPwithSCETbasis}

The final result of the previous section expresses the current operator in terms of background fields and is only valid at the first order in the strong coupling. The above expression is subject to higher-order corrections in the strong coupling that are not further power-suppressed. It is also not manifestly gauge invariant, as gauge invariance is only restored by including a tower of higher-order terms that sum into Wilson lines. In this section, we wish to write the effective current operator in terms of the gauge invariant building block operators from SCET and include Wilson coefficients that account for higher-order corrections. This way, assuming that our basis of operators is complete, we obtain an expression for the effective current operator valid to all orders in the coupling.

First, let us introduce the gauge invariant building blocks of SCET which involve collinear and soft Wilson lines. The collinear Wilson lines are defined as
\begin{align}
    \label{eq:CollinearWilsonLines}\grn{W(y)}&=
    \mathcal{P}\biggl\{\exp\biggl[\img g\int_{L}^0\df w^-\,
    \grn{A^+(y + w^- n)}\biggr]\biggr\}\,,
    \\
    \blu{W(y)}&=
    \mathcal{P}\biggl\{ \exp\biggl[\img g\int_{\bar L}^0\df w^+\,
    \blu{A^-(y + w^+ \nbar)}\biggr]\biggr\}\,,
\end{align}
and the soft Wilson lines are defined as,
\begin{align}
    \orn{S_n(y)}&=
    \mathcal{P}\biggl\{\exp\biggl[\img g\int_{L}^0\df w^-\,
    \orn{A^+(y + w^- n)}\biggr]\biggr\}\,,
    \\
    \orn{S_\nbar(y)}&=
    \mathcal{P}\biggl\{\exp\biggl[\img g\int_{\bar L}^0\df w^+\,
    \orn{A^-(y + w^+ \nbar)}\biggr]\biggr\}\,.\label{eq:SoftWilsonLines}
\end{align}
Here, the boundary conditions of the integrals depend on the process, and in the case of SIDIS with the initial hadron in the anti-collinear direction, we have $(L,\bar L)=(-\infty,+\infty)$. Using these Wilson lines, one can write down the following gauge invariant operators for collinear quark and gluon fields,
\begin{align}
    &\blu{\chi}=\frac{\gamma^+\gamma^-}{2}\blu{W^\dagger\psi}\,,&
    &\blu{\mathcal{A}_T^\rho}=\blu{W^\dagger\bigl[\img D_T^\rho,W\bigr]}\,,
    \\
    &\grn{\chi}=\frac{\gamma^-\gamma^+}{2}\grn{W^\dagger\psi}\,,&
    &\grn{\mathcal{A}_T^\rho}=\grn{W^\dagger\bigl[\img D_T^\rho,W\bigr]}\,.
\end{align}
This set of building blocks is, in fact, complete, as all other components can be eliminated via the field equations of motion. For the soft operators, we need two sets of building blocks, as gauge invariant operators can be constructed both with $\orn{S_n}$ and $\orn{S_\nbar}$,
\begin{align}
    &\orn{\Psi_n}=\orn{S_n^\dagger\psi}\,,&
    &\orn{\mathcal{A}_{T,n}^\rho}=\orn{S_n^\dagger\bigl[\img D_T^\rho,S_n\bigr]}\,,
    \\
    \label{eq:SoftAntiCollinearFieldsSCET}&\orn{\Psi_\nbar}=\orn{S_\nbar^\dagger\psi}\,,&
    &\orn{\mathcal{A}_{T,\nbar}^\rho}=
    \orn{S_\nbar^\dagger\bigl[\img D_T^\rho,S_\nbar\bigr]}\,.
\end{align}

Next, let us match the SCET building block operators to the background field operators that appear in the current operator. To do this, we first expand the above operators to first order in the coupling. For the Wilson lines, we have
\begin{align}
    &\blu{W}=1-\blu{\frac{1}{\img\partial^-}gA^-}+\mathcal{O}(g^2)\,,&
    &\orn{S_n}=1-\orn{\frac{1}{\img\partial^+}gA^+}+\mathcal{O}(g^2)\,,
    \\
    &\grn{W}=1-\grn{\frac{1}{\img\partial^+}gA^+}+\mathcal{O}(g^2)\,,&
    &\orn{S_\nbar}=1-\orn{\frac{1}{\img\partial^-}gA^-}+\mathcal{O}(g^2)\,,
\end{align}
and for the gluon fields, we have
\begin{align}
    &\blu{\mathcal{A}_T^\rho}=\blu{gA_T^\rho}
        -\blu{\frac{\img\partial_T^\rho}{\img\partial^-}gA^-}
        +\mathcal{O}(g^2)\,,&
    &\orn{\mathcal{A}_{T,n}^\rho}=\orn{gA_T^\rho}
        -\orn{\frac{\img\partial_T^\rho}{\img\partial^+}gA^+}
        +\mathcal{O}(g^2)\,,
    \\
    &\grn{\mathcal{A}_T^\rho}=\grn{gA_T^\rho}
        -\grn{\frac{\img\partial_T^\rho}{\img\partial^+}gA^+}
        +\mathcal{O}(g^2)\,,&
    &\orn{\mathcal{A}_{T,\nbar}^\rho}=\orn{gA_T^\rho}
        -\orn{\frac{\img\partial_T^\rho}{\img\partial^-}gA^-}
        +\mathcal{O}(g^2)\,.
\end{align}
To relate the collinear quark background fields $\psi$ to the SCET building blocks $\chi$, we use the equations of motion for the collinear quark field to write
\begin{align}
    \blu{W^\dagger\psi}
    &=
    \blu{\chi}
    -\frac{1}{2}\gamma^-\blu{\frac{\img\fsl{\partial}_T}{\img\partial^-}\chi}
    -\frac{1}{2}\gamma^-\blu{\frac{1}{\img\partial^-}
        \bigl(\fsl{\mathcal{A}}_T\chi\bigr)}\,,
    \\
    \grn{W^\dagger\psi}
    &=
    \grn{\chi}
    -\frac{1}{2}\gamma^+\grn{\frac{\img\fsl{\partial}_T}{\img\partial^+}\chi}
    -\frac{1}{2}\gamma^+\grn{\frac{1}{\img\partial^+}
        \bigl(\fsl{\mathcal{A}}_T\chi\bigr)}\,.
\end{align}
For the soft quark field operator, no expansion in the coupling is needed at this point. Using these expressions, we can write the effective current operator as
\begin{align}\label{eq:NLPcurrentSCET}
    J^\mu_{\text{eff}}(0)
    &=
    \blu{\bar\chi}\gamma_T^\mu\orn{S_n^\dagger S_\nbar}\grn{\chi}
    +(\blu{n}\leftrightarrow\grn{\nbar})
    \\ \nn
    &\quad+\biggl\{
    -n^\mu\blu{\bar\chi}\orn{S_n^\dagger S_\nbar}
        \grn{\frac{\img\fsl{\partial}_T}{\img\partial^+}\chi}
    -n^\mu\blu{\bar\chi}\orn{S_n^\dagger S_\nbar}
        \grn{\frac{1}{\img\partial^+}\bigl(\fsl{\mathcal{A}}_T\chi\bigr)}
    \\ \nn
    &\quad\qquad
    -n^\mu\blu{\bar\chi\fsl{\mathcal{A}}_T}
        \orn{S_n^\dagger S_\nbar}
        \grn{\frac{1}{\img\partial^+}\chi}
    -n^\mu\blu{\bar\chi}\orn{S_n^\dagger S_\nbar}
        \orn{\fsl{\mathcal{A}}_{T,\nbar}}
        \grn{\frac{1}{\img\partial^+}\chi}
    \\ \nn
    &\quad\qquad
    -\blu{\bar\chi}\gamma_T^\mu\orn{S_n^\dagger S_\nbar}
    \orn{\frac{1}{\img\partial^-}\mathcal{A}_{T,\nbar}^\rho}
    \grn{\frac{\img\partial_T^{\rho}}{\img\partial^+}\chi}
    -\frac{1}{2}\blu{\bar\chi}\gamma_T^\mu\orn{S_n^\dagger S_\nbar}
    \orn{\frac{\img\fsl{\partial}_T}{\img\partial^-}\fsl{\mathcal{A}}_{T,\nbar}}
    \grn{\frac{1}{\img\partial^+}\chi}
    \\ \nn
    &\quad\qquad
    -\frac{1}{2}\blu{\bar\chi}\gamma_T^\mu\gamma^-\orn{S_n^\dagger S_\nbar}
    \grn{\fsl{\mathcal{A}}_T}\orn{\frac{1}{\img\partial^-}\Psi_\nbar}
    -\frac{1}{2}\blu{\bar\chi}\gamma_T^\mu\gamma^-\orn{S_n^\dagger S_\nbar}
    \grn{\fsl{\mathcal{A}}_T}\orn{S_\nbar^\dagger S_n}\blu{\frac{1}{\img\partial^-}\chi}
    \\ \nn
    &\quad\qquad
    +(\blu{n}\leftrightarrow\grn{\nbar})+\text{h.c.}\biggr\}
    +\mathcal{O}\bigl(g^2,\lambda^{7/2}\bigr)\,.
\end{align}

The above result is now manifestly gauge invariant, but it is still incomplete as higher-order terms in the coupling can contribute at the same order in the power counting. To include these higher-order corrections, we provide each operator with a Wilson coefficient. These higher-order corrections can result in convolutions in the lightcone position arguments in the fields, even for soft fields, due to internal hard-collinear lines. The general way of writing the effective current operator is then,
\begin{align}
    J_\text{eff}^\mu(0)
    &=
    \sum_i\Gamma_i^{\mu,\alpha_1\dotsm\,,\beta_1\dotsm\,,\gamma_1\dotsm}    
    \int\blu{\df y_1^+\,\dotsm}\,\grn{\df y_1^-\,\dotsm}\,
    \orn{\df z_1^+\,\dotsm\df z_1^-\dotsm}\\ \nn
    &\quad\times
    C_i\bigl(\{y_1^+,\dotso\},\{z_1^+,\dotso\},\{z_1^-,\dotso\},\{y_1^-,\dotso\}\bigr)
    \\
    &\quad\times
    \blu{\mathcal{O}_i^{\alpha_1\dotsm}(y_1^+\nbar+\dotso)}\,
    \orn{\mathcal{O}_i^{\beta_1\dotsm}(z_1^+ \nbar + z_1^- n + \dotso)}\,
    \grn{\mathcal{O}_i^{\gamma_1\dotsm}(y_1^- n + \dotso)}\,,
    \nonumber
\end{align}
where the sum over $i$ runs over all terms that appear in eq.~\eqref{eq:NLPcurrentSCET} and $\alpha$, $\beta$ and $\gamma$ are placeholders for spin, color and Lorentz indices. We refrain from explicitly writing out all convolutions at this stage, as the upcoming subsection will demonstrate that the number of independent operators can be substantially reduced.

\subsection{Constraints from symmetries}\label{sec:SymmetryConstraints}

Not all Wilson coefficients that we introduced above are independent of one another. Constraints from reparametrization invariance, current conservation and charge, parity, and time reversal can greatly reduce the number of coefficients that must be determined. As we will show in this section, only five independent Wilson coefficients remain after applying these symmetries.

\subsubsection*{C, P, T conjugation}

We start by discussing the constraints coming from charge, parity, and time conjugation. These transformations have a somewhat trivial effect, as they only relate the Wilson coefficients operators that are related by swapping $\blu{n}\leftrightarrow\grn{\nbar}$ and by hermitian conjugation. The only meaningful result that can be concluded from these transformations is that the Wilson coefficients of operators that are related by time conjugation are related by complex conjugation. This constraint, however, can also be found by simply demanding that the hadronic tensor in eq.~\eqref{eq:HadronicTensorDefinitionSIDIS} satisfies $\mathcal{W}_{\mu\nu}^\dagger=\mathcal{W}_{\nu\mu}$. In the following, we analyze other symmetries which relate some of our operators to each other.

\subsubsection*{Current conservation}

Next, we consider current conservation and we demand that it holds on the operator level, up to power corrections,
\begin{align}\label{eq:CurrentConservation}
    \img\partial_\mu J_\text{eff}^\mu=\mathcal{O}(\lambda^{\frac{7}{2}})\,.
\end{align}
This only results in constraints for a subset of the operators that appear in eq.~\eqref{eq:NLPcurrentSCET}, because when $\img\partial_\mu$ is contracted with a transverse vector $v_T^\mu$ it automatically results in a power suppression by $\lambda$. The only terms that result in a constraint are when $\img\partial_T^\mu$ acts on a leading-power operator or when $\img\partial^\pm$ acts on a next-to-leading-power operator. To simplify the analysis below, we ignore the convolutions in the lightcone momentum fractions. This is allowed, as the derivative in eq.~\eqref{eq:CurrentConservation} is taken with respect to the global position and therefore doesn't act on the convolution variables.

Acting with the derivative on the leading power operator, which corresponds to the first line in eq.~\eqref{eq:NLPcurrentSCET}, we find
\begin{align*}
    \img\partial_\mu\bigl(\blu{\bar\chi}\gamma_T^\mu\orn{S_n^\dagger S_\nbar}\grn{\chi}\bigr)
    &=
    \blu{\img\partial_T^\rho\bar\chi\gamma_T^\rho}\orn{S_n^\dagger S_\nbar}\grn{\chi}
    +\blu{\bar\chi}\orn{S_n^\dagger S_\nbar}\grn{\img\fsl{\partial}_T\chi}
    -\blu{\bar\chi}\orn{\fsl{\mathcal{A}}_{n,T} S_n^\dagger S_\nbar}\grn{\chi}
    +\blu{\bar\chi}\orn{S_n^\dagger S_\nbar \fsl{\mathcal{A}}_{\nbar,T}}\grn{\chi}\,,
\end{align*}
up to corrections of order $\lambda^4$. Here, we have made use of the following identity for transverse derivatives of Wilson lines,
\begin{align}\label{eq:CurrentConstraint1}
    \orn{\bigl(\img\partial_T^\rho S_n^\dagger\bigr)S_\nbar}
    +\orn{S_n^\dagger\bigl(\img\partial_T^\rho S_\nbar\bigr)}
    &=
    -\orn{\mathcal{A}_{n,T}^\rho S_n^\dagger S_\nbar}
    +\orn{S_n^\dagger S_\nbar \mathcal{A}_{\nbar,T}^\rho}\,.
\end{align}
For the next-to-leading power terms in eq.~\eqref{eq:NLPcurrentSCET} that do not receive further power suppression when acted upon with $\partial^\mu$ we find
\begin{align}\label{eq:CurrentConstraint2}
    \img\partial_\mu\biggl(-n^\mu\blu{\bar\chi}\orn{S_n^\dagger S_\nbar}
        \grn{\frac{\img\fsl{\partial}_T}{\img\partial^+}\chi}
        +(\blu{n}\leftrightarrow\grn{\nbar})+\text{h.c.}\biggr)
    &=
    -\blu{\bar\chi}\orn{S_n^\dagger S_\nbar}\grn{\img\fsl{\partial}_T\chi}
    +(\blu{n}\leftrightarrow\grn{\nbar})-\text{h.c.}\,,
    \\\nonumber
    \img\partial_\mu\biggl(-n^\mu\blu{\bar\chi}\orn{S_n^\dagger S_\nbar}
        \grn{\frac{1}{\img\partial^+}\bigl(\fsl{\mathcal{A}}_T\chi\bigr)}
    +(\blu{n}\leftrightarrow\grn{\nbar})+\text{h.c.}\biggr)
    &=
    -\blu{\bar\chi}\orn{S_n^\dagger S_\nbar}\grn{\fsl{\mathcal{A}}_T\chi}
    +(\blu{n}\leftrightarrow\grn{\nbar})-\text{h.c.}\,,
    \\\nonumber
    \img\partial_\mu\biggl(-n^\mu\blu{\bar\chi\fsl{\mathcal{A}}_T}
        \orn{S_n^\dagger S_\nbar}\grn{\frac{1}{\img\partial^+}\chi}
    +(\blu{n}\leftrightarrow\grn{\nbar})+\text{h.c.}\biggr)
    &=
    -\blu{\bar\chi\fsl{\mathcal{A}}_T}
        \orn{S_n^\dagger S_\nbar}\grn{\chi}
    +(\blu{n}\leftrightarrow\grn{\nbar})-\text{h.c.}\,,
    \\\nonumber
    \img\partial_\mu\biggl(-n^\mu\blu{\bar\chi}\orn{S_n^\dagger S_\nbar}
        \orn{\fsl{\mathcal{A}}_{T,\nbar}}\grn{\frac{1}{\img\partial^+}\chi}
    +(\blu{n}\leftrightarrow\grn{\nbar})+\text{h.c.}\biggr)
    &=
    -\blu{\bar\chi}\orn{S_n^\dagger S_\nbar}
        \orn{\fsl{\mathcal{A}}_{T,\nbar}}\grn{\chi}
    +(\blu{n}\leftrightarrow\grn{\nbar})-\text{h.c.}\,.
\end{align}
Comparing the different terms in eqs.~\eqref{eq:CurrentConstraint1} and \eqref{eq:CurrentConstraint2}, we find that current conservation implies that the Wilson coefficients of the operators
\begin{align*}
    &-n^\mu\blu{\bar\chi}\orn{S_n^\dagger S_\nbar}
        \grn{\frac{\img\fsl{\partial}_T}{\img\partial^+}\chi}\,,&
    &-n^\mu\blu{\bar\chi}\orn{S_n^\dagger S_\nbar}
        \orn{\fsl{\mathcal{A}}_{T,\nbar}}\grn{\frac{1}{\img\partial^+}\chi}\,,
\end{align*}
and the operators that are obtained from these by swapping $\blu{n}\leftrightarrow\grn{\nbar}$ or taking the hermitian conjugate, are all related to the Wilson coefficient of the leading-power operator. As we will show in sec.~\ref{sec:HadronicTensor}, these contributions combine into kinematic power corrections~\cite{Vladimirov:2023aot}: NLP terms that share the same operator structure and hard coefficient as the LP contribution.

Additionally, we find that the Wilson coefficients of NLP operators involving both $\blu{\mathcal{A}_T}$ and $\grn{\mathcal{A}_T}$ are related among themselves but are independent of the LP coefficient. These define the genuine power corrections, which introduce new operator structures and therefore new twist-3 TMD distributions.

\subsubsection*{RPI constraints}

With the introduction of the background fields corresponding to collinear, soft, and anti-collinear modes, we have implicitly broken Lorentz invariance by introducing reference vectors $n$ and $\nbar$. However, this Lorentz invariance is restored order-by-order in the power counting. This restoration of Lorentz invariance happens only if effective operators are reparameterization invariant (RPI)~\cite{Manohar:2002fd,Marcantonini:2008qn}. RPI transformation describes a small change in the choice of the reference vectors $n$ and $\nbar$. In summary, there are three RPI transformations to consider,
\begin{alignat*}{3}
\text{I:}&\qquad n \to n+\Delta_T\,,&\qquad &\nbar \to\, \nbar\,,\\
\text{II:}&\qquad n \to n\,&\qquad  &\nbar \to\, \nbar+\bar\Delta_T\,,\\
\text{III:}&\qquad n \to n\,e^\alpha\,,&\quad\,\qquad &\nbar \to\, \nbar\,e^{-\alpha}\,.
\end{alignat*}
Here, the RPI parameters $\Delta_T$ and $\bar\Delta_T$ are of order $\lambda$, while the parameter $\alpha$ is order one not to break the power counting. In this subsection, we will consider the constraints on the Wilson coefficients in the effective operator that come from demanding that the effective current is RPI up to higher-order power corrections. 

In our case, the transformations type I and II are completely equivalent as the current operator is symmetric under $n\leftrightarrow\nbar$. Also, all operators that appear in eq.~\eqref{eq:NLPcurrentSCET}, are already invariant under the type-III transformations, and so we only need to consider the type-I transformations. Under type-I transformations, the components of a vector transform as
\begin{align}
    &\text{I:}\quad v^+\to v^+ + \Delta_T\cdot v_T\,,&v^-\to v^-\,,&
    &v_T^\rho\to v_T^\rho - (\Delta_T\cdot v_T)\nbar^\rho 
    - v^-\Delta_T^\rho\,.
\end{align}

To work out the transformations of the operators that appear in the effective current, we use the transformation properties of the SCET building blocks. Under type-I transformations, collinear Wilson lines are invariant to all orders in the power counting, while anti-collinear Wilson lines are invariant up to power corrections,
\begin{align}
    &\delta_I\blu{W}=0\,,&&\delta_I\grn{W}=\mathcal{O}(\lambda^2)\,.
\end{align}
Soft Wilson lines, on the other hand, do possess a non-trivial transformation given by
\begin{align}
    \delta_I \orn{S_n}&=-\Delta_T^\rho
    \orn{S_n\frac{1}{\img\partial^+}\mathcal{A}_{nT}^\rho}
    +\mathcal{O}(\lambda^2)\,,
    \\
    \delta_I \orn{S_\nbar}&=0\,.
\end{align}
With these transformation properties, we can derive the transformation of the building block operators. At this order in the power counting, we only need to consider the collinear and anti-collinear quark fields, as all other operators already carry a power suppression by themselves. For the collinear and anti-collinear quark operators, we have
\begin{align}
    &\delta_I\blu{\chi}=\frac{\fsl{\Delta}_T\gamma^-}{2}\blu{\chi}\,,&
    &\delta_I\grn{\chi}=\mathcal{O}(\lambda^3)\,.
\end{align}

Using these transformations, we can relate some of the Wilson coefficients of the sub-leading-power operators to the leading-power one. To do this, we demand that the transformations cancel between the different operators. The LP operator transforms as
\begin{align}
    \delta_I\biggl(\blu{\bar\chi}\gamma^\mu\orn{S_n^\dagger S_\nbar}\grn{\chi}
    +(\blu{n}\leftrightarrow\grn{\nbar})\biggr)
    &=
    -\nbar^\mu\blu{\bar\chi}\fsl{\Delta}_T\orn{S_n^\dagger S_\nbar}\grn{\chi}
    +\Delta_T^\rho\blu{\bar\chi}\gamma_T^\mu\orn{
    \frac{1}{\img\partial^+}\mathcal{A}_{nT}^\rho S_n^\dagger S_\nbar}\grn{\chi}
    \\ \nn
    &\quad
    -\nbar^\mu\grn{\bar\chi}\orn{S_\nbar^\dagger S_n}\fsl{\Delta}_T\blu{\chi}
    -\Delta_T^\rho
        \grn{\bar\chi}\gamma_T^\mu\orn{S_\nbar^\dagger S_n \frac{1}{\img\partial^+}\mathcal{A}_{nT}^\rho}\blu{\chi}\,,
\end{align}
up to sub-leading power terms. For the NLP operators, the only operators for which the RPI type-I transformations have an effect at this order in the power counting is where a transverse derivative acts on a collinear or anti-collinear field. This is because while the parameter $\Delta_T$ introduces another power of $\lambda$, changing $\partial_T$ to $\partial^\pm$ removes one power of $\lambda$. The relevant transformations are
\begin{align}
    \delta_I\biggl(-n^\mu\blu{\bar\chi}\orn{S_n^\dagger S_\nbar}
    \grn{\frac{\img\fsl{\partial}_T}{\img\partial^+}\chi}
    +(\blu{n}\leftrightarrow\grn{\nbar})+\text{h.c.}\biggr)
    =&\,
    \nbar^\mu\blu{\bar\chi}\fsl{\Delta}_T\orn{S_n^\dagger S_\nbar}\grn{\chi}
    \\
    &\nn+\nbar^\mu\grn{\bar\chi}\orn{S_\nbar^\dagger S_n}\fsl{\Delta}_T\blu{\chi}\,,
    \\
    \delta_I\biggl(-\blu{\bar\chi}\gamma_T^\mu\orn{S_n^\dagger S_\nbar}
    \orn{\frac{1}{\img\partial^-}\mathcal{A}_{T,\nbar}^\rho}
    \grn{\frac{\img\partial_T^{\rho}}{\img\partial^+}\chi}
    +(\blu{n}\leftrightarrow\grn{\nbar})+\text{h.c.}\biggr)
    =&
    -\Delta_T^\rho\blu{\bar\chi}
    \gamma_T^\mu\orn{\frac{1}{\img\partial^+}\mathcal{A}_{T,n}^\rho}
    \orn{S_n^\dagger S_\nbar}\grn{\chi}\\
    &\nn
    +\Delta_T^\rho\grn{\bar\chi}\gamma_T^\mu\orn{S_\nbar^\dagger S_n}
    \orn{\frac{1}{\img\partial^+}\mathcal{A}_{T,n}^\rho}\blu{\chi}\,.
\end{align}
For the transformations of these operators to exactly cancel that of the leading-power operator, their Wilson coefficients must be the same.

\subsection{Final result and cross check}\label{sec:NLPresult}

We now combine all the constraints on the Wilson coefficients to obtain the full next-to-leading power result for the SCET-II effective current operator. First, we split up the current operator according to the power counting,
\begin{align}\label{eq:FinalEffCurrent}
    J^\mu_{\text{eff}}(0)
    &=
    \bigl[J_\text{eff}^\mu(0)\bigr]^{(2)}
    +\bigl[J_\text{eff}^\mu(0)\bigr]^{(2.5)}
    +\bigl[J_\text{eff}^\mu(0)\bigr]^{(3)}\,
    +\mathcal{O}(\lambda^{7/2})\,,
\end{align}
where
\begin{align}
    \bigl[J_\text{eff}^\mu(0)\bigr]^{(k)}\sim\lambda^k\,.
\end{align}
Our final results then read 
\begin{align}
    \label{eq:EffCurrentLP}\bigl[J_\text{eff}^\mu(0)\bigr]^{(2)}
    &=
    (\gamma_T^\mu)_{\alpha\beta}
    \int\df y^+\,\df y^-\,C_1(y^+,y^-)\,
    \blu{\bar\chi_\alpha(y^+ \nbar)}\,
    \orn{S_n^\dagger S_\nbar}\,
    \grn{\chi_\beta(y^- n)}\,
    +\text{h.c.}
    \\[2ex]
    \label{eq:EffCurrentNLP2.5}\bigl[J_\text{eff}^\mu(0)\bigr]^{(2.5)}
    &=
    -\frac{1}{2}(\gamma_T^\mu\gamma^-\gamma_T^\rho)_{\alpha\beta}
    \int\df y^+\,\df z^+\, \df y^-\, C_4(y^+,z^+,y^-)\,
    \\
    &\qquad\times
    \blu{\bar\chi_\alpha(y^+ \nbar)}\,
    \orn{S_n^\dagger S_\nbar}\,
    \grn{\mathcal{A}_T^\rho(y^- n)}\,
    \orn{\frac{1}{\img\partial^-}\Psi_{\nbar,\beta}(z^+ \nbar)}
    +(\blu{n}\leftrightarrow\grn{\nbar})+\text{h.c.}\nonumber
    \\[2ex]
    \label{eq:EffCurrentNLP3}\bigl[J_\text{eff}^\mu(0)\bigr]^{(3)}
    &=
    -n^\mu\int\df y^+\,\df y^-\, C_1(y^+,y^-)\,
    \blu{\bar\chi(y^+ \nbar)}\,
    \orn{S_n^\dagger S_\nbar}\,
    \grn{\frac{\img\fsl{\partial}_T}{\img\partial^+}\chi(y^- n)}
    \\
    &\quad
    -\int\df y^+\,\df y^-\, C_1(y^+,y^-)\,
    \blu{\bar\chi(y^+ \nbar)}\,
    \orn{S_n^\dagger S_\nbar \fsl{\mathcal{A}}_{T,\nbar}}\,
    \grn{\frac{n^\mu}{\img\partial^+}\chi(y^- n)}\,
    \nonumber
    \\
    &\quad
    -(\gamma_T^\mu)_{\alpha\beta}\int\df y^+\,\df y^-\, C_1(y^+,y^-)\,
    \blu{\bar\chi_\alpha(y^+ \nbar)}\,
    \orn{S_n^\dagger S_\nbar \frac{1}{\img\partial^-}\mathcal{A}_{T,\nbar}^\rho}\,
    \grn{\frac{\img\partial_T^{\rho}}{\img\partial^+}\chi_\beta(y^- n)}
    \nonumber
    \\
     &\quad
    +\int\df y_1^+\,\df y_2^+\,\df y^-\,
    C_2\bigl(\{y_1^+,y_2^+\},y^-\bigr)\,
    \nonumber\\
    &\qquad\times
    \biggl(\blu{\frac{\nbar^\mu}{\img\partial^-}}
        -\grn{\frac{n^\mu}{\img\partial^+}}\biggr)
    \blu{\bar\chi(y_1^+ \nbar)\fsl{\mathcal{A}}_T(y_2^+ \nbar)}\,
    \orn{S_n^\dagger S_\nbar}\,
    \grn{\chi(y^- n)}
    \nonumber
    \\
    &\quad
    -\frac{1}{2}(\gamma_T^\mu\gamma_T^\rho\gamma_T^\sigma)_{\alpha\beta}
    \int\df y^+\,\df y^-\,\df z^+\,C_3(y^+,z^+,y^-)\,
    \nonumber
    \\
    &\qquad\times
    \blu{\bar\chi_\alpha(y^+ \nbar)}\,
    \orn{S_n^\dagger S_\nbar \frac{\img\partial_T^\rho}{\img\partial^-}\mathcal{A}_{T,\nbar}^\sigma(z^+ \nbar)}\,
    \grn{\frac{1}{\img\partial^+}\chi_\beta(y^- n)}
    \nonumber
    \\
    &\quad
    -\frac{1}{2}(\gamma_T^\mu\gamma^-\gamma_T^\rho)_{\alpha\beta}
    \int\df y_1^+\,\df y_2^+\, \df y^-\,
    C_5\bigl(\{y_1^+,y_2^+\},y^-\bigr)\,
    \nonumber
    \\
    &\qquad\times
    \blu{\bar\chi_\alpha(y_1^+ \nbar)}\,
    \orn{S_n^\dagger S_\nbar}\,
    \grn{\mathcal{A}_T^\rho(y^-)}\,
    \orn{S_\nbar^\dagger S_n}\,
    \blu{\frac{1}{\img\partial^-}\chi_\beta(y_2^+ \nbar)}
    \nonumber
    \\
    &\quad
    +(\blu{n}\leftrightarrow\grn{\nbar})+\text{h.c.}\nonumber
\end{align}
Let us emphasize that colored derivatives only act on the fields of the corresponding mode. The terms in eq.~\eqref{eq:EffCurrentNLP2.5} are identified as the soft quark contribution. We refer to the first line of eq.~\eqref{eq:EffCurrentNLP3} as the kinematic power correction, and to the fourth line of the same equation, associated with the Wilson coefficient $C_2$, as the genuine higher-twist correction. The last line of eq.~\eqref{eq:EffCurrentNLP3}, involving $C_5$, corresponds to the collinear quark–anti-quark contribution. The remaining terms in eq.~\eqref{eq:EffCurrentNLP3}, which contain insertions of soft gluon fields, are grouped together as soft gluon contributions. These can be further classified according to whether the inverse derivative acts on the soft gluon field itself, or on a collinear or anti-collinear field.

\subsection{Comparison with the literature}\label{sec:ComparisonCurrentNLP}

This is not the first work related to TMD factorization where the electromagnetic current operator is studied at next-to-leading power. In this subsection, we compare our result for the effective current operator to those of refs.~\cite{Ebert:2021jhy,Vladimirov:2021hdn}. Before doing so, we note that there are some substantial differences between our works: First, since there are no soft background fields in ref.~\cite{Vladimirov:2021hdn}, there are no soft Wilson lines present in their expression for the effective current operator. Second, ref.~\cite{Ebert:2021jhy} applies the label-momentum formulation of SCET. While it is possible to translate between the position-space formalism and the label formalism, it complicates a direct comparison between our results.

First, let us focus on the terms where we find agreement between both refs.~\cite{Ebert:2021jhy,Vladimirov:2021hdn} and our result. These are the kinematic power correction, the genuine higher-twist correction, and the collinear quark-anti-quark contribution, which correspond to operators of the form $\blu{\bar\chi}\grn{\partial_T\chi}$, $\blu{\bar\chi}\grn{\mathcal{A}_T\chi}$, and $\blu{\bar\chi}\grn{\mathcal{A}_T}\blu{\chi}$ respectively. Discrepancies in notation mainly come from the absence of soft Wilson lines and an alternate parameterization of the convolution in ref.~\cite{Vladimirov:2021hdn}, and the use of the label-momentum formulation of SCET in ref.~\cite{Ebert:2021jhy}. After accounting for these differences in notation, the contributions of these operators to the effective current are identical. Moreover, we also find agreement between the one-loop results for the Wilson coefficient corresponding to the genuine higher-twist correction, which is denoted by $C_2$ in our work and in ref.~\cite{Vladimirov:2021hdn} and by $C^{(1)}$ in ref.~\cite{Ebert:2021jhy}.

Next, let us focus on the soft power correction in the second line of eq.~\eqref{eq:EffCurrentNLP3}, where we find an apparent disagreement between ref.~\cite{Ebert:2021jhy} and our work. This correction corresponds to operators of the form $\blu{\bar\chi}\orn{\mathcal{A}_T}\grn{\chi}$. While this operator appears in both our work and theirs, we find different results for the Wilson coefficient that accompanies this operator. To elaborate, let us first note that the most general way of writing the contribution of this operator is
\begin{align}
    J_\text{eff}^\mu(0)\supset
    \int\df y^+\,\df z^+\,\df y^-\,
    C_{\blu{\qbar}\orn{g}\grn{q}}(y^+,z^+,y^-)\,
    \blu{\bar\chi(y^+ \nbar)}\,
    \orn{S_n^\dagger S_\nbar \fsl{\mathcal{A}}_{T,\nbar}(z^+\bar n)}\,
    \grn{\frac{n^\mu}{\img\partial^+}\chi(y^- n)}\,.
\end{align}
In our work, specifically in sec.~\ref{sec:SymmetryConstraints}, we concluded from current conservation that the above Wilson coefficient is constrained to be equal to
\begin{align}\label{eq:DifferentWilson1}
    C_{\blu{\qbar}\orn{g}\grn{q}}(y^+,z^+,y^-)
    &\stackrel{\text{this work}}{=}
    \delta(z^+)\,C_1(y^+,y^-)\,.
\end{align}
In contrast, in ref.~\cite{Ebert:2021jhy} it is argued on the basis of RPI that the Wilson coefficient of the above operator should have an additional term that involves the Wilson coefficient of the genuine higher-twist correction. Their expression for the Wilson coefficient, translated to a position-space convolution to make it easier to compare with our own result, reads
\begin{align}\label{eq:DifferentWilson2}
    C_{\blu{\qbar}\orn{g}\grn{q}}(y^+,z^+,y^-)
    &\stackrel{\text{ref.~\cite{Ebert:2021jhy}}}{=}
    \delta(z^+)\,C_1(y^+,y^-)
    +
    \int\df \tilde{y}^-\,C_2(y^+,\{y^-\!-\tilde{y}^-,\tilde{y}^-\})\,
    \mathcal{K}(z^+,y^- - \tilde{y}^-)\,,
\end{align}
where $\mathcal{K}$ is referred to as a hard-collinear matching coefficient, and is not further constrained by any symmetries. This additional hard-collinear contribution comes from the two-step matching, first onto \textcolor{Cyan}{hard-collinear} and \textcolor{LimeGreen}{hard-anti-collinear} fields: $\textcolor{Cyan}{\bar\chi}\textcolor{LimeGreen}{\mathcal{A}\chi}$, and then onto collinear, soft, and anti-collinear: $\blu{\bar\chi}\orn{\mathcal{A}_T}\grn{\chi}$. We present a possible resolution for this apparent disagreement in sec.~\ref{sec:HadronicTensorComparison}.

All remaining terms involve contributions from operators that are absent in both ref.~\cite{Vladimirov:2021hdn} and ref.~\cite{Ebert:2021jhy}. For ref.~\cite{Vladimirov:2021hdn}, this is expected as no soft background fields are introduced. For ref.~\cite{Ebert:2021jhy}, which does include soft dynamic degrees of freedom, these operators are also absent. This is likely due to the fact that inverse derivatives of soft fields are not part of the operator basis in the label formalism of SCET. For one of the operators however, we found that its Wilson coefficient is constrained by the RPI transformations of soft Wilson lines. While ref.~\cite{Ebert:2021jhy} does study the constraints from RPI, they do not discuss the RPI transformation of soft Wilson lines. The absence of these operators, however, will not be a problem, as we will show in the next section that the resulting soft matrix elements vanish identically.

To motivate the presence of the sub-leading soft operators, we compare our expression for the effective current operator to matrix elements in full QCD in appropriate limits. Since all operators involve fields in the combination $\bar\chi\mathcal{A}_T\chi$ we use $\qbar gq$ external states. The full-QCD matrix element reads
\begin{align}
    \bra{q(p)}J^\mu\ket{g(k) q(l)}
    &=
    g\,\bar{u}(p)
    \biggl[
     \gamma^\nu\frac{\fsl{p}-\fsl{k}}{2p\cdot k}\gamma^\mu
    -\gamma^\mu\frac{\fsl{l}+\fsl{k}}{2l\cdot k}\gamma^\nu
    \biggr]\epsilon_\nu(k) u(l)\,.
\end{align}
Expanded in the soft-gluon and soft-quark limit, while keeping the total momentum hard, we obtain the following expansion up to next-to-leading power,
\begin{align} \label{eq:jnlp}
    \bra{\blu{q(p)}}J^\mu\ket{\orn{g(k)}\grn{q(l)}}
    &=
    -g\,\blu{\bar{u}(p)}
    \biggl[
     \gamma_T^\mu\frac{\nbar^\nu}{\orn{k^-}}
    -\gamma_T^\mu\frac{n^\nu}{\orn{k^+}}
    \\\nonumber
    &\qquad\quad
    -\gamma_T^\mu\biggl(\frac{\gamma_T^\nu\orn{\fsl{k}_T}}{\orn{k^-}}+2n^\nu\biggr)
        \frac{1}{\grn{l^+}}
    +2\gamma_T^\mu\biggl(\frac{g_T^{\rho\nu}}{\orn{k^-}}
            -\frac{\orn{k_T^\rho}\nbar^\nu}{(\orn{k^-})^2}\biggr)
        \frac{\grn{l_T^\rho}}{\grn{l}^+}
    \\\nonumber
    &\qquad\quad
    -\frac{1}{\blu{p^-}}\biggl(\frac{\orn{\fsl{k}_T}\gamma_T^\nu}{\orn{k^+}}
        +2\nbar^\nu\biggr)\gamma_T^\mu
    -2\gamma_T^\mu\frac{\blu{p_T^\rho}}{\blu{p^-}}\biggl(\frac{g_T^{\rho\nu}}{\orn{k^+}}
        -\frac{\orn{k_T^\rho}n^\nu}{(\orn{k^+})^2}\biggr)
    \\\nonumber
    &\qquad\quad
    +2n^\mu\frac{1}{\blu{p^-}}\biggl(\gamma_T^\nu
        -\frac{\orn{\fsl{k}_T}n^\nu}{\orn{k^+}}\biggr) 
    +2\nbar^\mu\biggl(\gamma_T^\nu
        -\nbar^\nu\frac{\orn{\fsl{k}_T}}{\orn{k^-}}\biggr)\frac{1}{\grn{l^+}}
    \biggr]\orn{\epsilon_\nu(k)}\grn{u(l)}\,,
    \\[2ex]
    \bra{\blu{q(p)}}J^\mu\ket{\grn{g(k)}\orn{q(l)}}
    &=
    -g\,\blu{\bar{u}(p)}
    \biggl[
    \frac{\gamma_T^\mu\gamma^-\gamma_T^\nu}{2\orn{l^-}}
    \biggr]\grn{\epsilon_\nu(k)}\orn{u(l)}\,.
\end{align}
This can be directly checked against the result for the effective current operator. We find full agreement between the power-expanded full-QCD calculation and the result that is obtained by using the effective current operator of eq.~\eqref{eq:FinalEffCurrent}. 
Each term in eq.~\eqref{eq:jnlp} arises from a single soft operator. Note that for the soft operator associated with $C_3$, the following identity was used to check agreement with the above expression,
\begin{align}
    (\gamma_T^\mu\gamma_T^\rho\gamma_T^\sigma)
    \biggl(\frac{k_T^\rho}{k^-}g_T^{\sigma\nu}
        -\frac{k_T^\rho}{k^-}\frac{k_T^\sigma}{k^-}\nbar^\nu\biggr)
    \epsilon_\nu(k)
    =
    -\gamma_T^\mu
    \biggl(\frac{\gamma_T^\nu\fsl{k}_T}{k^-}+2n^\nu\biggr)
    \epsilon_\nu(k)\,.
\end{align}

\section{The hadronic tensor}\label{sec:HadronicTensor}

In the previous section, we derived an expression for the effective current operator that holds up to next-to-leading power. In this section, we will derive the NLP hadronic tensor for SIDIS. To derive the hadronic tensor, we start by inserting the expression for the effective current operator into the definition of the hadronic tensor. This results in a large number of different terms, which we summarize in section \ref{sec:UnsubtractedHadronicTensor}. Next, we address the presence of an overlap between the modes in section \ref{sec:OverlapSubtraction}$\,$: We first discuss our subtraction method in section \ref{sec:SubtractionMethod}, then summarize the distinct terms in the subtracted hadronic tensor in \ref{sec:OverlapSubtractionNLP}. In section \ref{sec:PowerMixing}, we show that these subtractions can mix different orders in the power counting. We demonstrate the cancellation of endpoint divergences in section \ref{sec:CancellationEndpointDivergences}. Finally, in section \ref{sec:SubtractedHadronicTensor} we define physical TMD parton distribution functions (TMDPDFs) and TMD fragmentation functions (TMDFFs), with open spin indices, and write the hadronic tensor in terms of these distributions. When we apply this to SIDIS in sec.~\ref{sec:JetSIDISFactorizationNLP}, we will replace the TMDFFs by jet functions.

\subsection{The unsubtracted hadronic tensor}\label{sec:UnsubtractedHadronicTensor}

With the NLP effective current operator at hand, we can now look at the object of interest for TMD factorization: the hadronic tensor. The hadronic tensor depends on the specific process under consideration; in what follows we focus exclusively on the SIDIS case. The corresponding expressions for other processes can be obtained by appropriately modifying the external states and measurements. As discussed in sec.~\ref{sec:SoftBFM}, the hadronic tensor for SIDIS is given by
\begin{align}\label{eq:HadronicTensorDefinitionSIDIS}
    W^{\mu\nu}
    &=
    \int\frac{\df^4 b}{(2\pi)^4}\,e^{+\img q\cdot b}\,
    \bra{\grn{P}} J_\text{eff}^\mu(b) \ket{\blu{p},X}
    \bra{\blu{p},X} J_\text{eff}^\nu(0) \ket{\grn{P}}\,,
\end{align}
$P$ and $p$ denote the momenta of the incoming hadron and outgoing hadron or jet, respectively. The colors in the external states encode the scaling of the momenta. To organize our results, we expand the hadronic tensor in power counting as
\begin{align}
    \bigl[\mathcal{W}^{\mu\nu}\bigr]^{(\kappa,\eta)}
    &=
    \int\frac{\df^4 b}{(2\pi)^4}\,e^{+\img q\cdot b}\,
    \bra{\grn{P}} \bigl[J_\text{eff}^\mu(b)\bigr]^{(\kappa)} \ket{\blu{p},X}
    \bra{\blu{p},X} \bigl[J_\text{eff}^\nu(0)\bigr]^{(\eta)} \ket{\grn{P}}\,,
\end{align}
where the two superscripts $(\kappa, \eta)$ indicate the respective orders in the $\lambda$ expansion of the current insertions, reflecting the fact that the hadronic tensor involves two such insertions at potentially different powers.

The sub-leading operators in the effective current sometimes involve an inverse derivative acting on a field. When the inverse derivative acts on all fields inside a matrix element, we can replace the inverse derivative with the momenta of the external states by virtue of the following identity,
\begin{align}\label{eq:InverseDerivativeIdentity}
    \bra{P_a}\frac{1}{\img\partial^+}\mathcal{O}(x)\ket{P_b}=
    \frac{1}{P_b^+ - P_a^+}\bra{P_a}\mathcal{O}(x)\ket{P_b}\,.
\end{align}
This identity can be derived from the definition of the inverse derivative in eq.~\eqref{eq:PlusInverseDerivative} and by translating the position argument of $O(x)$ using the momentum operator. In constructing the hadronic tensor, eq.~\eqref{eq:InverseDerivativeIdentity} allows us to replace inverse derivatives acting on collinear fields by explicit factors of $q^+$ and $q^-$.

The large number of operators in the effective current and the fact that we have two insertions of the current operator result in a large number of different terms in the hadronic tensor. Many contributions, however, vanish immediately due to violation of the fermion number, as was discussed for the leading power in section \ref{sec:NLPwithSoftBckg}. At next-to-leading power, we encounter, for example, a vacuum matrix element with a single quark field,
\begin{align}
    \orn{\bra{0}\bigl[S_n^\dagger S_\nbar\bigr]_{ij}
    \biggl[\frac{1}{\img\partial^-}\Psi_{\nbar}\biggr]_{k,\alpha}
    \ket{X}\bra{X}\bigl[S_\nbar^\dagger S_n\bigr]_{lm}\ket{0}}=0,
\end{align}
which vanishes. Similarly, we encounter collinear matrix elements with a single quark field and a single gluon field, which vanish for the same reason. To keep our results manageable, we discard all such contributions from the start.

Even without these terms, a large number of non-vanishing contributions remain. However, many of them are related to each other through some form of charge conjugation. Therefore, to reduce the amount of contributions to the hadronic tensor that we write out explicitly, we omit all terms that are related by charge conjugation. To consistently make this separation we only consider the terms in the hadronic tensor where the current operator at position $b$ contributes an operator of the form $\grn{\bar\chi}\blu{\chi}$ and where the current at position $0$ contributes an operator of the form $\blu{\bar\chi}\grn{\chi}$. This means that for each contribution labeled $X$ that we explicitly write out, there is another contribution labeled $\bar{X}$ that is related by complex conjugation. Schematically we write 
\begin{align}
    &X:
    &J(b)\,\to\,\grn{\bar\chi}\blu{\chi}&
    &J(0)\,\to\,\blu{\bar\chi}\grn{\chi}\,,&
    \\
    &\bar{X}:
    &J(b)\,\to\,\blu{\bar\chi}\grn{\chi}&
    &J(0)\,\to\,\grn{\bar\chi}\blu{\chi}\,,&
    \nonumber
\end{align}
Additionally, we only consider insertions of the sub-leading power current at the position $b$. Those with insertions at $0$ can be obtained straightforwardly from those with insertions at $b$ by swapping the operator content between the amplitude and the conjugate amplitude.

Below, we list all remaining contributions to the hadronic tensor, labeled from $A$ to $H$, with subscripts $n$ or $\bar n$ indicating whether the additional fields beyond the leading-power term reside in the collinear or anti-collinear sector, respectively. Note that these results are for the \emph{unsubtracted} hadronic tensor, as we will address the subtraction of mode overlap, crucial for obtaining physical results free of rapidity divergences, in the next section.

\subsubsection*{LP contribution}

We begin with the contributions from the LP effective current operator to the hadronic tensor. The product of two LP currents gives the first contribution,
\begin{align}\label{eq:HadronicTensorSIDIS_LP}
    \bigl[\mathcal{W}^{\mu\nu}\bigr]_{A}^{(2,2)}
    &=
    \int\frac{\df^2 b}{(2\pi)^2}\,e^{+\img q_T \cdot b_T}\,
    \int\frac{\df b^+}{2\pi}\,e^{+\img q^- b^+}\,
    \int\frac{\df b^-}{2\pi}\,e^{+\img q^+ b^-}\,
    \\\nonumber
    &\quad\times
    (\gamma_T^\mu)_{\alpha\beta}
    (\gamma_T^\nu)_{\gamma\delta}
    \int\df y^+\,\df y^-\,C_1(y^+,y^-)\,
    \int\df z^+\,\df z^-\,C_1^*(z^+,z^-)\,
    \\\nonumber
    &\quad\times
    \grn{\bra{P}\bar\chi_{i,\alpha}(b_T + b^- n + y^- n)\ket{X}}
    \grn{\bra{X}\chi_{l,\delta}(z^- n)\ket{P}}
    \\\nonumber
    &\quad\times
    \blu{\bra{0}\chi_{j,\beta}(b_T + b^+ \nbar + y^+ \nbar)\ket{p,X}}
    \blu{\bra{p,X}\bar\chi_{k,\gamma}(z^+ \nbar)\ket{0}}
    \\\nonumber
    &\quad\times
    \orn{\bra{0}\bigl[S_\nbar^\dagger S_n (b_T)\bigr]_{ij}\ket{X}}
    \orn{\bra{X}\bigl[S_n^\dagger S_\nbar (0)\bigr]_{kl}\ket{0}}\,.
\end{align}

In addition, we account for corrections arising from the multipole expansion. At NLP, two additional terms arise from expanding the soft Wilson lines in the light-cone directions $b^+$ and $b^-$ around $(b^+,b^-,b_T) = (0,0,b_T)$. Through the Fourier transformations, the explicit factors of $b^+$ and $b^-$ can be translated into derivatives with respect to $q^-$ and $q^+$. These contributions then read
\begin{align}
    \bigl[\mathcal{W}^{\mu\nu}\bigr]_{Bn}^{(2,2)}
    &=
    -\frac{\partial}{\partial q^+}
    \int\frac{\df^2 b}{(2\pi)^2}\,e^{+\img q_T \cdot b_T}\,
    \int\frac{\df b^+}{2\pi}\,e^{+\img q^- b^+}\,
    \int\frac{\df b^-}{2\pi}\,e^{+\img q^+ b^-}\,
    \\\nonumber
    &\quad\times
    (\gamma_T^\mu)_{\alpha\beta}
    (\gamma_T^\nu)_{\gamma\delta}
    \int\df y^+\,\df y^-\,C_1(y^+,y^-)\,
    \int\df z^+\,\df z^-\,C_1^*(z^+,z^-)\,
    \\\nonumber
    &\quad\times
    \grn{\bra{P}\bar\chi_{i,\alpha}(b_T + b^- n + y^- n)\ket{X}}
    \grn{\bra{X}\chi_{l,\delta}(z^- n)\ket{P}}
    \\\nonumber
    &\quad\times
    \blu{\bra{0}\chi_{j,\beta}(b_T + b^+ \nbar + y^+ \nbar)\ket{p,X}}
    \blu{\bra{p,X}\bar\chi_{k,\gamma}(z^+ \nbar)\ket{0}}
    \\\nonumber
    &\quad\times
    \orn{\bra{0}\img\partial^+
    \bigl[S_\nbar^\dagger S_n (b_T)\bigr]_{ij}\ket{X}}
    \orn{\bra{X}\bigl[S_n^\dagger S_\nbar (0)\bigr]_{kl}\ket{0}}\,,
    \\[3ex]
    \label{eq:MultipoleCorrection}\bigl[\mathcal{W}^{\mu\nu}\bigr]_{B\nbar}^{(2,2)}
    &=
    -\frac{\partial}{\partial q^-}
    \int\frac{\df^2 b}{(2\pi)^2}\,e^{+\img q_T \cdot b_T}\,
    \int\frac{\df b^+}{2\pi}\,e^{+\img q^- b^+}\,
    \int\frac{\df b^-}{2\pi}\,e^{+\img q^+ b^-}\,
    \\\nonumber
    &\quad\times
    (\gamma_T^\mu)_{\alpha\beta}
    (\gamma_T^\nu)_{\gamma\delta}
    \int\df y^+\,\df y^-\,C_1(y^+,y^-)\,
    \int\df z^+\,\df z^-\,C_1^*(z^+,z^-)\,
    \\\nonumber
    &\quad\times
    \grn{\bra{P}\bar\chi_{i,\alpha}(b_T + b^- n + y^- n)\ket{X}}
    \grn{\bra{X}\chi_{l,\delta}(z^- n)\ket{P}}
    \\\nonumber
    &\quad\times
    \blu{\bra{0}\chi_{j,\beta}(b_T + b^+ \nbar + y^+ \nbar)\ket{p,X}}
    \blu{\bra{p,X}\bar\chi_{k,\gamma}(z^+ \nbar)\ket{0}}
    \\\nonumber
    &\quad\times
    \orn{\bra{0}\img\partial^-
    \bigl[S_\nbar^\dagger S_n (b_T)\bigr]_{ij}\ket{X}}
    \orn{\bra{X}\bigl[S_n^\dagger S_\nbar (0)\bigr]_{kl}\ket{0}}\,.
\end{align}

\subsubsection*{NLP: kinematic power correction}

We now consider kinematic power corrections, originating from terms in the effective current operator eq.~\eqref{eq:EffCurrentNLP3} where a transverse derivative acts on a quark field. These corrections contribute to the hadronic tensor as follows:
\begin{align}
    \label{eq:HadronicTensorCn}
    \bigl[\mathcal{W}^{\mu\nu}\bigr]_{Cn}^{(3,2)}
    &=
    \biggl(-\frac{\nbar^\mu}{q^-}\biggr)
    \int\frac{\df^2 b}{(2\pi)^2}\,e^{+\img q_T \cdot b_T}\,
    \int\frac{\df b^+}{2\pi}\,e^{+\img q^- b^+}\,
    \int\frac{\df b^-}{2\pi}\,e^{+\img q^+ b^-}\,
    \\\nonumber
    &\quad\times
    (\gamma_T^\rho)_{\alpha\beta}
    (\gamma_T^\nu)_{\gamma\delta}
    \int\df y^+\,\df y^-\,C_1(y^+,y^-)\,
    \int\df z^+\,\df z^-\,C_1^*(z^+,z^-)\,
    \\\nonumber
    &\quad\times
    \grn{\bra{P}\bar\chi_{i,\alpha}(b_T + b^- n + y^- n)\ket{X}}
    \grn{\bra{X}\chi_{l,\delta}(z^- n)\ket{P}}
    \\\nonumber
    &\quad\times
    \blu{\bra{0}\img\partial_T^\rho
    \chi_{j,\beta}(b_T + b^+ \nbar + y^+ \nbar)\ket{p,X}}
    \blu{\bra{p,X}\bar\chi_{k,\gamma}(z^+ \nbar)\ket{0}}
    \\\nonumber
    &\quad\times
    \orn{\bra{0}\bigl[S_\nbar^\dagger S_n (b_T)\bigr]_{ij}\ket{X}}
    \orn{\bra{X}\bigl[S_n^\dagger S_\nbar (0)\bigr]_{kl}\ket{0}}\,.
    \\[3ex]
    \label{eq:HadronicTensorCnbar}
    \bigl[\mathcal{W}^{\mu\nu}\bigr]_{C\nbar}^{(3,2)}
    &=
    \biggl(-\frac{n^\mu}{q^+}\biggr)
    \int\frac{\df^2 b}{(2\pi)^2}\,e^{+\img q_T \cdot b_T}\,
    \int\frac{\df b^+}{2\pi}\,e^{+\img q^- b^+}\,
    \int\frac{\df b^-}{2\pi}\,e^{+\img q^+ b^-}\,
    \\\nonumber
    &\quad\times
    (\gamma_T^\rho)_{\alpha\beta}
    (\gamma_T^\nu)_{\gamma\delta}
    \int\df y^+\,\df y^-\,C_1(y^+,y^-)\,
    \int\df z^+\,\df z^-\,C_1^*(z^+,z^-)\,
    \\\nonumber
    &\quad\times
    \grn{\bra{P}\img\partial_T^\rho
    \bar\chi_{i,\alpha}(b_T + b^- n + y^- n)\ket{X}}
    \grn{\bra{X}\chi_{l,\delta}(z^- n)\ket{P}}
    \\\nonumber
    &\quad\times
    \blu{\bra{0}\chi_{j,\beta}(b_T + b^+ \nbar + y^+ \nbar)\ket{p,X}}
    \blu{\bra{p,X}\bar\chi_{k,\gamma}(z^+ \nbar)\ket{0}}
    \\\nonumber
    &\quad\times
    \orn{\bra{0}\bigl[S_\nbar^\dagger S_n (b_T)\bigr]_{ij}\ket{X}}
    \orn{\bra{X}\bigl[S_n^\dagger S_\nbar (0)\bigr]_{kl}\ket{0}}\,,
\end{align}
In sec.~\ref{sec:OverlapSubtraction}, we will define the subtracted kinematic contribution by combining this term with those in eqs.~\eqref{eq:HadronicTensorEn}–\eqref{eq:HadronicTensorEnbar}.

\subsubsection*{NLP: genuine higher-twist correction}

Next, we examine the contribution from operators in eq.~\eqref{eq:EffCurrentNLP3} that involve an additional collinear or anti-collinear gluon field. Their contribution reads
\begin{align}
    \label{eq:HadronicTensorDn}\bigl[\mathcal{W}^{\mu\nu}\bigr]_{Dn}^{(3,2)}
    &=
    \biggl(\frac{n^\mu}{q^+}-\frac{\nbar^\mu}{q^-}\biggr)
    \int\frac{\df^2 b}{(2\pi)^2}\,e^{+\img q_T \cdot b_T}\,
    \int\frac{\df b^+}{2\pi}\,e^{+\img q^- b^+}\,
    \int\frac{\df b^-}{2\pi}\,e^{+\img q^+ b^-}\,
    \\\nonumber
    &\quad\times
    (\gamma_T^\rho)_{\alpha\beta}
    (\gamma_T^\nu)_{\gamma\delta}
    \int\df y^-\,\df y_1^+\df y_2^+\,C_2(\{y_1^+,y_2^+\},y^-)\,
    \int\df z^-\,\df z^+\,C_1^*(z^-,z^+)\,
    \\\nonumber
    &\quad\times
    \grn{\bra{P}\bar\chi_{i,\alpha}(b_T + b^- n + y^- n)\ket{X}}
    \grn{\bra{X}\chi_{l,\delta}(z^- n)\ket{P}}
    \\\nonumber
    &\quad\times
    \blu{\bra{0}\bigl[\mathcal{A}_T^\rho(b_T + b^+ \nbar + y_2^+ \nbar)
    \chi(b_T + b^+ \nbar + y_1^+ \nbar)\bigr]_{j,\beta}\ket{p,X}}
    \blu{\bra{p,X}\bar\chi_{k,\gamma}(z^+ \nbar)\ket{0}}
    \\\nonumber
    &\quad\times
    \orn{\bra{0}\bigl[S_\nbar^\dagger S_n(b_T)\bigr]_{ij}\ket{X}}
    \orn{\bra{X}\bigl[S_n^\dagger S_\nbar (0)\bigr]_{kl}\ket{0}}\,,
    \\[3ex]
    \label{eq:HadronicTensorDnbar}\bigl[\mathcal{W}^{\mu\nu}\bigr]_{D\nbar}^{(3,2)}
    &=
    \biggl(\frac{n^\mu}{q^+}-\frac{\nbar^\mu}{q^-}\biggr)
    \int\frac{\df^2 b}{(2\pi)^2}\,e^{+\img q_T \cdot b_T}\,
    \int\frac{\df b^+}{2\pi}\,e^{+\img q^- b^+}\,
    \int\frac{\df b^-}{2\pi}\,e^{+\img q^+ b^-}\,
    \\\nonumber
    &\quad\times
    (\gamma_T^\rho)_{\alpha\beta}
    (\gamma_T^\nu)_{\gamma\delta}
    \int\df y_1^-\,\df y_2^-\df y^+\,C_2(\{y_1^-,y_2^-\},y^+)\,
    \int\df z^-\,\df z^+\,C_1^*(z^-,z^+)\,
    \\\nonumber
    &\quad\times
    \grn{\bra{P}\bigl[\bar\chi(b_T + b^- n + y_1^- n)
    \mathcal{A}_T^\rho(b_T + b^- n + y_2^- n)\bigr]_{i,\alpha}\ket{X}}
    \grn{\bra{X}\chi_{l,\delta}(z^- n)\ket{P}}
    \\\nonumber
    &\quad\times
    \blu{\bra{0}\chi_{j,\beta}(b_T + b^+ \nbar + y^+ \nbar)\ket{p,X}}
    \blu{\bra{p,X}\bar\chi_{k,\gamma}(z^+ \nbar)\ket{0}}
    \\\nonumber
    &\quad\times
    \orn{\bra{0}\bigl[S_\nbar^\dagger S_n (b_T)\bigr]_{ij}\ket{X}}
    \orn{\bra{X}\bigl[S_n^\dagger S_\nbar (0)\bigr]_{kl}\ket{0}}\,,
\end{align}
After proper subtraction of rapidity divergences, these terms form the genuine twist-3 contribution at NLP.

\subsubsection*{NLP: soft gluon contribution}

We now consider the contribution of the soft gluon operators where the inverse derivative acts on a collinear or anti-collinear field. In this case the inverse derivative can be replaced by $q^\pm$ by virtue of eq.~\eqref{eq:InverseDerivativeIdentity}. We then find the following contributions
\begin{align}
    \label{eq:HadronicTensorEn}\bigl[\mathcal{W}^{\mu\nu}\bigr]_{En}^{(3,2)}
    &=
    \biggl(-\frac{\nbar^\mu}{q^-}\biggr)
    \int\frac{\df^2 b}{(2\pi)^2}\,e^{+\img q_T \cdot b_T}\,
    \int\frac{\df b^+}{2\pi}\,e^{+\img q^- b^+}\,
    \int\frac{\df b^-}{2\pi}\,e^{+\img q^+ b^-}\,
    \\\nonumber
    &\quad\times
    (\gamma_T^\rho)_{\alpha\beta}
    (\gamma_T^\nu)_{\gamma\delta}
    \int\df y^+\,\df y^-\,C_1(y^+,y^-)\,
    \int\df z^+\,\df z^-\,C_1^*(z^+,z^-)\,
    \\\nonumber
    &\quad\times
    \grn{\bra{P}\bar\chi_{i,\alpha}(b_T + b^- n + y^- n)\ket{X}}
    \grn{\bra{X}\chi_{l,\delta}(z^- n)\ket{P}}
    \\\nonumber
    &\quad\times
    \blu{\bra{0}\chi_{j,\beta}(b_T + b^+ \nbar + y^+ \nbar)\ket{p,X}}
    \blu{\bra{p,X}\bar\chi_{k,\gamma}(z^+ \nbar)\ket{0}}
    \\\nonumber
    &\quad\times
    \orn{\bra{0}\bigl[
    S_\nbar^\dagger S_n \mathcal{A}_{T,n}^{\rho}(b_T)\bigr]_{ij}\ket{X}}
    \orn{\bra{X}\bigl[S_n^\dagger S_\nbar (0)\bigr]_{kl}\ket{0}}\,,
    \\[3ex]
    \label{eq:HadronicTensorEnbar}\bigl[\mathcal{W}^{\mu\nu}\bigr]_{E\nbar}^{(3,2)}
    &=
    \biggl(+\frac{n^\mu}{q^+}\biggr)
    \int\frac{\df^2 b}{(2\pi)^2}\,e^{+\img q_T \cdot b_T}\,
    \int\frac{\df b^+}{2\pi}\,e^{+\img q^- b^+}\,
    \int\frac{\df b^-}{2\pi}\,e^{+\img q^+ b^-}\,
    \\\nonumber
    &\quad\times
    (\gamma_T^\rho)_{\alpha\beta}
    (\gamma_T^\nu)_{\gamma\delta}
    \int\df y^+\,\df y^-\,C_1(y^+,y^-)\,
    \int\df z^+\,\df z^-\,C_1^*(z^+,z^-)\,
    \\\nonumber
    &\quad\times
    \grn{\bra{P}\bar\chi_{i,\alpha}(b_T + b^- n + y^- n)\ket{X}}
    \grn{\bra{X}\chi_{l,\delta}(z^-n)\ket{P}}
    \\\nonumber
    &\quad\times
    \blu{\bra{0}\chi_{j,\beta}(b_T + b^+ \nbar + y^+ \nbar)\ket{p,X}}
    \blu{\bra{p,X}\bar\chi_{k,\gamma}(z^+ \nbar)\ket{0}}
    \\\nonumber
    &\quad\times
    \orn{\bra{0}\bigl[\mathcal{A}_{T,\nbar}^{\rho}S_\nbar^\dagger S_n (b_T)\bigr]_{ij}\ket{X}}
    \orn{\bra{X}\bigl[S_n^\dagger S_\nbar (0)\bigr]_{kl}\ket{0}}\,.
\end{align}
In the next section, the above contributions will be combined with eqs.~\eqref{eq:HadronicTensorCn} and \eqref{eq:HadronicTensorCnbar} to define a kinematic power correction that is free of rapidity divergences.

\subsubsection*{NLP: remaining contributions}

We now consider the contributions of all remaining sub-leading soft operators, and will argue that these vanish due to boost invariance. The two remaining soft gluon corrections read
\begin{align}
    \bigl[\mathcal{W}^{\mu\nu}\bigr]_{Fn}^{(3,2)}
    &=
    -\frac{1}{q^-}
    \int\frac{\df^2 b}{(2\pi)^2}\,e^{+\img q_T \cdot b_T}\,
    \int\frac{\df b^+}{2\pi}\,e^{+\img q^- b^+}\,
    \int\frac{\df b^-}{2\pi}\,e^{+\img q^+ b^-}\,
    \\\nonumber
    &\quad\times
    (\gamma_T^\mu)_{\alpha\beta}
    (\gamma_T^\nu)_{\gamma\delta}
    \int\df y^+\,\df y^-\,C_1(y^+,y^-)\,
    \int\df z^+\,\df z^-\,C_1^*(z^+,z^-)\,
    \\\nonumber
    &\quad\times
    \grn{\bra{P}
    \bar\chi_{i,\alpha}(b_T + b^- n + y^- n)\ket{X}}
    \grn{\bra{X}\chi_{l,\delta}(z^- n)\ket{P}}
    \\\nonumber
    &\quad\times
    \blu{\bra{0}\img\partial_T^\rho\chi_{j,\beta}(b_T + b^+ \nbar + y^+ \nbar)\ket{p,X}}
    \blu{\bra{p,X}\bar\chi_{k,\gamma}(z^+ \nbar)\ket{0}}
    \\\nonumber
    &\quad\times
    \orn{\bra{0}\biggl[
    S_\nbar^\dagger S_n\frac{1}{\img\partial^+}\mathcal{A}_{T,n}^\rho(b_T)\biggr]_{ij}\ket{X}}
    \orn{\bra{X}\bigl[S_n^\dagger S_\nbar (0)\bigr]_{kl}\ket{0}}\,,
    \\[3ex]
    \bigl[\mathcal{W}^{\mu\nu}\bigr]_{F\nbar}^{(3,2)}
    &=
    \frac{1}{q^+}
    \int\frac{\df^2 b}{(2\pi)^2}\,e^{+\img q_T \cdot b_T}\,
    \int\frac{\df b^+}{2\pi}\,e^{+\img q^- b^+}\,
    \int\frac{\df b^-}{2\pi}\,e^{+\img q^+ b^-}\,
    \\\nonumber
    &\quad\times
    (\gamma_T^\mu)_{\alpha\beta}
    (\gamma_T^\nu)_{\gamma\delta}
    \int\df y^+\,\df y^-\,C_1(y^+,y^-)\,
    \int\df z^+\,\df z^-\,C_1^*(z^+,z^-)\,
    \\\nonumber
    &\quad\times
    \grn{\bra{P}\img\partial_T^\rho\bar\chi_{i,\alpha}(b_T + b^- n + y^- n)\ket{X}}
    \grn{\bra{X}\chi_{l,\delta}(z^- n)\ket{P}}
    \\\nonumber
    &\quad\times
    \blu{\bra{0}
    \chi_{j,\beta}(b_T + b^+ \nbar + y^+ \nbar)\ket{p,X}}
    \blu{\bra{p,X}\bar\chi_{k,\gamma}(z^+ \nbar)\ket{0}}
    \\\nonumber
    &\quad\times
    \orn{\bra{0}\biggl[\frac{1}{\img\partial^-}\mathcal{A}_{T,\nbar}^\rho S_\nbar^\dagger S_n (b_T)
    \biggr]_{ij}\ket{X}}
    \orn{\bra{X}\bigl[S_n^\dagger S_\nbar(0)\bigr]_{kl}\ket{0}}\,,
\end{align}
and
\begin{align}
    \bigl[\mathcal{W}^{\mu\nu}\bigr]_{Gn}^{(3,2)}
    &=
    -\frac{1}{2q^-}
    \int\frac{\df^2 b}{(2\pi)^2}\,e^{+\img q_T \cdot b_T}\,
    \int\frac{\df b^+}{2\pi}\,e^{+\img q^- b^+}\,
    \int\frac{\df b^-}{2\pi}\,e^{+\img q^+ b^-}\,
    \\\nonumber
    &\quad\times
    (\gamma_T^\mu\gamma_T^\rho\gamma_T^\sigma)_{\alpha\beta}
    (\gamma_T^\nu)_{\gamma\delta}
    \int\df y^+\,\df w^-\,\df y^-\,C_3(y^+,w^-,y^-)\,
    \int\df z^+\,\df z^-\,C_1^*(z^+,z^-)\,
    \\\nonumber
    &\quad\times
    \grn{\bra{P}\bar\chi_{i,\alpha}(b_T + b^- n + y^- n)\ket{X}}
    \grn{\bra{X}\chi_{l,\delta}(z^- n)\ket{P}}
    \\\nonumber
    &\quad\times
    \blu{\bra{0}\chi_{j,\beta}(b_T + b^+ \nbar + y^+ \nbar)\ket{p,X}}
    \blu{\bra{p,X}\bar\chi_{k,\gamma}(z^+ \nbar)\ket{0}}
    \\\nonumber
    &\quad\times
    \orn{\bra{0}\biggl[
    S_\nbar^\dagger S_n (b_T)\,\frac{\img\partial_T^\rho}{\img\partial^+}
    \mathcal{A}_{T,n}^\sigma(b_T + w^- n)\biggr]_{ij}\ket{X}}
    \orn{\bra{X}\bigl[S_n^\dagger S_\nbar (0)\bigr]_{kl}\ket{0}}\,,
    \\[3ex]
    \bigl[\mathcal{W}^{\mu\nu}\bigr]_{G\nbar}^{(3,2)}
    &=
    \frac{1}{2q^+}
    \int\frac{\df^2 b}{(2\pi)^2}\,e^{+\img q_T \cdot b_T}\,
    \int\frac{\df b^+}{2\pi}\,e^{+\img q^- b^+}\,
    \int\frac{\df b^-}{2\pi}\,e^{+\img q^+ b^-}\,
    \\\nonumber
    &\quad\times
    (\gamma_T^\sigma\gamma_T^\rho\gamma_T^\mu)_{\alpha\beta}
    (\gamma_T^\nu)_{\gamma\delta}
    \int\df y^+\,\df w^+\,\df y^-\,C_3(y^+,w^+,y^-)\,
    \int\df z^+\,\df z^-\,C_1^*(z^+,z^-)\,
    \\\nonumber
    &\quad\times
    \grn{\bra{P}\bar\chi_{i,\alpha}(b_T + b^- n + y^- n)\ket{X}}
    \grn{\bra{X}\chi_{l,\delta}(z^- n)\ket{P}}
    \\\nonumber
    &\quad\times
    \blu{\bra{0}\chi_{j,\beta}(b_T + b^+ \nbar + y^+ \nbar)\ket{p,X}}
    \blu{\bra{p,X}\bar\chi_{k,\gamma}(z^+ \nbar)\ket{0}}
    \\\nonumber
    &\quad\times
    \orn{\bra{0}\biggl[\frac{\img\partial_T^\rho}{\img\partial^-}
    \mathcal{A}_{T,\nbar}^\sigma(b_T + w^+ \nbar)S_\nbar^\dagger S_n (b_T)\,
    \biggr]_{ij}\ket{X}}
    \orn{\bra{X}\bigl[S_n^\dagger S_\nbar (0)\bigr]_{kl}\ket{0}}\,.
\end{align}
For the soft quark contribution, we need an insertion in both the current at $b$ and the current at the origin. This gives
\begin{align}
    &\bigl[\mathcal{W}^{\mu\nu}\bigr]_{Hn}^{(2.5,2.5)}=
    \\\nonumber
    &-\frac{1}{4}
    (\gamma_T^\mu\gamma^-\gamma_T^\rho)_{\alpha\beta}
    (\gamma_T^\sigma\gamma^-\gamma_T^\nu)_{\gamma\delta}
    \int\frac{\df^2 b}{(2\pi)^2}\,e^{+\img q_T \cdot b_T}\,
    \int\frac{\df b^+}{2\pi}\,e^{+\img q^- b^+}\,
    \int\frac{\df b^-}{2\pi}\,e^{+\img q^+ b^-}\,
    \\\nonumber
    &\quad\times
    \int\df y^+\,\df v^+\,\df y^-\,C_4(y^+,v^+,y^-)\,
    \int\df z^+\,\df w^+\,\df z^-\,C_4^*(z^+,w^+,z^-)\,
    \\\nonumber
    &\quad\times
    \grn{\bra{P}\bar\chi_{i,\alpha}(b_T + b^- n + y^- n)\ket{X}}
    \grn{\bra{X}\chi_{n,\delta}(z^- n)\ket{P}}
    \\\nonumber
    &\quad\times
    \blu{\bra{0}\bigl[\mathcal{A}_T^\rho(b_T + y^+ \nbar)\bigr]_{jk}\ket{p,X}}
    \blu{\bra{p,X}\bigl[\mathcal{A}_T^\sigma(z^+ \nbar)\bigr]_{lm}\ket{0}}
    \\\nonumber
    &\quad\times
    \orn{\bra{0}\bigl[S_\nbar^\dagger S_n (b_T)\bigr]_{ij}
    \biggl[\frac{1}{\img\partial^+}
    \Psi_n(b_T + v^- n)\biggr]_{k,\beta}\ket{X}}
    \orn{\bra{X}\biggl[\frac{1}{\img\partial^+}
    \bar\Psi_{n}(w^- n)\biggr]_{l,\gamma}
    \bigl[S_n^\dagger S_\nbar (0)\bigr]_{mn}\ket{0}},
    \\[3ex]
    \label{eq:HadronicTensorHnbar}&\bigl[\mathcal{W}^{\mu\nu}\bigr]_{H\nbar}^{(2.5,2.5)}=
    \\\nonumber
    &-\frac{1}{4}(\gamma_T^\mu\gamma^+\gamma_T^\rho)_{\alpha\beta}
    (\gamma_T^\sigma\gamma^+\gamma_T^\nu)_{\gamma\delta}
    \int\frac{\df^2 b}{(2\pi)^2}\,e^{+\img q_T \cdot b_T}\,
    \int\frac{\df b^+}{2\pi}\,e^{+\img q^- b^+}\,
    \int\frac{\df b^-}{2\pi}\,e^{+\img q^+ b^-}
    \\\nonumber
    &\quad\times
    \int\df y^+\,\df v^+\,\df y^-\,C_4^*(y^+,v^-,y^-)\,
    \int\df z^+\,\df w^+\,\df z^-\,C_4(z^+,w^-,z^-)\,
    \\\nonumber
    &\quad\times
    \grn{\bra{P}\bigl[\mathcal{A}_T^\rho(b_T + y^- n)\bigr]_{jk}\ket{X}}
    \grn{\bra{X}\bigl[\mathcal{A}_T^\sigma(z^- n)\bigr]_{lm}\ket{P}}
    \\\nonumber
    &\quad\times
    \blu{\bra{0}\bar\chi_{i,\alpha}(b_T + b^+ \nbar + y^+\nbar)\ket{p,X}}
    \blu{\bra{p,X}\chi_{n,\delta}(z^+ \nbar)\ket{0}}
    \\\nonumber
    &\quad\times
    \orn{\bra{0}\bigl[S_n^\dagger S_\nbar (b_T)\bigr]_{ij}
    \biggl[\frac{1}{\img\partial^-}
    \Psi_{\nbar}(b_T + v^+ \nbar)\biggr]_{k,\beta}\ket{X}}
    \orn{\bra{X}\biggl[\frac{1}{\img\partial^-}
    \bar\Psi_{\nbar}(w^+ \nbar)\biggr]_{l,\gamma}
    \bigl[S_\nbar^\dagger S_n (0)\bigr]_{mn}\ket{0}},
\end{align}

We will now argue that the $F$, $G$, and $H$ contributions to the hadronic tensor vanish based on boost invariance. To do this, we start by analyzing the soft matrix elements corresponding to these terms at tree-level. The soft matrix elements are given by,
\begin{align}
    S_F^\rho(b_T)&=
    \orn{\bra{0} \frac{1}{\img\partial^-}\mathcal{A}_{T,\nbar}^\rho (b_T)\,S_\nbar^\dagger S_n (b_T)
    \ket{X}}
    \orn{\bra{X} S_n^\dagger S_\nbar (0)\ket{0}}\,,
    \\
    S_G^{\rho\sigma}(b_T)&=
    \orn{\bra{0} \frac{\img\partial_T^\sigma}{\img\partial^-}
    \mathcal{A}_{T,\nbar}^\rho(b_T)\,S_\nbar^\dagger S_n (b_T)\ket{X}}
    \orn{\bra{X} S_n^\dagger S_\nbar (0)\ket{0}}\,,
    \\
    \bigl[S_H(b_T)\bigr]_{\beta\gamma}&=
    \orn{\bra{0} S_n^\dagger S_\nbar (b_T) 
    \biggl[\frac{1}{\img\partial^-}\gamma^-
    \Psi_{\nbar}(b_T)\biggr]_\beta\ket{X}}
    \orn{\bra{X}\biggl[\frac{1}{\img\partial^-}
    \bar\Psi_{\nbar}(0)\gamma^-\biggr]_{\gamma}
    S_\nbar^\dagger S_n (0)\ket{0}}\,,
\end{align}
where the light-cone coordinates of all fields have been set to zero. 
This is justified by the fact that the tree-level Wilson coefficient is a trivial delta function, which sets the light-cone coordinate of the field to zero. What is crucial to proving that the above soft functions vanish is that all matrix elements contain more inverse derivatives associated with one light-cone direction than with the other. This, combined with the fact that the vacuum, the TMD measurement, and all other operators in the matrix element transform trivially under a boost, makes the above matrix elements vanish to all orders in perturbation theory. These same soft functions have appeared in ref.~\cite{Beneke:2018gvs} but with a different measurement breaking the boost invariance and yielding a nonzero result.

This conclusion still holds in the presence of a rapidity regulator. For example, for the $\delta$-regulator, described in more detail in sec.~\ref{sec:OverlapSubtraction}, the leading-order (LO) soft functions evaluate to
\begin{align}
    \label{eq:IntegralsLOforFSoftOperators}[S_F^\rho(b_T)]^{\text{LO}}&=
    g^2\int\frac{\df^d k}{(2\pi)^d}\,
    (2\pi)\delta(k^2)\,\theta(k^0)\,
    e^{-\img b_T \cdot k_T}\,
    \frac{k_T^\rho}{(k^-+\img\delta^-)^2}\frac{1}{k^+-\img\delta^+}\,,
    \\
    [S_G^{\rho\sigma}(b_T)]^{\text{LO}}&=g^2
    \int\frac{\df^d k}{(2\pi)^d}\,
    (2\pi)\delta(k^2)\,\theta(k^0)\,
    e^{-\img b_T \cdot k_T}\,
    \frac{k_T^\rho k_T^\sigma}{(k^-+\img\delta^-)^2} \frac{1}{k^+-\img\delta^+}\,,
    \\
    \label{eq:IntegralLOforHSoftOperators}\bigl[S_H(b_T)\bigr]^{\text{LO}}_{\beta\gamma}&=
    -2(\gamma^-)_{\beta\gamma}
    \int\frac{\df^d k}{(2\pi)^d}\,
    (2\pi)\delta(k^2)\,\theta(k^0)\,
    e^{-\img b_T \cdot k_T}\,
    \frac{1}{k^-}\,.
\end{align}
The last soft function, $S_H$, results in a scaleless integral and therefore vanishes. For $S_F$, an explicit evaluation yields:
\begin{align}
    [S_F^\rho(b_T)]^{\text{LO}}
    &=
    -g^2\frac{b_T^\rho\delta^+}{(4\pi)^{1-\epsilon}}
    \biggl\{\bigl(2\delta^+\delta^-\bigr)^{-\epsilon}\Gamma(1-\epsilon)\Gamma^2(\epsilon)
    \\\nonumber
    &\quad\qquad
    -\biggl(\frac{b_T^2}{4}\biggr)^\epsilon\Gamma(-\epsilon)
    \biggl[(\ln\biggl(\frac{\delta^+\delta^-b_T^2}{2e^{-2\gamma_E}}\biggr)
    +\psi(-\epsilon)-\gamma_E\biggr]\biggr\}\,\propto\,\delta^+\,,
\end{align}
where $\psi(z)$ is the digamma function and $\gamma_E$ is Euler’s constant. Similarly, one finds $S_G \propto \delta^+$. All contributions thus vanish in the $\delta^\pm \to 0$ limit. For other rapidity regulators, such as the $\eta$-regulator, all these integrals would be scaleless and vanish identically. These soft matrix elements, therefore, appear to be artifacts of the rapidity regulators.

\subsection{Overlap subtraction}\label{sec:OverlapSubtraction}

The factorization formula for the hadronic tensor presented in the previous section is not quite complete yet: we need to address the presence of an overlap between the different modes, also referred to as the zero-bin~\cite{Manohar:2006nz,Lee:2006fn,Idilbi:2007ff}. The only thing that separates collinear, soft, and anti-collinear modes is the ratio between the $k^+$ and $k^-$ components, which is related to the rapidity. For the rapidities of the different modes, we have the following hierarchy,
\begin{align*}
    &e^{\blu{Y}}\ll 1\,,&e^{\orn{Y}}\sim1\,,&&e^{\grn{Y}}\gg1\,,&
\end{align*}
where the rapidity Y is defined by
\begin{align}
    Y=\frac{1}{2}\ln\frac{k^+}{k^-}\,.
\end{align}
Because of this separation in rapidity, a clean formulation of the mode decomposition would require a cut in rapidity for each mode. In principle, one could introduce an explicit cut-off on the rapidity, making the separation between the modes clear and removing any overlap. This would introduce a cut-off dependence in the collinear, soft, and anti-collinear functions that drops out of the final cross section, as the cross section itself is independent of how we separate the modes. In practice, however, introducing such a cut-off is tedious. Instead, the approach taken in SCET is to have no cut-off at all and subtract the double-counted region. 

To regularize rapidity divergences, we employ the so-called $\delta$-regulator \cite{Echevarria:2012js,Echevarria:2016scs,Vladimirov:2017srds}, which is motivated by the fact that rapidity divergences come from the integration over the lightcone coordinate that one finds in Wilson lines. The $\delta$-regulator regularizes rapidity divergences by introducing a damping factor in the definition in eqs.~\eqref{eq:CollinearWilsonLines}-\eqref{eq:SoftWilsonLines} of Wilson lines,
\begin{align}\label{eq:DeltaWilsonLines}
    \grn{W(x)}&=
    \lim_{\delta^+\to0}\mathcal{P}\bigg\{ \exp\biggl[
    \img g\int_{L}^0\df y^+\,e^{-s\delta^+|y^-|}\,
    \grn{A^+(x + y^- n)}\biggr]\bigg\}\,,
    \\
    \blu{W(x)}&=
    \lim_{\delta^-\to0}\mathcal{P}\bigg\{ \exp\biggl[
    \img g\int_{\bar L}^0\df y^+\,e^{-\bar s\delta^-|y^+|}\,
    \blu{A^-(x + y^+ \nbar)}\biggr]\bigg\}\,,
    \\
    \orn{S_n(x)}&=
    \lim_{\delta^+\to0}\mathcal{P}\bigg\{ \exp\biggl[
    \img g\int_{L}^0\df y^+\,e^{-s\delta^+|y^-|}\,
    \orn{A^+(x + y^- n)}\biggr]\bigg\}\,,
    \\
    \orn{S_\nbar(x)}&=
    \lim_{\delta^-\to0}\mathcal{P}\bigg\{ \exp\biggl[
    \img g\int_{\bar L}^0\df y^+\,e^{-\bar s\delta^-|y^+|}\,
    \orn{A^-(x + y^+ \nbar)}\biggr]\bigg\}\,,
\end{align}
where $s=\text{sign}(L)$ and $\bar s=\text{sign}(\bar L)$. Note that we have introduced two regulators. This is, in fact, quite natural, as there are two cuts in rapidity: one between soft and collinear and one between soft and anti-collinear. The divergences in $\delta^+$ and $\delta^-$ cancel independently. In calculating radiative corrections, this change in definition leads to the replacement
\begin{align}
    \frac{k^+}{k^-}\to\frac{k^+}{k^- + \img\delta^-}\,.
\end{align}
However, this replacement should be treated with care at higher orders in perturbation theory~\cite{Echevarria:2015byo}. This regularization prescription makes the relevant rapidity integral well-defined and leads to logarithmic divergences in $\delta^\pm$. 

Rapidity divergences cancel in the cross section, but the structure of this cancellation is somewhat non-trivial. First, note that rapidity divergences are found in both the collinear, soft, and anti-collinear ingredients, as well as in the overlap region. The rapidity divergences cancel only upon combining the three sectors and subtracting the overlap. At next-to-leading power, the structure of the subtraction of the overlap region and the cancellation of divergences becomes even more involved. This is because the presence of inverse derivatives can lead to power-mixing, i.e., the overlap subtraction of order $\mathcal{O}(\lambda^2)$-suppressed term can become of order $\mathcal{O}(\lambda)$, as discussed in sec.~\ref{sec:PowerMixing}.

Another subtlety at next-to-leading power is the presence of an additional type of divergence, referred to as an endpoint divergence~\cite{Liu:2019oav,Liu:2020tzd,Beneke:2022zkz,Vladimirov:2021hdn,Vladimirov:2023aot,delCastillo:2023rng}. Endpoint divergences arise from the next-to-leading power operators that contain an additional collinear gluon field. Their contribution leads to a term in the hadronic tensor that involves a convolution between a hard function and a parton distribution that both depend on an additional momentum fraction $\xi$, e.g.
\begin{align}\label{eq:HadronicTensorDnbar2}
    \bigl[\mathcal{W}^{\mu\nu}\bigr]_{D\nbar}^{(3,2)}
    \quad\Rightarrow\quad
    \int\df\xi\,H(\xi,Q)\,\grn{F(\xi,b_T)}\,\blu{D(b_T)}\,,
\end{align}
where the function $F(\xi,b_T)$ contains the operator with the collinear gluon field. The additional momentum fraction $\xi$ is associated with the collinear gluon field and originates from the convolution between the collinear matrix element and the Wilson coefficient involving $y_2^-$ in eq.~\eqref{eq:HadronicTensorDnbar}. Endpoint divergences then arise in eq.~\eqref{eq:HadronicTensorDnbar2} as $\xi\to0$, which physically can be associated with the collinear gluon becoming soft. These divergences are not unique to TMD factorization but are rather a characteristic property of sub-leading-power factorization in general. While endpoint divergences can be consistently regularized and eventually canceled in the cross section, the structure of the cancellation of these divergences remains to be understood~\cite{Liu:2019oav,Liu:2020tzd,Beneke:2022zkz}.

\subsubsection{Subtraction method}\label{sec:SubtractionMethod}

In this work, we subtract the soft-collinear overlap by introducing a background field for the overlap modes, following the (leading-power) approach of ref.~\cite{Idilbi:2007ff}. This subtraction is done in four steps:
\begin{enumerate}
    \item Split the collinear background fields up into a pure collinear field and a background field that contains the overlap with the soft region.
    \item Redefine the pure collinear fields such that the leading-power interactions with the overlap background field are removed. This is similar to how leading-power soft-collinear interactions are removed in SCET-I~\cite{Bauer:2001yt,Bauer:2001ct,Bauer:2002nz,Beneke:2002ph} via a decoupling transformation.
    \item Factorize the collinear matrix elements into a pure collinear part and an overlap part. To do this, take each collinear matrix element in the hadronic tensor and apply the field transformations of step (1) and (2) to separate the contributions of the pure-collinear modes and the overlap modes.
    \item Invert the relations from step (3) to obtain the pure-collinear matrix elements.
\end{enumerate}
Below, we will demonstrate this process in more detail.

For the first step, we split the collinear background fields into a pure collinear background field that has no overlap with the soft region and a background field that contains the overlap region. As a result of this decomposition, the action for the collinear background fields now contains interactions between the pure-collinear modes and the soft overlap modes. For simplicity, we focus only on the quark-gluon interaction terms and ignore gluon self-interactions. We start by taking the collinear Lagrangian and splitting off a soft component from the collinear fields,
\begin{align}\label{eq:QCDActionOverlap}
    S_\text{QCD}[\blu{\phi}+\orn{\phi}]
    &=
    \int\df^d x\,
    \biggl[
    \blu{\bar\psi(x)}
    \Bigl(\img\fsl{\partial}+\blu{g\fsl{A}(x)}+\orn{g\fsl{A}(x)}\Bigr)
    \blu{\psi(x)}
    \\\nonumber
    &\quad
    +\orn{\bar\psi(x)}\blu{g\fsl{A}(x)}\blu{\psi(x)}
    +\blu{\bar\psi(x)}\blu{g\fsl{A}(x)}\orn{\psi(x)}
    \biggr]
    +\text{field strength}+S_\text{QCD}[\orn{\phi}]
\end{align}
This action now contains interactions between collinear and soft modes. Note that we do not discard terms that are not allowed by momentum conservation, as in principle the \orn{soft} fields live in the overlap region. In terms of the power counting, only the collinear-quark soft-gluon interaction is leading power\footnote{Note that in the interaction terms the position integration measure scales as $\df^4 x\sim\lambda^{-3}$ as opposed to $\df^4 x\sim\lambda^{-4}$ for the pure-collinear terms.}. In fact, only the interaction with $\orn{A^+}$ component is leading power as this component is enhanced compared to $\blu{A^-}$.

In the second step, we perform a field redefinition that makes all the interactions between soft and collinear power-suppressed, similar to the decoupling transformation in SCET-I, but without multipole expanding the soft Wilson lines at this stage. We define the transformation as follows:
\begin{align}\label{eq:DecouplingTransformation}
    \blu{\psi(x)}&\rightarrow\orn{S_n(x)}\blu{\psi(x)}\,,\\
    \nn\blu{A^\mu(x)}&\rightarrow
    \orn{S_n(x)}\blu{A^\mu(x)}\orn{S_n^\dagger(x)}\,.
\end{align}
This results in the following transformation for the first line in eq.~\eqref{eq:QCDActionOverlap}:
\begin{align}
    &\blu{\bar\psi(x)}\orn{S_n^\dagger(x)}
    \Bigl(\img\fsl{\partial}
    +\orn{S_n(x)}\blu{g\fsl{A}(x)}\orn{S_n^\dagger(x)}
    +\orn{g\fsl{A}(x)}\Bigr)
    \orn{S_n(x)}\blu{\psi(x)}
    \\
    \nn&=
    \blu{\bar\psi(x)}
    \Bigl(\img\fsl{\partial}+\blu{g\fsl{A}(x)}
    +\orn{S_n^\dagger(x)}[\img\fsl{\partial},\orn{S_n(x)}]
    +\orn{S_n^\dagger(x)}\orn{g\fsl{A}(x)}\orn{S_n(x)}
    \Bigr)
    \blu{\psi(x)}
    \\
    \nn&=
    \blu{\bar\psi(x)}
    \Bigl(\img\fsl{\partial}+\blu{g\fsl{A}(x)}
    +\orn{g\fsl{\mathcal{A}}_n(x)}\Bigr)
    \blu{\psi(x)}\,.
\end{align}
This transformation removes the leading-power overlap interaction with $\orn{A^+}$. The transformation of the remaining terms is trivial and leads to
\begin{align}
    S_\text{QCD}[\orn{S_n}\blu{\phi}+\orn{\phi}]
    &=
    S_\text{QCD}[\blu{\phi}]+S_\text{QCD}[\orn{\phi}]
    \\\nonumber
    &\quad\qquad
    +\int\df^d x\,
    \biggl[
    \blu{\bar\psi}\orn{\fsl{\mathcal{A}_n}}\blu{\psi}
    +\orn{\bar\Psi_n}\blu{g\fsl{A}}\blu{\psi}
    +\blu{\bar\psi}\blu{g\fsl{A}}\orn{\Psi_n}
    +\text{gluon interactions}
    \biggr]\,,
\end{align}
where all terms on the second line are power suppressed. This suppression allows us to separate the Hilbert spaces of the pure collinear modes and the soft overlap mode.

To subtract the overlap region, starting from the full hadronic tensor derived in the previous section, we take the following approach: First, for each collinear and anti-collinear matrix element, we explicitly add the overlap by splitting off a soft component from all collinear field operators. Second, we apply the decoupling transformation of eq.~\eqref{eq:DecouplingTransformation} to the collinear operators. These two steps can be achieved by making the following replacement inside all collinear matrix elements,
\begin{align}\label{eq:OverlapTransformation}
    \blu{\psi(x)}&\to\orn{S_n(x)}\blu{\psi(x)}+\orn{\psi(x)}\,,
    \\
    \nn \blu{A_T^\mu(x)}&\to\orn{S_n(x)}\blu{A_T^\mu(x)}\orn{S_n^\dagger(x)}+\orn{A_T^\mu(x)}\,.
\end{align}
The result of this transformation on a collinear matrix element gives the same collinear matrix element but with the soft overlap region split off. Finally, to subtract the overlap, one inverts the relation between the original and the transformed matrix elements and replaces the collinear matrix elements in the unsubtracted hadronic tensor by the pure-collinear ones given by the aforementioned relation.

Before applying this scheme to the full next-to-leading-power hadronic tensor, let us provide a simple example at leading power. First, we need to work out the transformation of the building blocks. For the gauge invariant SCET building blocks, the transformations of eq.~\eqref{eq:OverlapTransformation} become
\begin{align}\label{eq:OverlapTransformationSCET}
    \blu{\chi}&\to
    \orn{S_\nbar^\dagger S_n}\blu{\chi}+\orn{\Psi_\nbar}\,,
    \\
    \nn\blu{\mathcal{A}_T^\mu}&\to
    \orn{S_\nbar^\dagger S_n}\blu{\mathcal{A}_T^\mu}\orn{S_n^\dagger S_\nbar}+\orn{\mathcal{A}_{T,\nbar}^\mu}\,,
\end{align}
where the soft quark and gluon fields are defined in eq.~\eqref{eq:SoftAntiCollinearFieldsSCET}. The leading power collinear matrix element that appears in eq.~\eqref{eq:HadronicTensorLP} transforms as
\begin{align}
    &\blu{\bra{0}\chi_{j,\beta}(b_T + b^+\nbar)\ket{p,X}\bra{p,X}\bar\chi_{k,\gamma}(0)\ket{0}}
    \\
    &\to
    \blu{\bra{0}\chi_{j,\beta}(b_T + b^+\nbar)\ket{p,X}\bra{p,X}\bar\chi_{k,\gamma}(0)\ket{0}}
    \orn{\frac{1}{N_c}\tr\Bigl[\bra{0}\bigl[S_\nbar^\dagger S_n (b_T)\bigr]\ket{X}
    \bra{X}\bigl[S_n^\dagger S_\nbar (0)\bigr]\ket{0}\Bigr]}\,.
    \nonumber
\end{align}
Here, the color trace over the soft matrix element is the only surviving color structure in the decomposition of representations that one would get from eq.~\eqref{eq:OverlapTransformationSCET}. Inverting this relation then gives the collinear matrix element with the overlap region subtracted,
\begin{align*}
    \blu{\bra{0}\chi_{j,\beta}(b_T + b^+\nbar)\ket{p,X}\bra{p,X}\bar\chi_{k,\gamma}(0)\ket{0}_\text{sub.}}
    &=
    \frac{\blu{\bra{0}\chi_{j,\beta}(b_T + b^+\nbar)\ket{p,X}\bra{p,X}\bar\chi_{k,\gamma}(0)\ket{0}}}
    {\orn{\frac{1}{N_c}\tr\Bigl[\bra{0}\bigl[S_\nbar^\dagger S_n (b_T)\bigr]\ket{X}
    \bra{X}\bigl[S_n^\dagger S_\nbar (0)\bigr]\ket{0}\Bigr]}}\,.
    \nonumber
\end{align*}

\subsubsection{Overlap subtraction at NLP}\label{sec:OverlapSubtractionNLP}

At next-to-leading power, the overlap subtraction and the structure of the cancellation of rapidity divergences become more involved. In this case, the overlap subtraction not only involves multiplying by soft Wilson lines but also additional additive terms.

\subsubsection*{LP contribution}

First, we look at the overlap subtraction for the leading-power contribution. We first add the overlap by inserting the transformation of eq.~\eqref{eq:OverlapTransformationSCET} into the collinear and anti-collinear matrix elements. This results in
\begin{align}\label{eq:OverlapMultipoleCorrection}
    &\blu{\bra{0}\chi_\beta(b_T + b^+ \nbar)\ket{p,X}
    \bra{p,X}\bar\chi_\gamma(0)\ket{0}}
    \\
    &\to
    \blu{\bra{0}}\blu{\chi_\beta(b_T + b^+\nbar)} 
    \blu{\ket{p,X}}\blu{\bra{p,X}}\blu{\bar\chi_\gamma(0)}\blu{\ket{0}}
    \orn{\bra{0}}\orn{\bigl[S_\nbar^\dagger S_n (b_T)\bigr]}\orn{\ket{X}}
    \orn{\bra{X}}\orn{\bigl[S_n^\dagger S_\nbar (0)\bigr]}\orn{\ket{0}}
    \nonumber
    \\
    &\quad+
    \blu{\bra{0}}\blu{\chi_\beta(b_T + b^+\nbar)} 
    \blu{\ket{p,X}}\blu{\bra{p,X}}\blu{\bar\chi_\gamma(0)}\blu{\ket{0}}
    \orn{\bra{0}}\orn{b^+ \partial^-\bigl[S_\nbar^\dagger S_n (b_T)\bigr]}\orn{\ket{X}}
    \orn{\bra{X}}\orn{\bigl[S_n^\dagger S_\nbar (0)\bigr]}\orn{\ket{0}}
    \nonumber
    \\
    &\quad+\text{overlap interactions}\,.
    \nonumber
\end{align}
The first line is again just multiplication by the soft function, while the second line is a multipole correction to the first line from expanding the soft Wilson lines in light-cone components. In principle, one also needs to consider higher-order overlap interaction terms, but we omit those in the present analysis. To remove the overlap, we need to invert the above relation and plug that into the hadronic tensor. Doing this, we find that the multipole corrections of the overlap removal of eq.~\eqref{eq:OverlapMultipoleCorrection} completely cancel the multipole corrections of the unsubtracted hadronic tensor in eq.~\eqref{eq:MultipoleCorrection}. We therefore have
\begin{align}
    &\bigl[\mathcal{W}_\text{sub.}^{\mu\nu}\bigr]_{A}^{(2,2)}
    +\bigl[\mathcal{W}_\text{sub.}^{\mu\nu}\bigr]_{Bn}^{(2,2)}
    +\bigl[\mathcal{W}_\text{sub.}^{\mu\nu}\bigr]_{B\nbar }^{(2,2)}
    \\
    \nn&=
    \int\frac{\df^2 b}{(2\pi)^2}\,e^{+\img q_T \cdot b_T}\,
    \int\frac{\df b^+}{2\pi}\,e^{+\img q^- b^+}\,
    \int\frac{\df b^-}{2\pi}\,e^{+\img q^+ b^-}\,
    \\
    \nn&\quad\times
    (\gamma_T^\mu)_{\alpha\beta}
    (\gamma_T^\nu)_{\gamma\delta}
    \int\df y^+\,\df y^-\,C_1(y^+,y^-)\,
    \int\df z^+\,\df z^-\,C_1^*(z^+,z^-)\,
    \\
    \nn&\quad\times
    \grn{\bra{P}\bar\chi_\alpha(b_T + b^- n + y^- n)\ket{X}}
    \grn{\bra{X}\chi_\delta(z^-n)\ket{P}}
    \\
    \nn&\quad\times
    \blu{\bra{0}\chi_\beta(b_T + b^+ \nbar + y^+ \nbar)\ket{p,X}}
    \blu{\bra{p,X}\bar\chi_\gamma(z^+ \nbar)\ket{0}}
    \\
    \nn&\quad\div
    \orn{\bra{0}\bigl[S_\nbar^\dagger S_n (b_T)\bigr]\ket{X}}
    \orn{\bra{X}\bigl[S_n^\dagger S_\nbar (0)\bigr]\ket{0}}\,.
\end{align}

\subsubsection*{NLP: kinematic correction + soft gluon}

Next, we consider the overlap subtraction for the kinematic corrections. The transverse derivative acting on the quark field results in an additional term in the overlap transformation,
\begin{align}
    &\blu{\bra{0}\img\partial_T^\rho\chi_\beta(b_T + b^+ \nbar)\ket{p,X}
    \bra{p,X}\bar\chi_\gamma(0)\ket{0}}
    \\
    &\to
    \blu{\bra{0}\img\partial_T^\rho\chi_\beta(b_T + b^+\nbar)} 
    \blu{\ket{p,X}}\blu{\bra{p,X}}\blu{\bar\chi_\gamma(0)}\blu{\ket{0}}
    \orn{\bra{0}}\orn{\bigl[S_\nbar^\dagger S_n(b_T)\bigr]}\orn{\ket{X}}
    \orn{\bra{X}}\orn{\bigl[S_n^\dagger S_\nbar (0)\bigr]}\orn{\ket{0}}
    \nonumber
    \\
    &\quad+
    \blu{\bra{0}}\blu{\chi_\beta(b_T + b^+\nbar)} 
    \blu{\ket{p,X}}\blu{\bra{p,X}}\blu{\bar\chi_\gamma(0)}\blu{\ket{0}}
    \orn{\bra{0}}\orn{\img\partial_T^\rho
    \bigl[S_\nbar^\dagger S_n (b_T)\bigr]}\orn{\ket{X}}
    \orn{\bra{X}}\orn{\bigl[S_n^\dagger S_\nbar (0)\bigr]}\orn{\ket{0}}\,.
    \nonumber
\end{align}
Inverting this relation and plugging it into the hadronic tensor then gives us
\begin{align}\label{eq:OverlapKinematicCorrection}
    \bigl[\mathcal{W}_\text{sub.}^{\mu\nu}\bigr]_{Cn}^{(3,2)}
    &=
    \biggl(-\frac{\nbar^\mu}{q^-}\biggr)
    \int\frac{\df^2 b}{(2\pi)^2}\,e^{+\img q_T \cdot b_T}\,
    \int\frac{\df b^+}{2\pi}\,e^{+\img q^- b^+}\,
    \int\frac{\df b^-}{2\pi}\,e^{+\img q^+ b^-}\,
    \\\nonumber
    &\quad\times
    (\gamma_T^\rho)_{\alpha\beta}
    (\gamma_T^\nu)_{\gamma\delta}
    \int\df y^+\,\df y^-\,C_1(y^+,y^-)\,
    \int\df z^+\,\df z^-\,C_1^*(z^+,z^-)\,
    \\\nonumber
    &\quad\times
    \grn{\bra{P}
    \bar\chi_\alpha(b_T + b^- n + y^- n)\ket{X}}
    \grn{\bra{X}\chi_\delta(z^- n)\ket{P}}
    \\\nonumber
    &\quad\times
    \blu{\bra{0}\img\partial_T^\rho\chi_\beta(b_T + b^+ \nbar + y^+ \nbar)\ket{p,X}}
    \blu{\bra{p,X}\bar\chi_\gamma(z^+\nbar)\ket{0}}
    \\\nonumber
    &\quad\div
    \orn{\bra{0} S_\nbar^\dagger S_n (b_T) \ket{X}}
    \orn{\bra{X} S_n^\dagger S_\nbar (0) \ket{0}}
    \\\nonumber
    &-\biggl(-\frac{\nbar^\mu}{q^-}\biggr)
    \int\frac{\df^2 b}{(2\pi)^2}\,e^{+\img q_T \cdot b_T}\,
    \int\frac{\df b^+}{2\pi}\,e^{+\img q^- b^+}\,
    \int\frac{\df b^-}{2\pi}\,e^{+\img q^+ b^-}\,
    \\\nonumber
    &\quad\times
    (\gamma_T^\rho)_{\alpha\beta}
    (\gamma_T^\nu)_{\gamma\delta}
    \int\df y^+\,\df y^-\,C_1(y^+,y^-)\,
    \int\df z^+\,\df z^-\,C_1^*(z^+,z^-)\,
    \\\nonumber
    &\quad\times
    \grn{\bra{P}\bar\chi_\alpha(b_T + b^- n + y^- n)\ket{X}}
    \grn{\bra{X}\chi_\delta(z^-n)\ket{P}}
    \\\nonumber
    &\quad\times
    \blu{\bra{0}\chi_\beta(b_T + b^+\nbar + y^+ \nbar)\ket{p,X}}
    \blu{\bra{p,X}\bar\chi_\gamma(z^+\nbar)\ket{0}}
    \\\nonumber
    &\quad\times
    \orn{\bra{0} \img\partial_T^\rho\bigl[S_\nbar^\dagger S_n (b_T)\bigr] \ket{X}}
    \orn{\bra{X} S_n^\dagger S_\nbar (0) \ket{0}}
    \\\nonumber
    &\quad\div
    \orn{\bra{0} S_\nbar^\dagger S_n (b_T) \ket{X}}
    \orn{\bra{X} S_n^\dagger S_\nbar (0) \ket{0}}
\end{align}

The above expression can be combined with the soft gluon contribution of eq.~\eqref{eq:HadronicTensorEn} to reduce the number of terms in the hadronic tensor. To do this we first use that the derivative of the soft Wilson lines in eq.~\eqref{eq:OverlapKinematicCorrection} can be rewritten in terms of the gauge invariant gluon field operators,
\begin{align}
    \orn{\img\partial_T^\rho \bigl[S_\nbar^\dagger S_n (b_T)\bigr]}
    &=
    -\orn{\mathcal{A}_{T,\nbar}^\rho S_\nbar^\dagger S_n}
    +\orn{S_\nbar^\dagger S_n \mathcal{A}_{T,n}^\rho}\,.
\end{align}
Multiplying the above identity by a factor of a half and subtracting another $\orn{\mathcal{A}_{T,n}}$ to both sides then gives us
\begin{align}
    \orn{S_\nbar^\dagger S_n\mathcal{A}_{T,n}^\rho }
    &=
    \frac{1}{2}\orn{\img\partial_T^\rho \bigl[S_\nbar^\dagger S_n\bigr]}
    +\frac{1}{2}\orn{\bigl[\orn{\mathcal{A}_{T,\nbar}^\rho S_\nbar^\dagger S_n}
        + S_\nbar^\dagger S_n\mathcal{A}_{T,n}^\rho \bigr]}\,.
\end{align}
The second term, which involves a combination of $\orn{\mathcal{A}_{T,n}}$ and $\orn{\mathcal{A}_{T,\nbar}}$, can be shown to vanish inside a soft matrix element using charge and parity conjugation,
\begin{align}
    \orn{\bra{0}\Bigl[\mathcal{A}_{T,\nbar}^\rho S_\nbar^\dagger S_n(b_T)
        + S_\nbar^\dagger S_n\mathcal{A}_{T,n}^\rho(b_T)\Bigr]\ket{X}
    \bra{X} S_n^\dagger S_\nbar(0)\ket{0}}
    &=0\,.
\end{align}
This allows us to combine the subtracted kinematic correction and the soft gluon contributions in the following way,
\begin{align}\label{eq:HadronicTensorSubtractedKinematicCorrections}
    &\bigl[\mathcal{W}_\text{sub.}^{\mu\nu}\bigr]_{Cn}^{(3,2)}
    +\bigl[\mathcal{W}_\text{sub.}^{\mu\nu}\bigr]_{En}^{(3,2)}
    \\\nonumber
    &=
    \biggl(-\frac{\nbar^\mu}{q^-}\biggr)
    \int\frac{\df^2 b}{(2\pi)^2}\,e^{+\img q_T \cdot b_T}\,
    \int\frac{\df b^+}{2\pi}\,e^{+\img q^- b^+}\,
    \int\frac{\df b^-}{2\pi}\,e^{+\img q^+ b^-}\,
    \\\nonumber
    &\quad\times
    (\gamma_T^\rho)_{\alpha\beta}
    (\gamma_T^\nu)_{\gamma\delta}
    \int\df y^+\,\df y^-\,C_1(y^+,y^-)\,
    \int\df z^+\,\df z^-\,C_1^*(z^+,z^-)\,
    \\\nonumber
    &\quad\times
    \grn{\bra{P}\img\partial_T^\rho
    \bar\chi_\alpha(b_T + b^- n + y^- n)\ket{X}}
    \grn{\bra{X}\chi_\delta(z^- n)\ket{P}}
    \\\nonumber
    &\quad\times
    \blu{\bra{0}\chi_\beta(b_T + b^+ \nbar + y^+ \nbar)\ket{p,X}}
    \blu{\bra{p,X}\bar\chi_\gamma(z^+\nbar)\ket{0}}
    \\\nonumber
    &\quad\div
    \orn{\bra{0} S_\nbar^\dagger S_n (b_T) \ket{X}}
    \orn{\bra{X} S_n^\dagger S_\nbar (0) \ket{0}}
    \\\nonumber
    &-\frac{1}{2}\biggl(-\frac{\nbar^\mu}{q^-}\biggr)
    \int\frac{\df^2 b}{(2\pi)^2}\,e^{+\img q_T \cdot b_T}\,
    \int\frac{\df b^+}{2\pi}\,e^{+\img q^- b^+}\,
    \int\frac{\df b^-}{2\pi}\,e^{+\img q^+ b^-}\,
    \\\nonumber
    &\quad\times
    (\gamma_T^\rho)_{\alpha\beta}
    (\gamma_T^\nu)_{\gamma\delta}
    \int\df y^+\,\df y^-\,C_1(y^+,y^-)\,
    \int\df z^+\,\df z^-\,C_1^*(z^+,z^-)\,
    \\\nonumber
    &\quad\times
    \grn{\bra{P}\bar\chi_\alpha(b_T + b^- n + y^- n)\ket{X}}
    \grn{\bra{X}\chi_\delta(z^- n)\ket{P}}
    \\\nonumber
    &\quad\times
    \blu{\bra{0}\chi_\beta(b_T + b^+ \nbar + y^+ \nbar)\ket{p,X}}
    \blu{\bra{p,X}\bar\chi_\gamma(z^+ \nbar)\ket{0}}
    \\\nonumber
    &\quad\times
    \orn{\bra{0} \img\partial_T^\rho\bigl[S_\nbar^\dagger S_n (b_T)\bigr] \ket{X}}
    \orn{\bra{0} S_n^\dagger S_\nbar (0) \ket{0}}
    \\\nonumber
    &\quad\div
    \orn{\bra{0} S_\nbar^\dagger S_n (b_T) \ket{X}}
    \orn{\bra{X} S_n^\dagger S_\nbar (0) \ket{0}}\,.
\end{align}
The extra factor of $-\frac{1}{2}$ will prove incredibly useful later on: it will allow us to replace the derivative on the unsubtracted TMD distribution, which contains rapidity divergences, with a TMD derivative which is manifestly rapidity-finite as defined in eq.~\eqref{eq:Fprime}.

\subsubsection*{NLP: genuine higher-twist correction}

Next, we examine the power correction, which involves an additional collinear gluon. The overlap transformation reads,
\begin{align}
    &\blu{\bra{0}
    \mathcal{A}_T^\rho(b_T + b^+ \nbar + y_2^+ \nbar)\chi_\beta(b_T + b^+ \nbar + y_1^+ \nbar)\ket{p,X}}
    \blu{\bra{p,X}\bar\chi_\gamma(z^+ \nbar)\ket{0}}
    \\\nonumber
    &\to
    \blu{\bra{0}\mathcal{A}_T^\rho(b_T + b^+ \nbar + y_2^+ \nbar)\chi_\beta(b_T + b^+ \nbar + y_1^+ \nbar)
    \ket{p,X}}
    \blu{\bra{p,X}\bar\chi_{\gamma}(z^+ \nbar)\ket{0}}
    \\\nonumber
    &\quad\qquad\times
    \orn{\bra{0} S_\nbar^\dagger S_n (b_T)
    \ket{X}\bra{X} S_n^\dagger S_\nbar (0)\ket{0}}
    \\\nonumber
    &\quad+
    \blu{\bra{0}\chi_\beta(b_T + b^+ \nbar + y_1^+ \nbar)\ket{p,X}}
    \blu{\bra{p,X}\bar\chi_{\gamma}(z^+ \nbar)\ket{0}}
    \\\nonumber
    &\quad\qquad\times
    \orn{\bra{0}\mathcal{A}_{T,\nbar}^\rho(b_T + y_2^+ \nbar)\, S_\nbar^\dagger S_n (b_T)
    \ket{X}\bra{X} S_n^\dagger S_\nbar (0)\ket{0}}\,,
\end{align}
where we find an additive term because $\orn{\mathcal{A}_T}$ is not suppressed compared to $\blu{\mathcal{A}_T}$. Inverting this relation to subtract the overlap gives us
\begin{align}\label{eq:OverlapHadronicTensorDn}
    \bigl[\mathcal{W}_\text{sub.}^{\mu\nu}\bigr]_{Dn}^{(3,2)}
    &=
    \biggl(\frac{n^\mu}{q^+}-\frac{\nbar^\mu}{q^-}\biggr)
    \int\frac{\df^2 b}{(2\pi)^2}\,e^{+\img q_T \cdot b_T}\,
    \int\frac{\df b^+}{2\pi}\,
    e^{+\img q^- b^+}\,
    \int\frac{\df b^-}{2\pi}\,
    e^{+\img q^+ b^-}\,
    \\\nonumber
    &\quad\qquad\times
    (\gamma_T^\rho)_{\alpha\beta}
    (\gamma_T^\nu)_{\gamma\delta}
    \int\df y_1^+\,\df y_2^+\df y^-\,C_2(\{y_1^+,y_2^+\},y^-)\,
    \int\df z^+\,\df z^-\,C_1^*(z^+,z^-)\,
    \\\nonumber
    &\quad\qquad\times
    \grn{\bra{P}
    \bar\chi_{i,\alpha}(b_T + y^- n)\ket{X}}
    \grn{\bra{X}\chi_{l,\delta}(z^- n)\ket{P}}
    \\\nonumber
    &\quad\qquad\times
    \bigl\{
    \blu{\bra{0}\bigl[\mathcal{A}_T^\rho(b_T + b^+ \nbar + y_2^+ \nbar)\chi(b_T + y_1^+ \nbar)
    \bigr]_{j,\beta}\ket{p,X}}
    \blu{\bra{p,X}\bar\chi_{k,\gamma}(z^+ \nbar)\ket{0}}
    \\\nonumber
    &\quad\qquad\qquad
    -\blu{\bra{0}\chi_{j,\beta}(b_T + b^+ \nbar + y_1^+ \nbar)\ket{p,X}}
    \blu{\bra{p,X}\bar\chi_{k,\gamma}(z^+ \nbar)\ket{0}}
    \\\nonumber
    &\quad\qquad\qquad\quad\times
    \orn{\bra{0}\mathcal{A}_{T,\nbar}^\rho(b_T + y_2^+ \nbar)\, S_\nbar^\dagger S_n (b_T)
    \ket{X}\bra{X} S_n^\dagger S_\nbar (0)\ket{0}}\bigr\}
    \\\nonumber
    &\quad\qquad\div
    \orn{\bra{0} S_\nbar^\dagger S_n (b_T) \ket{X}}
    \orn{\bra{X} S_n^\dagger S_\nbar (0) \ket{0}}\,.
\end{align}
In section \ref{sec:CancellationEndpointDivergences}, we will show that the second term within the curly brackets exactly renders the above contribution free of endpoint divergences.

\subsubsection{Power mixing}\label{sec:PowerMixing}

In the background field approach to factorization presented in this work, the presence of inverse derivatives on soft fields can lead to mixing between overlap subtraction terms of a higher-power contribution with those of a lower power. This can be seen as follows: for each operator that scales as $\lambda^k$ where an inverse derivative acts on a soft field, there is a counterpart where the field and its inverse derivative are collinear or anti-collinear that scales as $\lambda^{k+1}$. Consequently, the new terms that arise from the overlap subtraction of a power-suppressed operator can result in a power-enhanced contribution. 

To demonstrate this point, let us consider the overlap subtraction of one specific higher-order contribution. We choose to study the contribution of the operator in the last line of eq.~\eqref{eq:EffCurrentNLP3},
\begin{align}\label{eq:InverseDerivativeCollinearFildsContribution}
    \bigl[J_\text{eff}^\mu(0)\bigr]^{(3)}
    &\supset
    -\frac{1}{2}(\gamma_T^\mu\gamma^-\gamma_T^\rho)_{\alpha\beta}
    \int\df y_1^+\,\df y_2^+\, \df y^-\,
    C_5\bigl(\{y_1^+,y_2^+\},y^-\bigr)\,
    \\\nonumber
    &\qquad\times
    \blu{\bar\chi_\alpha(y_1^+ \nbar)}\,
    \orn{S_n^\dagger S_\nbar}\,
    \grn{\mathcal{A}_T^\rho(y^-)}\,
    \orn{S_\nbar^\dagger S_n}\,
    \blu{\frac{1}{\img\partial^-}\chi_\beta(y_2^+ \nbar)}+\text{h.c.}\,,\,,
\end{align}
which is the collinear analog of the soft quark contribution
\begin{align}
    \bigl[J_\text{eff}^\mu(0)\bigr]^{(2.5)}
    &=
    -\frac{1}{2}(\gamma_T^\mu\gamma^-\gamma_T^\rho)_{\alpha\beta}
    \int\df y^+\,\df z^+\, \df y^-\, C_4(y^+,z^+,y^-)\,
    \\\nonumber
    &\qquad\times
    \blu{\bar\chi_\alpha(y^+ \nbar)}\,
    \orn{S_n^\dagger S_\nbar}\,
    \grn{\mathcal{A}_T^\rho(y^- n)}\,
    \orn{\frac{1}{\img\partial^-}\Psi_{\nbar,\beta}(z^+ \nbar)}
    +(\blu{n}\leftrightarrow\grn{\nbar})+\text{h.c.}\,.
\end{align}
To keep the analysis below simple we consider only the abelian case. There are several ways that the operator in eq.~\eqref{eq:InverseDerivativeCollinearFildsContribution} can contribute to the hadronic tensor, however, for the current purposes we only consider the following contribution,
\begin{align}
    &\bigl[\mathcal{W}^{\mu\nu}\bigr]^{(3,3)}\supset
    \\\nonumber
    &-\frac{1}{4}
    (\gamma_T^\mu\gamma^-\gamma_T^\rho)_{\alpha\beta}
    (\gamma_T^\sigma\gamma^-\gamma_T^\nu)_{\gamma\delta}
    \int\frac{\df^2 b}{(2\pi)^2}\,e^{+\img q_T \cdot b_T}\,
    \int\frac{\df b^+}{2\pi}\,e^{+\img q^- b^+}\,
    \int\frac{\df b^-}{2\pi}\,e^{+\img q^+ b^-}\,
    \\\nonumber
    &\quad\times
    \int\df y_1^+\,\df y_2^+\,\df y^-\,C_5\bigl(\{y_1^+,y_2^+\},y^-\bigr)\,
    \int\df z_1^+\,\df z_2^+\,\df z^-\,C_5^*\bigl(\{y_1^+,y_2^+\},y^-\bigr)\,
    \\\nonumber
    &\quad\times
    \grn{\bra{P} \mathcal{A}_T^\rho(b_T + y^- n)\ket{X}}
    \grn{\bra{X} \mathcal{A}_T^\sigma(z^- n)\ket{P}}
    \\\nonumber
    &\quad\times
    \blu{\bra{0}\bar\chi_\alpha(b_T + b^+ \nbar + y_1^+ \nbar)
    \frac{1}{\img\partial^-}\chi_\beta(b_T + b^+ \nbar + y_2^+ \nbar)\ket{p,X}}
    \blu{\bra{p,X}\frac{1}{\img\partial^-}\bar\chi_\gamma(z_2^+ \nbar)
    \chi_\delta(z_1^+ \nbar)\ket{0}}\,.
\end{align}
Note that in the abelian case, the soft factor for this contribution is unity, as the soft Wilson lines in eq.~\eqref{eq:InverseDerivativeCollinearFildsContribution} can be commuted through $\grn{\mathcal{A}}$ upon which they cancel out.

Consider now applying the overlap transformation of eq.~\eqref{eq:OverlapTransformationSCET} to the above collinear matrix element. Due to the presence of the inverse derivative, the overlap transformation can enhance the scaling of the collinear quark field,
\begin{align}\label{eq:OverlapInverseCollinearDerivative}
    \blu{\frac{1}{\img\partial^-}\chi}&\to
    \orn{S_\nbar^\dagger S_n}\blu{\frac{1}{\img\partial^-}\chi}
    +\orn{\frac{1}{\img\partial^-}\Psi_\nbar}\,.
\end{align}
Here, the first term on the r.h.s.~scales as $\lambda$ while the second term scales as $\lambda^{\frac12}$. As a result of this power enhancement, the overlap subtraction of \eqref{eq:OverlapInverseCollinearDerivative} gets lifted in the power counting and must be accounted for. In fact, at leading order where the Wilson coefficients are trivial delta functions, we find that the contribution of this subtraction term completely cancels that of the soft quark operator in eq.~\eqref{eq:HadronicTensorHnbar}.

On a final note, one may be concerned about the fact that the subtraction of the overlap region can lift the power suppression of certain operators. It indeed makes the derivation of factorization formulas at sub-leading powers more cumbersome, as at each order in the power counting, one also needs to consider the overlap subtraction for higher-power terms. At next-to-leading power, however, this mixing between power corrections does not pose a problem. This is because all subtraction terms of higher-power operators result in soft matrix elements that vanish due to boost invariance, similar to what we discussed in the previous section. Moreover, this apparent mixing of power counting can potentially be avoided entirely by choosing a convenient rapidity regulator.

\subsubsection{Cancellation of endpoint divergences}\label{sec:CancellationEndpointDivergences}

In the introduction of this section, we discussed that next-to-leading power factorization formulas contain a new type of divergence: endpoint divergences. In this section, we briefly highlight how these divergences appear, and show how the cancellation of these divergences can be made explicit.

First, let us illustrate how endpoint divergences arise. They arise from the contributions to the hadronic tensor that involve a sub-leading collinear operator with an additional gluon field, as given in eqs.~\eqref{eq:HadronicTensorDn}-\eqref{eq:HadronicTensorDnbar}. This contribution to the hadronic tensor has a convolution of the form, 
\begin{align}\label{eq:ConvolutionHardWithTwist3}
    \int\df\xi\,H_2(\xi,q^2)\,\grn{F_{21}(x,\xi,b_T)}\,\blu{D_{11}(z,b_T)}\,,
\end{align}
where $D_{11}$ is the standard unpolarized TMDFF (which can be replaced by the corresponding jet function). The two new ingredients in the above equation are the hard function $H_2$ and the sub-leading power parton distribution $F_{21}$. The hard function $H_2$ can be expressed in terms of the Wilson coefficients that appear in the effective current operator by,
\begin{align}\label{eq:H2Definition}
    H_2(\xi,q^2)
    &=
    \int\df y_1^+\,\df y_2^+\,\df y^-\,
    e^{+\img y_1^+ \bar\xi q^-}\,e^{+\img y_2^+ \xi q^-}\,e^{+\img y^- q^+}\,
    C_2^*\bigl(\{y_1^+,y_2^+\},y^-\bigr)
    \\\nonumber
    &\quad\times
    \int\df z^+\,\df z^-\,
    e^{-\img z^+ q^-}\,e^{-\img z^- q^+}\,
    C_1(z^+,z^-)\,.
\end{align}
The naive (i.e.~without a proper overlap subtraction) parton distribution $F_{21}$ can be expressed in terms of a matrix element involving the sub-leading power anti-collinear operator,
\begin{align}
    \grn{\bigl[F_{21}^\text{naive}(x,\xi,b_T)\bigr]_{\delta\alpha}}
    &=
    \int\frac{\df b_1^-}{2\pi}\,\frac{\df b_2^-}{2\pi}\,
    e^{-\img \bar\xi x b_1^- P^+}\,e^{-\img \xi x b_2^- P^+}\,
    \\\nonumber
    &\quad\times
    \grn{\bra{P}\bigl[\bar\chi(b_T + b_1^- n)
    \mathcal{A}_T^\rho(b_T + b_2^- \nbar)\bigr]_{i,\alpha}\ket{X}}
    \grn{\bra{X}\chi_{i,\delta}(0)\ket{P}}\,.
\end{align}
To keep the discussion general we work with open spin indices on the parton distributions, avoiding the need for repeating the analysis below for the different types of distributions that can arise after applying Fierz relations.

To illustrate the presence of divergences in the convolution in eq.~\eqref{eq:ConvolutionHardWithTwist3} using $F_{21}^{\rm naive}$, we evaluate this parton distribution function to leading order in a partonic state. For simplicity, we choose an external quark state with momentum $P=(P^+,0^-,0_T)$. From its definition, we find the following expression,
\begin{align}
    &\grn{\bigl[F_{21}^\text{naive}(x,\xi,b_T)\bigr]^{\text{LO}}_{\delta\alpha}}
    \\\nonumber
    &=
    \img x P^+\int\frac{\df b_1^-}{2\pi}\,\frac{\df b_2^-}{2\pi}
    e^{-\img \bar\xi x b_1^- P^+}e^{\img \xi x b_2^- P^+}\!
    \int\frac{\df^d k}{(2\pi)^d}(2\pi)\delta(k^2)\,\theta(k^0)
    e^{-\img b_T \cdot (k_T - p_T)}
    e^{+\img b_1^- P^+}e^{-\img b_2^- k^+}
    \\\nonumber
    &\quad\times
    \biggl[
    \bar{u}(P) \frac{\gamma^+\gamma^-}{2}
    \biggl(g_T^{\rho\mu} - \frac{k_T^\rho n^\mu}{k^+ + \img\delta^+}\biggr)
    \epsilon_\mu(k)\biggr]_\alpha
    \biggl[\epsilon_\nu^*(k) 
    \frac{\gamma^-\gamma^+}{2}
    \frac{\img(\fsl{p}-\fsl{k})}{(p-k)^2+\img0} 
    (\img g \gamma^\nu) u(P)\biggr]_\delta\,.
\end{align}
The above expression can be evaluated by first summing and averaging over spins, then eliminating the integrals over $k^+$ and $k^-$ by using the delta functions from the Fourier transformation and the on-shell condition respectively, and then performing the remaining integration over $k_T$. This results in
\begin{align}\label{eq:F21naivePole}
    \grn{\bigl[F_{21}^\text{naive}(x,\xi,b_T)\bigr]^{\text{LO}}_{\delta\alpha}}
    &=
    a_s\,\frac{b_T^\sigma}{b_T^2}
    \theta\bigl(-\xi x\bigr)\,
    \biggl[
    \gamma_T^\sigma\gamma_T^\rho\gamma^- x
    -2(1+\xi x)
    \frac{g_T^{\rho\sigma}}{\xi-\img\delta^+/xP^+}\gamma^-
    \biggr]_{\delta\alpha}\,,
\end{align}
which contains a singularity at $\xi\to 0$. Upon convolving with the hard function, which also depends on $\xi$, this singularity leads to a so-called endpoint divergence. While these divergences can be regularized (in the above formula by $\delta^+$) and cancel in the final cross section, their presence prohibits us from defining physical factorization ingredients that are free from divergences upon convolution.

In making the cancellation of endpoint divergences explicit in the factorization formula, we use the terms arising from the overlap subtraction process to define a physical parton distribution function. To demonstrate this approach, we first consider the subtraction term that appears in eq.~\eqref{eq:OverlapHadronicTensorDn}, but for the anti-collinear distribution, to define 
\begin{align}
    \grn{\bigl[F_{21}^\text{o.s.}(x,\xi,b_T)\bigr]_{\delta\alpha}}
    &=\nonumber
    \img x P^+\int\frac{\df b_1^-}{2\pi}\,\frac{\df b_2^-}{2\pi}\,
    e^{-\img \bar\xi x b_1^- P^+}\,e^{-\img \xi x b_2^- P^+}\,
    \grn{\bra{P}\bar\chi_\alpha(b_T + b_1^- n)\ket{X}}
    \grn{\bra{X}\chi_\delta(0)\ket{P}}
    \\
    &\quad\times
    \orn{\bra{0} S_\nbar^\dagger S_n (b_T)\,
    \mathcal{A}_{T,n}^\rho(b_T + b_2^- n)
    \ket{X}\bra{X} S_n^\dagger S_\nbar (0)\ket{0}}\,.
\end{align}
This overlap subtraction (indicated by the superscript o.s.) needs to be done to obtain the physical distribution.
As for the naive parton distribution, we now calculate the subtraction term to leading order in the coupling. From its definition, we obtain,
\begin{align}
    &\grn{\bigl[F_{21}^\text{o.s.}(x,\xi,b_T)\bigr]^{\text{LO}}_{\delta\alpha}}
    \\\nonumber
    &=
    \img x P^+\int\frac{\df b_1^-}{2\pi}\frac{\df b_2^-}{2\pi}
    e^{-\img \bar\xi x b_1^- P^+}e^{-\img \xi x b_2^- P^+}\!
    \int\frac{\df^d k}{(2\pi)^d}(2\pi)\delta(k^2)\theta(k^0)
    e^{-\img b_T \cdot (k_T - p_T)}
    e^{+\img b_1^- P^+}\,e^{-\img b_2^- k^+}
    \\\nonumber
    &\quad\times
    \biggl[\bar{u}(P) \frac{\gamma^+\gamma^-}{2}\biggr]_\alpha
    \biggl[\frac{\gamma^-\gamma^+}{2} u(P)\biggr]_\delta
    \biggl(-\frac{k_T^\rho n^\mu}{k^+ + \img\delta^+}\biggr)
    \epsilon_\mu(k)\epsilon^*_\nu(k)
    \biggl(\frac{n^\nu}{-k^+ + \img\delta^+} 
    -\frac{\bar n^\nu}{-k^- + \img\delta^-}\biggr),
\end{align}
which evaluates to
\begin{align*}
    \grn{\bigl[F_{21}^\text{o.s.}(x,\xi,b_T)\bigr]^{\text{LO}}_{\delta\alpha}}
    &=
    -2a_s\,\frac{b_T^\rho}{b_T^2}
    \theta\bigl(-\xi x\bigr)\,
    \biggl[\frac{1}{\xi-\img\delta^+/xP^+}\gamma^+\biggr]_{\delta\alpha}\,.
\end{align*}

Comparing the subtraction term to the leading-order result for the naive parton distribution in \eqref{eq:F21naivePole}, we see that the singular terms in the $\xi\to0$ limit are identical. This means that, at least to leading order in the coupling, we have 
\begin{align}
    \lim_{\xi\to0}
    \xi \bigg(\grn{\bigl[F_{21}^\text{naive}(x,\xi,b_T)\bigr]^{\text{LO}}_{\delta\alpha}}
    -\grn{\bigl[F_{21}^\text{o.s.}(x,\xi,b_T)\bigr]^{\text{LO}}_{\delta\alpha}}\bigg)
    &=
    0\,,
\end{align}
i.e.~the expression in brackets is non-singular in the $\xi \to 0 $ limit.
We expect that this relation holds to all orders in the coupling and on a non-perturbative level as well. This is because the subtraction term, as defined in eq.~\eqref{eq:OverlapHadronicTensorDn}, corresponds exactly to the limit where the collinear gluon field becomes soft, which in turn can be associated with the limit where $\xi\to0$. Therefore, we conclude that the convolution between the hard function and the subtracted parton distribution is free of endpoint divergences,
\begin{align}
    \int\df\xi\,H_2(\xi,Q^2)\,
    \Bigl(\grn{F_{21}^\text{naive}(x,\xi,b_T)}
    -\grn{F_{21}^\text{o.s.}(x,\xi,b_T)}\Bigr)\,
    \blu{D_{11}(z,b_T)}
    &=\text{finite}\,.
\end{align}

\subsection{The subtracted hadronic tensor}\label{sec:SubtractedHadronicTensor}

In this section, we present our result for the (overlap-subtracted) hadronic tensor for SIDIS. Here, we define a set of sub-leading-power TMDPDFs and fragmentation/jet functions  (with open spin) indices that are free of rapidity divergences and do not result in endpoint divergences. The application of Fierz relations and the contraction of the spin indices is left to the next section. To organize the presentation of the results, we split up the hadronic tensor into three parts: a leading power contribution, a kinematic power correction, and a genuine higher-twist correction,
\begin{align}\label{eq:HadronicTensorLabels}
    W^{\mu\nu}&=
    W^{\mu\nu}_\text{LP}+W^{\mu\nu}_\text{kNLP}+W^{\mu\nu}_\text{gNLP}\,.
\end{align}

To define the physical distributions, we will exploit the exponentiation of the soft function. The soft function is defined as
\begin{align}\label{eq:SoftFunctionDefinition}
    \orn{S^\text{bare}(b_T,2\delta^+\delta^-)}
    &=
    \orn{\frac{1}{N_c}\tr\Bigl[\bra{0}\bigl[S_\nbar^\dagger S_n (b_T)\bigr]\ket{X}
    \bra{X}\bigl[S_n^\dagger S_\nbar (0)\bigr]\ket{0}\Bigr]}\,,
\end{align}
where the dependence on the $\delta$-regulator on the right-hand-side is hidden inside the Wilson lines according to eq.~\eqref{eq:DeltaWilsonLines}. With the $\delta$-regulator, the exponentiation of the soft function reads,
\begin{align}\label{eq:SoftFunctionExponentiation}
    \orn{S^\text{bare}(b_T,2\delta^+\delta^-)}&=
    \exp\Bigl[K^\text{bare}(b_T)\log(2\delta^+\delta^-)
    +s^\text{bare}(b_T)\Bigr]\,,
\end{align}
where $K^\text{bare}$ is the Collins-Soper kernel for bare TMD distributions and $s^\text{bare}$ is a rapidity independent remainder.

First, let us present the leading-power contribution to the hadronic tensor. To define the leading-power TMD distributions, we can use eq.~\eqref{eq:SoftFunctionDefinition} to absorb a square root of the soft function into the definitions of the TMD distributions,
\begin{align}
    \orn{S^\text{bare}(b_T,2\delta^+\delta^-)}
    &=
    \sqrt{\orn{S^\text{bare}\bigl(b_T,    \zeta(\delta^+/q^+)^2\bigr)}}\,\,\sqrt{\orn{S^\text{bare}\bigl(b_T,\bar\zeta(\delta^-/q^-)^2\bigr)}}\,.
\end{align}
where $\zeta$ and $\bar\zeta$ are referred to as rapidity scales and must satisfy
\begin{align}
    \zeta\bar\zeta=(2q^+ q^-)^2=Q^4+\mathcal{O}(q_T^2 Q^2)\,.
\end{align}
We then define rapidity-finite TMD parton distributions and fragmentation/jet functions as follows,
\begin{align}\label{eq:F_11_indices}
    \grn{\bigl[\mathcal{F}_{q,11}^\text{bare}(x,b_T,\zeta)\bigr]_{\delta\alpha}}
    &=
    \int\frac{\df b^-}{2\pi}\,e^{+\img b^- q^+}\,
    \frac{\grn{\bra{P} \bar\chi_{i,\alpha}(b_T + b^-n)\ket{X}
    \bra{X}\chi_{i,\delta}(0)\ket{P}}}
    {\sqrt{\orn{S\big(b_T,\zeta(\delta^+/q^+)^2\big)}}}\,,
    \\
    \grn{\bigl[\mathcal{F}_{\qbar,11}^\text{bare}(x,b_T,\zeta)\bigr]_{\beta\gamma}}
    &=
    \int\frac{\df b^-}{2\pi}\,e^{+\img b^- q^+}\,
    \frac{\grn{\bra{P}\chi_{i,\beta}(b_T + b^-n)\ket{X}
    \bra{X}\bar\chi_{i,\gamma}(0)\ket{P}}}
    {\sqrt{\orn{S\big(b_T,\zeta(\delta^+/q^+)^2\big)}}}\,,
    \\
    \blu{\bigl[\mathcal{D}_{q,11}^\text{bare}(z,b_T,\bar\zeta)\bigr]_{\beta\gamma}}
    &=
    \int\frac{\df b^+}{2\pi}\,e^{+\img b^+ q^-}\,
    \frac{\blu{\bra{0}\chi_{i,\beta}(b_T + b^+ \nbar)\ket{p,X}
    \bra{p,X}\bar\chi_{i,\gamma}(0)\ket{0}}}
    {\sqrt{\orn{S\big(b_T,\bar\zeta(\delta^-/q^-)^2\big)}}}\,,
    \\\label{eq:D_11_indices}
    \blu{\bigl[\mathcal{D}_{\qbar,11}^\text{bare}(z,b_T,\bar\zeta)\bigr]_{\delta\alpha}}
    &=
    \int\frac{\df b^+}{2\pi}\,e^{+\img b^+ q^-}\,
    \frac{\blu{\bra{0}\bar\chi_{i,\alpha}(b_T + b^+ \nbar)\ket{p,X}
    \bra{p,X}\chi_{i,\delta}(0)\ket{0}}}
    {\sqrt{\orn{S\big(b_T,\bar\zeta(\delta^-/q^-)^2\big)}}}\,,
\end{align}
and the (bare) hard function as
\begin{align}
    H_1(Q^2)\approx H_1(2q^+q^-)
    &=
    \int\df y^+\,\df y^-\,
    e^{+\img y^+ q^-}\,e^{+\img y^- q^+}\,C_1^*(y^+,y^-)
    \\\nonumber
    &\quad\times
    \int\df z^+\,\df z^-\,
    e^{-\img z^+ q^-}\,e^{-\img z^- q^+}\,
    C_1(z^+,z^-)\,.
\end{align}
With these definitions, the leading-power contribution to the hadronic tensor reads
\begin{align}\label{eq:WLP}
    W_{\text{LP}}^{\mu\nu}(q)=\bigl[\mathcal{W}_\text{sub.}^{\mu\nu}\bigr]_{A}^{(2,2)}
    &=
    \frac{1}{N_c} (\gamma_T^\mu)_{\alpha\beta} (\gamma_T^\nu)_{\gamma\delta}\,
    H_1(q^2)\int\frac{\df^2 b_T}{(2\pi)^2}\,e^{\img b_T \cdot q_T}\,
    \\\nonumber
    &\times\sum_q\Bigl\{
    \grn{\bigl[\mathcal{F}_{q,11}\bigr]_{\delta\alpha}}\,
    \blu{\bigl[\mathcal{D}_{q,11}\bigr]_{\beta\gamma}}
    +
    \grn{\bigl[\mathcal{F}_{\qbar,11}\bigr]_{\beta\gamma}}\,
    \blu{\bigl[\mathcal{D}_{\qbar,11}\bigr]_{\delta\alpha}}
    \Bigr\}\,. 
\end{align}

Next, we consider the kinematic power corrections. We first use eq.~\eqref{eq:SoftFunctionExponentiation} to rewrite the terms involving transverse derivatives of unsubtracted TMD distributions and the soft function in terms of derivatives of the Collins-Soper kernel via 
\begin{align*}
    \frac{\orn{\partial_T^\rho S^\text{bare}(b_T,\Delta_1^2)}}
    {\orn{S^\text{bare}(b_T,\Delta_1^2)}}
    -\frac{\orn{\partial_T^\rho S^\text{bare}(b_T,\Delta_2^2)}}
    {\orn{S^\text{bare}(b_T,\Delta_2^2)}}
    &=
    \partial_T^\rho K^\text{bare}(b_T)\ln\biggl(\frac{\Delta_1^2}{\Delta_2^2}\biggr)\,.
\end{align*}
The kinematic power correction is then written as,
\begin{align}\label{eq:WkNLP}
    W_{\text{kNLP}}^{\mu\nu}(q)
    &=
    \bigl[\mathcal{W}_\text{sub.}^{\mu\nu}\bigr]_{C+E}^{(3,2)}
    +\bigl[\mathcal{W}_\text{sub.}^{\mu\nu}\bigr]_{C+E}^{(2,3)}
    \\\nonumber
    &=
    -\frac{\img}{N_c} \frac{1}{q^+}
    \bigl[n^\mu (\gamma_T^\rho)_{\alpha\beta} (\gamma_T^\nu)_{\gamma\delta} 
        + n^\nu (\gamma_T^\mu)_{\alpha\beta} (\gamma_T^\rho)_{\gamma\delta}\bigr]\,
    H_1(q^2)\int\frac{\df^2 b_T}{(2\pi)^2}\,e^{\img b_T \cdot q_T}\,
    \\\nonumber
    &\quad\times\sum_q
    \biggl\{
    \grn{\bigl[\partial_\rho \mathcal{F}_{q,11}\bigr]_{\delta\alpha}}\,
    \blu{\bigl[\mathcal{D}_{q,11}\bigr]_{\beta\gamma}}
    +\frac{1}{2}\bigl[\partial_\rho K\bigr]\ln\biggl(\frac{\zeta}{\bar\zeta}\biggr)\,
    \grn{\bigl[\mathcal{F}_{q,11}\bigr]_{\delta\alpha}}\,
    \blu{\bigl[\mathcal{D}_{q,11}\bigr]_{\beta\gamma}}
    \\\nonumber
    &\quad\qquad+
    \grn{\bigl[\partial_\rho \mathcal{F}_{\qbar,11}\bigr]_{\beta\gamma}}\,
    \blu{\bigl[\mathcal{D}_{\qbar,11}\bigr]_{\delta\alpha}}
    +\frac{1}{2}\bigl[\partial_\rho K\bigr]\ln\biggl(\frac{\zeta}{\bar\zeta}\biggr)\,
    \grn{\bigl[\mathcal{F}_{\qbar,11}\bigr]_{\beta\gamma}}\,
    \blu{\bigl[\mathcal{D}_{\qbar,11}\bigr]_{\delta\alpha}}
    \biggr\}
    \\\nonumber
    &\quad
    -\frac{\img}{N_c}
    \frac{1}{q^-}
    \bigl[\bar{n}^\mu (\gamma_T^\rho)_{\alpha\beta} (\gamma_T^\nu)_{\gamma\delta} 
        + \bar{n}^\nu (\gamma_T^\mu)_{\alpha\beta} (\gamma_T^\rho)_{\gamma\delta}\bigr]\,
    H_1(q^2)\int\frac{\df^2 b_T}{(2\pi)^2}\,e^{\img b_T \cdot q_T}\,
    \\\nonumber
    &\quad\times\sum_q
    \biggl\{
    \grn{\bigl[\mathcal{F}_{q,11}\bigr]_{\delta\alpha}}\,
    \blu{\bigl[\partial_\rho \mathcal{D}_{q,11}\bigr]_{\beta\gamma}}
    +\frac{1}{2}\bigl[\partial_\rho K\bigr]\ln\biggl(\frac{\bar\zeta}{\zeta}\biggr)\,
    \grn{\bigl[\mathcal{F}_{q,11}\bigr]_{\delta\alpha}}\,
    \blu{\bigl[\mathcal{D}_{q,11}\bigr]_{\beta\gamma}}
    \\\nonumber
    &\quad\qquad+
    \grn{\bigl[\mathcal{F}_{\qbar,11}\bigr]_{\beta\gamma}}\,
    \blu{\bigl[\partial_\rho \mathcal{D}_{\qbar,11}\bigr]_{\delta\alpha}}
    +\frac{1}{2}\bigl[\partial_\rho K\bigr]\ln\biggl(\frac{\bar\zeta}{\zeta}\biggr)\,
    \grn{\bigl[\mathcal{F}_{\qbar,11}\bigr]_{\beta\gamma}}\,
    \blu{\bigl[\mathcal{D}_{\qbar,11}\bigr]_{\delta\alpha}}
    \biggr\}\,.
\end{align}

Finally, we present the result for the genuine higher-twist correction. We first define a set of open-spin-index TMDPDFs and TMDFFs as
\begin{align}
    \label{eq:F_21_indices}&\grn{\bigl[\mathcal{F}_{q,21}^\text{bare}(x,\xi,b_T,\zeta)\bigr]_{\delta\alpha}}
    \\\nonumber
    &=\img q^+\int\frac{\df b_1^-}{2\pi}\,\frac{\df b_2^-}{2\pi}\,
    e^{-\img \bar\xi b_1^- q^+}\,e^{-\img \xi b_2^- q^+}\,
    \\\nonumber
    &\qquad\times
    \biggl\{
    \frac{\grn{\bra{P}\bigl[\bar\chi(b_T + b_1^- n)
    \fsl{\mathcal{A}}_T(b_T + b_2^- n)\bigr]_{i,\alpha}\ket{X}
    \bra{X}\chi_{i,\delta}(0)\ket{P}}}
    {\sqrt{\orn{S(b_T,\zeta(\delta^+/q^+)^2})}}
    \\\nonumber
    &\qquad\qquad
    -\frac{\grn{\bra{P}\bigl[\bar\chi(b_T + b_1^- n)\,
    \gamma_T^\rho\bigr]_{i,\alpha}\ket{X}
    \bra{X}\chi_{i,\delta}(0)\ket{P}}}
    {\sqrt{\orn{S(b_T,\zeta(\delta^+/q^+)^2})}}
    \\\nonumber
    &\qquad\qquad\qquad\times
    \orn{\frac{1}{N_c}\tr\Bigl[\bra{0} S_\nbar^\dagger S_n (b_T)\,
    \mathcal{A}_{T,n}^\rho(b_T + b_2^- n)\ket{X}
    \bra{X} S_n^\dagger S_\nbar (0) \ket{0}\Bigr]}
    \biggr\}\,,
    \\[6ex]
    &\grn{\bigl[\mathcal{F}_{\qbar,21}^\text{bare}(x,\xi,b_T,\zeta)\bigr]_{\delta\alpha}}
    \\\nonumber
    &=\img q^+\int\frac{\df b_1^-}{2\pi}\,\frac{\df b_2^-}{2\pi}\,
    e^{-\img \bar\xi b_1^- q^+}\,e^{-\img \xi b_2^- q^+}\,
    \\\nonumber
    &\qquad\times
    \biggl\{
    \frac{\grn{\bra{P}\bigl[\fsl{\mathcal{A}}_T(b_T + b_2^- n)
    \chi(b_T + b_1^- n)\bigr]_{i,\delta}\ket{X}
    \bra{X}\chi_{i,\alpha}(0)\ket{P}}}
    {\sqrt{\orn{S(b_T,\zeta(\delta^+/q^+)^2})}}
    \\\nonumber
    &\qquad\qquad
    -\frac{\grn{\bra{P}\bigl[\gamma_T^\rho
    \chi(b_T + b_1^- n)\,\bigr]_{i,\delta}\ket{X}
    \bra{X}\chi_{i,\alpha}(0)\ket{P}}}
    {\sqrt{\orn{S(b_T,\zeta(\delta^+/q^+)^2})}}
    \\\nonumber
    &\qquad\qquad\qquad\times
    \orn{\frac{1}{N_c}\tr\Bigl[\bra{0} \mathcal{A}_{T,n}^\rho(b_T + b_2^- n)\,
    S_n^\dagger S_\nbar (b_T)\ket{X}
    \bra{X} S_\nbar^\dagger S_n (0) \ket{0}\Bigr]}
    \biggr\}\,,
    \\[6ex]
    &\grn{\bigl[\mathcal{F}_{q,12}^\text{bare}(x,\xi,b_T,\zeta)\bigr]_{\delta\alpha}}
    \\\nonumber
    &=-\img q^+\int\frac{\df b_1^-}{2\pi}\,\frac{\df b_2^-}{2\pi}\,
    e^{+\img \bar\xi b_1^- q^+}\,e^{+\img \xi b_2^- q^+}\,
    \\\nonumber
    &\qquad\times
    \biggl\{
    \frac{\grn{\bra{P}\bar\chi_{i,\alpha}(b_T)\ket{X}
    \bra{X}\bigl[\fsl{\mathcal{A}}_T(b_2^- n)
    \chi(b_1^- n)\bigr]_{i,\delta}\ket{P}}}
    {\sqrt{\orn{S(b_T,\zeta(\delta^+/q^+)^2})}}
    \\\nonumber
    &\qquad\qquad
    -\frac{\grn{\bra{P}\bar\chi_{i,\alpha}(b_T)\ket{X}
    \bra{X}\bigl[\gamma_T^\rho\chi(b_1^- n)\bigr]_{i,\delta}\ket{P}}}
    {\sqrt{\orn{S(b_T,\zeta(\delta^+/q^+)^2})}}
    \\\nonumber
    &\qquad\qquad\qquad\times
    \orn{\frac{1}{N_c}\tr\Bigl[\bra{0} S_\nbar^\dagger S_n (b_T)\ket{X}
    \bra{X}\mathcal{A}_{T,n}^\rho(b_2^- n)\,S_n^\dagger S_\nbar(0)\ket{0}\Bigr]}
    \biggr\}\,,
    \\[6ex]
    \label{eq:F_12_indices}&\grn{\bigl[\mathcal{F}_{\qbar,12}^\text{bare}(x,\xi,b_T,\zeta)\bigr]_{\delta\alpha}}
    \\\nonumber
    &=-\img q^+\int\frac{\df b_1^-}{2\pi}\,\frac{\df b_2^-}{2\pi}\,
    e^{+\img \bar\xi b_1^- q^+}\,e^{+\img \xi b_2^- q^+}\,
    \\\nonumber
    &\qquad\times
    \biggl\{
    \frac{\grn{\bra{P}\chi_{i,\delta}(b_T)\ket{X}
    \bra{X}\bigl[\bar\chi(b_1^- n)
    \fsl{\mathcal{A}}_T(b_2^- n)\bigr]_{i,\alpha}\ket{P}}}
    {\sqrt{\orn{S(b_T,\zeta(\delta^+/q^+)^2})}}
    \\\nonumber
    &\qquad\qquad
    -\frac{\grn{\bra{P}\chi_{i,\delta}(b_T)\ket{X}
    \bra{X}\bigl[\bar\chi(b_1^- n)\gamma_T^\rho\bigr]_{i,\alpha}\ket{P}}}
    {\sqrt{\orn{S(b_T,\zeta(\delta^+/q^+)^2})}}
    \\\nonumber
    &\qquad\qquad\qquad\times
    \orn{\frac{1}{N_c}\tr\Bigl[\bra{0} S_n^\dagger S_\nbar (b_T)\ket{X}
    \bra{X} S_\nbar^\dagger S_n(0)\,\mathcal{A}_{T,n}^\rho(b_2^- n)\ket{0}\Bigr]}
    \biggr\}\,,
    \\[6ex]
    \label{eq:D_21_indices}&\blu{\bigl[\mathcal{D}_{q,21}^\text{bare}(z,\xi,b_T,\zeta)\bigr]_{\delta\alpha}}
    \\\nonumber
    &=\img q^-\int\frac{\df b_1^+}{2\pi}\,\frac{\df b_2^+}{2\pi}\,
    e^{+\img \bar\xi b_1^+ q^-}\,e^{+\img \xi b_2^+ q^-}\,
    \\\nonumber
    &\qquad\times
    \biggl\{
    \frac{\blu{\bra{0}\bigl[\fsl{\mathcal{A}}_T(b_T + b_2^- n)
    \chi(b_T + b_1^- n)\bigr]_{i,\delta}\ket{p,X}
    \bra{p,X}\chi_{i,\alpha}(0)\ket{0}}}
    {\sqrt{\orn{S\big(b_T,\bar\zeta(\delta^-/q^-)^2\big)}}}
    \\\nonumber
    &\qquad\qquad
    -\frac{\blu{\bra{0}\bigl[\gamma_T^\rho
    \chi(b_T + b_1^- n)\,\bigr]_{i,\delta}\ket{p,X}
    \bra{p,X}\chi_{i,\alpha}(0)\ket{0}}}
    {\sqrt{\orn{S\big(b_T,\bar\zeta(\delta^-/q^-)^2\big)}}}
    \\\nonumber
    &\qquad\qquad\qquad\times
    \orn{\frac{1}{N_c}\tr\Bigl[\bra{0}\mathcal{A}_{T,n}^\rho(b_T + b_2^- n)\,
    S_n^\dagger S_\nbar (b_T)\ket{X}
    \bra{X} S_\nbar^\dagger S_n (0) \ket{0}\Bigr]}
    \biggr\}\,,
    \\[6ex]
    &\blu{\bigl[\mathcal{D}_{\qbar,21}^\text{bare}(z,\xi,b_T,\zeta)\bigr]_{\delta\alpha}}
    \\\nonumber
    &=\img q^-\int\frac{\df b_1^+}{2\pi}\,\frac{\df b_2^+}{2\pi}\,
    e^{+\img \bar\xi b_1^+ q^-}\,e^{+\img \xi b_2^+ q^-}\,
    \\\nonumber
    &\qquad\times
    \biggl\{
    \frac{\blu{\bra{0}\bigl[\bar\chi(b_T + b_1^+ \nbar)
    \fsl{\mathcal{A}}_T(b_T + b_2^+ \nbar)\bigr]_{i,\alpha}\ket{p,X}
    \bra{p,X}\chi_{i,\delta}(0)\ket{0}}}
    {\sqrt{\orn{S\big(b_T,\bar\zeta(\delta^-/q^-)^2\big)}}}
    \\\nonumber
    &\qquad\qquad
    -\frac{\blu{\bra{0}\bigl[\bar\chi(b_T + b_1^+ \nbar)\,
    \gamma_T^\rho\bigr]_{i,\alpha}\ket{p,X}
    \bra{p,X}\chi_{i,\delta}(0)\ket{0}}}
    {\sqrt{\orn{S\big(b_T,\bar\zeta(\delta^-/q^-)^2\big)}}}
    \\\nonumber
    &\qquad\qquad\qquad\times
    \orn{\frac{1}{N_c}\tr\Bigl[\bra{0} S_\nbar^\dagger S_n (b_T)\,
    \mathcal{A}_{T,n}^\rho(b_T + b_2^+ \nbar)\ket{X}
    \bra{X} S_n^\dagger S_\nbar(0)\ket{0}\Bigr]}
    \biggr\}\,,
    \\[6ex]
    &\blu{\bigl[\mathcal{D}_{q,12}^\text{bare}(z,\xi,b_T,\zeta)\bigr]_{\delta\alpha}}
    \\\nonumber
    &=-\img q^-\int\frac{\df b_1^+}{2\pi}\,\frac{\df b_2^+}{2\pi}\,
    e^{-\img \bar\xi b_1^+ q^-}\,e^{-\img \xi b_2^+ q^-}\,
    \\\nonumber
    &\qquad\times
    \biggl\{
    \frac{\blu{\bra{0}\chi_{i,\delta}(b_T)\ket{p,X}
    \bra{p,X}\bigl[\bar\chi(b_1^+ \nbar)
    \fsl{\mathcal{A}}_T(b_2^+ \nbar)\bigr]_{i,\alpha}\ket{0}}}
    {\sqrt{\orn{S\big(b_T,\bar\zeta(\delta^-/q^-)^2\big)}}}
    \\\nonumber
    &\qquad\qquad
    -\frac{\blu{\bra{0}\chi_{i,\delta}(b_T)\ket{p,X}
    \bra{p,X}\bigl[\bar\chi(b_1^+ \nbar)\gamma_T^\rho\bigr]_{i,\alpha}\ket{0}}}
    {\sqrt{\orn{S\big(b_T,\bar\zeta(\delta^-/q^-)^2\big)}}}
    \\\nonumber
    &\qquad\qquad\qquad\times
    \orn{\frac{1}{N_c}\tr\Bigl[\bra{0} S_n^\dagger S_\nbar (b_T)\ket{X}
    \bra{X} S_\nbar^\dagger S_n(0)\,\mathcal{A}_{T,n}^\rho(b_2^+ \nbar)\ket{0}\Bigr]}
    \biggr\}\,,
    \\[6ex]
    \label{eq:D_12_indices}&\blu{\bigl[\mathcal{D}_{\qbar,12}^\text{bare}(z,\xi,b_T,\zeta)\bigr]_{\delta\alpha}}
    \\\nonumber
    &=-\img q^-\int\frac{\df b_1^+}{2\pi}\,\frac{\df b_2^+}{2\pi}\,
    e^{-\img \bar\xi b_1^+ q^-}\,e^{-\img \xi b_2^+ q^-}\,
    \\\nonumber
    &\qquad\times
    \biggl\{
    \frac{\blu{\bra{0}\bar\chi_{i,\alpha}(b_T)\ket{p,X}
    \bra{p,X}\bigl[\fsl{\mathcal{A}}_T(b_2^+ \nbar)
    \chi(b_1^+ \nbar)\bigr]_{i,\delta}\ket{0}}}
    {\sqrt{\orn{S\big(b_T,\bar\zeta(\delta^-/q^-)^2\big)}}}
    \\\nonumber
    &\qquad\qquad
    -\frac{\blu{\bra{0}\bar\chi_{i,\alpha}(b_T)\ket{p,X}
    \bra{p,X}\bigl[\gamma_T^\rho\chi(b_1^+ \nbar)\bigr]_{i,\delta}\ket{0}}}
    {\sqrt{\orn{S\big(b_T,\bar\zeta(\delta^-/q^-)^2\big)}}}
    \\\nonumber
    &\qquad\qquad\qquad\times
    \orn{\frac{1}{N_c}\tr\Bigl[\bra{0} S_\nbar^\dagger S_n (b_T)\ket{X}
    \bra{X}\mathcal{A}_{T,n}^\rho(b_2^+ \nbar)\,S_n^\dagger S_\nbar(0)\ket{0}\Bigr]}
    \biggr\}\,,
\end{align}
These subtraction terms ensure that the hadronic tensor is free of endpoint divergences as seen in section \ref{sec:CancellationEndpointDivergences}. Additionally, we define a new hard function as
\begin{align}
    H_2(\xi,Q^2)\approx H_2(\xi,2 q^+ q^-)
    &=
    \int\df y_1^+\,\df y_2^+\,\df y^-\,
    e^{+\img y_1^+ \bar\xi q^-}\,e^{+\img y_2^+ \xi q^-}\,e^{+\img y^- q^+}\,
    C_2^*\bigl(\{y_1^+,y_2^+\},y^-\bigr)
    \nonumber\\
    &\quad\times
    \int\df z^+\,\df z^-\,
    e^{-\img z^+ q^-}\,e^{-\img z^- q^+}\,
    C_1(z^+,z^-)\,.
\end{align}
The genuine higher-twist correction comes from eq.~\eqref{eq:OverlapHadronicTensorDn} and its counterparts and is given by
\begin{align}\label{eq:WgenuineNLP}
    &W_{\text{gNLP}}^{\mu\nu}(q)=\bigl[\mathcal{W}_\text{sub.}^{\mu\nu}\bigr]_D^{(3,2)}
    +\bigl[\mathcal{W}_\text{sub.}^{\mu\nu}\bigr]_D^{(2,3)}
    \\\nonumber
    &=
    \frac{\img}{N_c}\,
    \biggl[\frac{\bar{n}^\mu}{q^-}-\frac{n^\mu}{q^+}\biggr]
    (\mathbb{1})_{\alpha\beta} (\gamma_T^\nu)_{\gamma\delta}
    \int\df\xi\,H_2(\xi,q^2)\int\frac{\df^2 b_T}{(2\pi)^2}\,e^{\img b_T \cdot q_T}\,
    \\\nonumber
    &\quad\times\sum_q\Bigl\{
    \grn{\bigl[\mathcal{F}_{q,21}\bigr]_{\delta\alpha}}\,\blu{\bigl[\mathcal{D}_{q,11}\bigr]_{\beta\gamma}}
    -\grn{\bigl[\mathcal{F}_{\qbar,21}\bigr]_{\beta\gamma}}\,\blu{\bigl[\mathcal{D}_{\qbar,11}\bigr]_{\delta\alpha}}
    \\\nonumber
    &\quad\qquad
    +\grn{\bigl[\mathcal{F}_{q,11}\bigr]_{\delta\alpha}}\,\blu{\bigl[\mathcal{D}_{q,21}\bigr]_{\beta\gamma}}
    -\grn{\bigl[\mathcal{F}_{\qbar,11}\bigr]_{\beta\gamma}}\,\blu{\bigl[\mathcal{D}_{\qbar,21}\bigr]_{\delta\alpha}}
    \Bigr\}
    \\\nonumber
    &\quad+
    \frac{\img}{N_c}\,
    \biggl[\frac{\bar{n}^\nu}{q^-}-\frac{n^\nu}{q^+}\biggr]
    (\gamma_T^\mu)_{\alpha\beta} (\mathbb{1})_{\gamma\delta}
    \int\df\xi\,H_2^*(\xi,q^2)\int\frac{\df^2 b_T}{(2\pi)^2}\,e^{\img b_T \cdot q_T}\,
    \\\nonumber
    &\quad\times\sum_q\Bigl\{
    \grn{\bigl[\mathcal{F}_{q,12}\bigr]_{\delta\alpha}}\,\blu{\bigl[\mathcal{D}_{q,11}\bigr]_{\beta\gamma}}
    -\grn{\bigl[\mathcal{F}_{\qbar,12,}\bigr]_{\beta\gamma}}\,\blu{\bigl[\mathcal{D}_{\qbar,11}\bigr]_{\delta\alpha}}
    \\\nonumber
    &\quad\qquad
    +\grn{\bigl[\mathcal{F}_{q,11}\bigr]_{\delta\alpha}}\,\blu{\bigl[\mathcal{D}_{q,12}\bigr]_{\beta\gamma}}
    -\grn{\bigl[\mathcal{F}_{\qbar,11}\bigr]_{\beta\gamma}}\,\blu{\bigl[\mathcal{D}_{\qbar,12}\bigr]_{\delta\alpha}}
    \Bigr\}\,,
\end{align}
The integration bounds for $\xi$ in eq.~\eqref{eq:WgenuineNLP} are determined by the distinct support regions of the different distributions, and we leave them unspecified for now. For more details on the bounds of integration, see ref.~\cite{Rodini:2022wki}.

\subsection{Comparison between methods}\label{sec:HadronicTensorComparison}

Having arrived at the final expression for the next-to-leading power factorized hadronic tensor, let us compare our approach and result to the different methods of TMD factorization that exist in the literature. Here we compare to two different approaches: the SCET approach of ref.~\cite{Ebert:2021jhy,Gao:2022ref,Gao:2023scet,Michel:2023esi} and the background field approach of ref.~\cite{Vladimirov:2021hdn,Rodini:2023plb,Vladimirov:2023aot}\footnote{There is an additional background field approach developed in ref.~\cite{Balitsky:2017gis}, that is quite similar in philosophy to ref.~\cite{Vladimirov:2021hdn}.}. Both these methods derive the formula for TMD factorization by first constructing an effective current operator, which is then used to obtain the hadronic tensor. We already compared the results for the effective current between the different works in sec.~\ref{sec:ComparisonCurrentNLP}. Here, for each method, we give an overview of the method, discuss how some differences in the methods are automatically resolved, and how some disagreements could remain. 

First, we compare our method and results to ref.~\cite{Vladimirov:2021hdn}. In the background field approach of ref.~\cite{Vladimirov:2021hdn}, background fields are introduced for collinear and anti-collinear modes, and the off-shell dynamical fields are integrated out of the theory. This results in an effective theory with only collinear and anti-collinear degrees of freedom. From here, the effective current is constructed by explicitly integrating out the dynamical field order-by-order in perturbation theory in position space. Notably, no soft background fields are introduced. Instead, soft functions arise from subtracting the overlap between the collinear and anti-collinear sectors. Because of this, no soft functions from sub-leading power soft operators appear at next-to-leading power.

Comparing our approach to that of the background field approach of ref.~\cite{Vladimirov:2021hdn}, the main difference is that we include a background field for soft modes. While at leading power the contribution from soft modes can be captured by the eikonal limits of the collinear and anti-collinear sectors, at sub-leading powers contributions could arise for which this no longer holds, though we do not find such contributions at the perturbative order that we are working.

A crucial difference between our approach and the background field method of ref.~\cite{Vladimirov:2021hdn} has to do with the subtraction of the overlap sector and its effect on rapidity divergences and endpoint divergences. In their work, they state that the overlap region can be subtracted entirely by just dividing by the soft function. The TMD distributions that enter the power corrections, however, contain what ref.~\cite{Vladimirov:2021hdn} refers to as special rapidity divergences. Moreover, the twist-3 TMD distributions defined in that work result in endpoint divergences upon convolution with the corresponding hard function. Although it is argued that these special rapidity divergences cancel, and the endpoint divergences can be removed by introducing subtraction terms, as done in ref.~\cite{Rodini:2022wki}, no operator definitions for the relevant ingredients are introduced. In our work, the subtraction terms are derived from studying the overlap between the collinear and soft sectors. We have demonstrated that this overlap subtraction results in operator matrix element definitions for physical twist-3 TMD distributions that are free of rapidity divergences and also free of endpoint divergences upon convolution with the hard function. In fact, our result provides a matrix element definition for the subtraction term of ref.~\cite{Rodini:2022wki}. Additionally, for the kinematic power corrections, we found that some of the subtraction terms combine with sub-leading soft fields to result in derivatives of the leading-power TMD distributions upon combining with the contributions of sub-leading soft operators. While the final results for the hadronic tensor agree, we believe that including a soft background field allows one to more rigorously define all ingredients of the factorization.

Next, we compare our method and results to  ref.~\cite{Ebert:2021jhy}. The approach of ref.~\cite{Ebert:2021jhy} employs soft-collinear effective theory (SCET-II) in the label formalism. In SCET, off-shell modes are integrated out of the theory, and collinear, soft, and anti-collinear modes are the effective degrees of freedom. In this approach, the effective action and operators are constructed order-by-order in the power counting in a bottom-up fashion. In particular, they construct the effective current operator by formulating a minimal basis of operators and constrain the corresponding Wilson coefficients using gauge invariance and reparameterization invariance.

Comparing our approach to that of SCET, the main differences are in the content of the operator basis and in the top-down versus bottom-up approach. The difference in the operator basis can be explained by the fact that we use a position-space operator basis while their work employs the label formalism of SCET. These different formulations of SCET can lead to different operators and different constraints on the Wilson coefficients that accompany these operators (for an example in the context of SCET-I, see~\cite{Beneke:2017ztn}). Their bottom-up approach results in an additional hard-collinear function besides the usual hard, collinear, and soft functions. 
As clarified in refs.~\cite{Gao:2023scet,Michel:2023esi}, the contribution of the hard-collinear-soft operator does not vanish in the bare SIDIS factorization theorem.
In our work, these hard-collinear functions hide in a further (re-)factorization of the hard functions or Wilson coefficients that we do not consider in this work. We find even more contributions from sub-leading-power soft operators than ref.~\cite{Ebert:2021jhy}. However, these additional contributions all vanish at the level of the hadronic tensor due to boost invariance of the vacuum and the transverse momentum measurement.

Crucially, there is one disagreement between this work and ref.~\cite{Ebert:2021jhy} at the level of the effective current that can propagate to a disagreement at the level of the hadronic tensor. This disagreement was discussed in sec.~\ref{sec:ComparisonCurrentNLP} and is related to the Wilson coefficient of the soft gluon contribution. In our work, it is constrained to be equal to $C_1$ through current conservation, while in their work, it is given by $C_1$ and an additional independent part, described by a hard-collinear function. 

It is possible that this apparent disagreement is resolved by a (yet unproven) relation between the unknown hard-collinear function of ref.~\cite{Ebert:2021jhy} and the Wilson coefficient $C_2$. This would work as follows. In this work, there are two places where the soft gluon operator contributes: directly from the effective current with Wilson coefficient $C_1$ and indirectly from the overlap subtraction of the higher-twist correction with Wilson coefficient $C_2$. This additional indirect term in our work could be in one-to-one correspondence with the additional term corresponding to the hard-collinear term arising from the Wilson coefficient of ref.~\cite{Ebert:2021jhy}. For this correspondence to work, however, the following relation must hold\footnote{We thank J.~Michel for discussions on this point.}, see eq.~\eqref{eq:DifferentWilson2}
\begin{align}\label{eq:HardCoefAgreementCondition}
    H_2(\xi,Q^2)
    &=
    H_1(Q^2)+ \bigl[H_2\otimes\mathcal{K}\bigr](\xi,Q^2)+\mathcal{O}(\xi)\,.
\end{align}
The contribution of the soft gluon operator can only be absorbed into the collinear and anti-collinear distributions if eq.~\eqref{eq:HardCoefAgreementCondition} is satisfied.

In conclusion, these three different approaches differ in the effective degrees of freedom, in the basis of operators and in the top-down or bottom-up approach and in the prescription for accounting for the overlap between modes. However, the final results for the hadronic tensor are quite close. The only possible source of disagreement lies in the contribution of a particular sub-leading soft operator, for which a condition of agreement can be formulated. It would be interesting to establish this relation. Whether such conditions can be formulated beyond NLP remains an open question.

\section{SIDIS with a jet at NLP}
\label{sec:JetSIDISFactorizationNLP}

The importance of jet measurements at EIC is reflected by the EIC Yellow report
 \cite{AbdulKhalek:2021gbh} and dedicated studies, e.g.~\cite{Arratia:2019vju,Arratia:2020nxw}.
The jets measurements enable the exploration of e.g.~small-$x$ physics \cite{Caucal:2024vbv} and TMD distributions \cite{Kang:2016ehg,Gutierrez-Reyes:2018qez}.
The possibility of higher twist effects in TMD jet measurements has been explored in ref.~\cite{delCastillo:2023rng}.
 
 In SIDIS  we have the scattering of an electron and a hadron, with momenta $\ell$ and $P$ respectively, and their interaction produces a scattered electron with momentum $\ell'$ and a jet with momentum $p$, as described by the process:
\begin{equation}\label{eq:DISwithJet}
    e(\ell)+h(P)\rightarrow e(\ell')+\text{Jet}(p)+X\,.
\end{equation}
where $X$ represents any unspecified or undetected particles produced in the interaction. At leading order in the electromagnetic coupling, the process proceeds via a photon with momentum $q=\ell-\ell'$, which subsequently interacts with a quark inside the initial hadron to produce the jet. The target hadron and the detected jet have the following masses:
\begin{align}
    M^2=P^2\,,\qquad\qquad m^2=p^2\,.&
\end{align}
Since we employ the Winner-Takes-All (WTA) jet algorithm \cite{Bertolini:2013iqa}, the resulting jet momentum satisfies $m = 0$. However, other jet algorithms can produce jets with nonzero mass. While we will ultimately also set $M = 0$, we keep both $M$ and $m$ as free parameters in this section for a more general analysis of DIS kinematics and for intermediate computational purposes. 

The standard kinematic variables used to describe the DIS cross section are:
\begin{align}
    &Q^2 = -q^2\,,\qquad x=\frac{Q^2}{2q\cdot P}\,,\qquad
    y=\frac{q\cdot P}{\ell\cdot P}\,.
\end{align}
Here, $Q^2$ is the virtuality of the exchanged photon and corresponds to the hard scale of the scattering process. The Bjorken scaling variable $x$ quantifies the fraction of the hadron's momentum carried by the struck parton in the infinite momentum frame. Lastly, $y$ is the so-called inelasticity, which describes the fraction of the electron's energy transferred to the target in the laboratory frame. These are related by the Mandelstam invariant $s=(P+\ell)^2$ by
\begin{equation}
    xy(s-M^2)=Q^2\,,
\end{equation}
which leads to the following relation between differentials:
\begin{equation}
    \frac{\text{d}x}{x}=\frac{\text{d}y}{y}=\frac{\text{d}Q^2}{Q^2}\,.
\end{equation}

Additional relevant kinematic variables are
\begin{align}
    z=\frac{p\cdot P}{q\cdot P}\,,\qquad
    \gamma=\frac{2x M}{Q}\,,\quad\text{and}\qquad
    \gamma_h=\frac{m}{zQ}\,,
\end{align}
where the fragmentation variable $z$ represents the fraction of the energy of the virtual photon carried by the detected jet. Note that, in the case of SIDIS, where a final-state hadron is detected, the momentum fraction $z$ can range from $0$ to $1$ (e.g., at HERA \cite{HERMES:2004zsh}, the measured range was $0.2 \leq z \leq 0.8$). However, we assume a large jet radius and $q_T^2\ll Q^2$, such that all collinear final-state particles are clustered within the jet, while any radiation outside, denoted by X in eq.~(\ref{eq:DISwithJet}), is soft; consequently $z=1$. The parameter $\gamma$ is a dimensionless variable proportional to the target mass and is relevant for higher power corrections beyond the scope of this work. Similarly, $\gamma_h$ highlights the effects of jet (or hadron, in the case of fragmentation) mass in the fragmentation process.

To facilitate our analysis, we take the light-like vectors $n$ and $\bar{n}$ such that the momentum of the produced jet and the momentum of the incoming hadron have no transverse components (and are not necessarily back-to-back),
\begin{equation}
    P^\mu=P^+\bar{n}^\mu+\frac{\gamma^2Q^2}{8x^2P^+}n^\mu\,, \hspace{0.5cm} p^\mu=p^-n^\mu+\frac{z^2\gamma_h^2Q^2}{2p^-}\bar{n}^\mu\,.
\end{equation}
These lightcone vectors can be expressed in terms of these kinematic variables as:
\begin{align}\label{eq:n-vector}
    n^\mu
    &=
    \frac{2x P^+}{z Q^2 \sqrt{1-\gamma_h^2\gamma^2}}
    \biggl[p^\mu-\frac{2xz(1-\sqrt{1-\gamma_h^2\gamma^2})}{\gamma^2}P^\mu\biggr]\,,
    \\
    \nbar^\mu
    &=
    \frac{1}{2P^+\sqrt{1-\gamma_h^2\gamma^2}}
    \biggl[P^\mu(1+\sqrt{1-\gamma_h^2\gamma^2})
    -\frac{\gamma^2}{2xz}p^\mu\biggr]\,.
\end{align}

To define our observables, we introduce two coordinate planes: the transverse plane, orthogonal to the initial hadron momentum $P$ and the jet momentum $p$ (or to $n$ and $\bar{n}$ as defined in eq.~\eqref{eq:VectorComponents}, and the perpendicular plane, orthogonal to $P$ and to the momentum of the intermediate photon $q$. These planes can be described in terms of the metric tensors that project onto the corresponding subspaces and read
\begin{align}
    g_T^{\mu\nu}
    &=
    g^{\mu\nu}
    -\frac{1}{Q^2}\frac{1}{1-\gamma^2\gamma_h^2}
    \biggl[
    -4x^2 \gamma_h^2 P^\mu P^\nu
    +\frac{2x}{z} (P^\mu p^\nu + p^\mu P^\nu)
    -\frac{\gamma^2}{z^2} p^\mu p^\nu
    \biggr]\,,\\[2ex]
    g_\perp^{\mu\nu}
    &=
    g^{\mu\nu}
    -\frac{1}{Q^2}\frac{1}{1+\gamma^2}
    \Bigl[4x^2 P^\mu P^\nu + 2x (P^\mu q^\nu + q^\mu P^\nu) - \gamma^2 q^\mu q^\nu\Bigr]\,.
\end{align}

The transverse and perpendicular components of a vector $v$ are defined as $v^\mu_T = g^{\mu\nu}_Tv_\nu$ and $v^\mu_\perp = g^{\mu\nu}_\perp v_\nu$, respectively. For clarity, two-dimensional transverse vectors are represented in boldface notation. Since transverse vectors are more frequently encountered in our theoretical analysis, we suppress the subscript $T$ where the context permits, e.g., $\bm{b}= \bm{b}_T$ or $\bm{q}= \bm{q}_T$. Additionally, we employ the Levi-Civita tensor, whose transverse and perpendicular forms are given by:
\begin{align}
    \epsilon_T^{\mu\nu}
    &=\epsilon^{\mu\nu\rho\sigma}\bar n_\rho n_\sigma=
    \frac{2x}{z Q^2 \sqrt{1-\gamma_h^2 \gamma^2}}
    \epsilon^{\mu\nu\rho\sigma} P_\rho p_\sigma\,,\\[1ex]
    \epsilon_\perp^{\mu\nu}
    &=
     \frac{2x}{Q^2\sqrt{1+\gamma^2}}
    \epsilon^{\mu\nu\rho\sigma} P_\rho q_\sigma\,.\label{eq:Levi-Civita-Perp}
\end{align}

Since spin-dependent observables will be considered later in the analysis, it is useful to express the spin vector of the target hadron in terms of either a transverse or perpendicular decomposition. These are given by:
\begin{equation}\label{eq:TransverseSpin}
    S^\mu=S_{L}\left(\frac{2x}{Q\gamma\sqrt{1-\gamma_h^2\gamma^2}}P^\mu-\frac{\gamma}{z\,Q\sqrt{1-\gamma_h^2\gamma^2}}p^\mu\right)+S^\mu_T\,,
\end{equation}
\begin{equation}\label{eq:PerpSpin}
    S^\mu=S_{||}\left(\frac{2x}{Q\gamma\sqrt{1+\gamma^2}}P^\mu-\frac{\gamma}{Q\sqrt{1+\gamma^2}}q^\mu\right)+S^\mu_\perp\,.
\end{equation}

\begin{figure}[t]
    \centering
\includegraphics[width=0.7\linewidth]{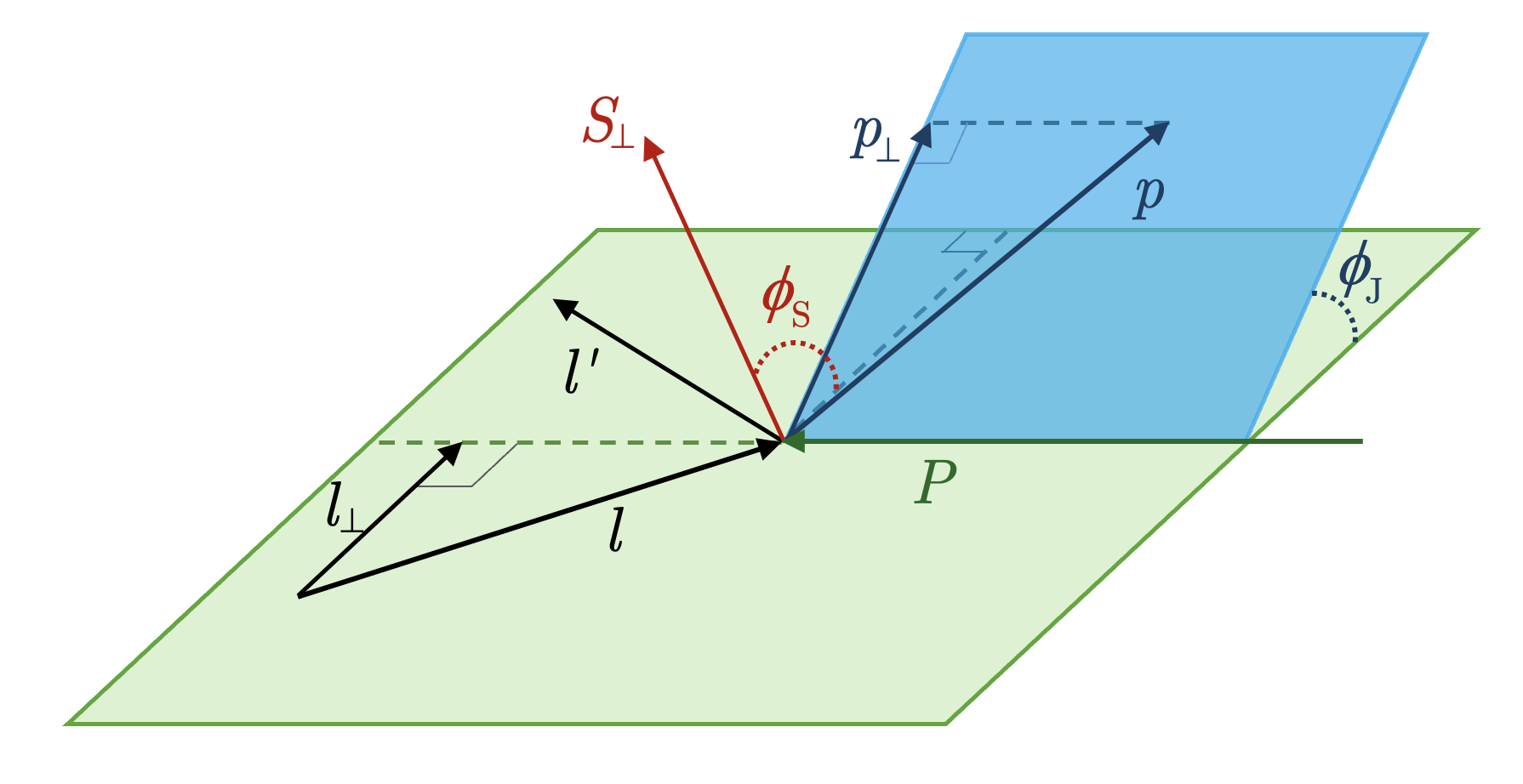}
    \caption{Kinematics of the jet production in SIDIS. The jet production plane is depicted in blue, and the lepton-hadron plane in green.}
    \label{fig:KinematicsSIDIS}
\end{figure}

We also introduce the azimuthal angle $\phi_J$, defined as the angle between the lepton-hadron plane and the jet production plane as shown in fig.~\ref{fig:KinematicsSIDIS},
\begin{equation}\label{eq:AnglePhiJ}
    \cos{\phi_J}=-\frac{p_\perp^\sigma \ell_{\perp,\sigma}}{|p_\perp|\,|\ell_\perp|},\quad\quad \sin{\phi_J}=-\frac{\ell_\mu p_\nu \epsilon_\perp^{\mu\nu}}{|p_\perp|\,|\ell_\perp|}\,.
\end{equation}
Similarly, one can define the azimuthal angle for the spin vector:
\begin{equation}\label{eq:AnglePhiS}
    \cos{\phi_S}=-\frac{S_\perp^\sigma \ell_{\perp,\sigma}}{|S_\perp|\,|\ell_\perp|},\quad\quad \sin{\phi_S}=-\frac{\ell_\mu S_\nu \epsilon_\perp^{\mu\nu}}{|S_\perp|\,|\ell_\perp|}\,.
\end{equation}

From this point onward, we assume that the jet algorithm yields massless jet momenta ($p^2=0$), as is the case for the WTA recombination scheme. The differential cross section can then be expressed in the limit $Q^2\gg M^2$ (see ref.~\cite{Rodini:2023plb}) as
\begin{equation}\label{eq:CrossSectionDIS}
    \frac{\text{d}\sigma}{\text{d}x\,\text{d}y\,\text{d}\phi_J\,\text{d}\phi_S\,\text{d}\bm{q}^2}=\frac{\alpha_{\text{em}}^2y}{8Q^4}L_{\mu\nu}W^{\mu\nu}\,,
\end{equation}
where $L_{\mu\nu}$ and $W^{\mu\nu}$ are the leptonic and hadronic tensors, respectively. The leptonic tensor can be directly obtained as
\begin{equation}\label{eq:LeptonicTensor}
    L_{\mu\nu}=2\big(\ell_\mu \ell'_\nu+\ell'_\mu \ell_\nu-(\ell\cdot \ell')g_{\mu\nu}\big)+2i\lambda_e\epsilon_{\mu\nu\rho\sigma}\ell^\rho \ell'^\sigma\,,
\end{equation}
where $\lambda_e$ is the lepton helicity. Our primary focus lies on the hadronic tensor, which encodes the QCD dynamics of the process. In the following sections, we build on the framework established in sec.~\ref{sec:NLP} to analyze its properties up to NLP corrections. 

\subsection{Hadronic tensor factorization}
To achieve a minimal set of bare TMDPDFs and jet functions with no open spinor or Lorentz indices, we apply the following Fierz transformations to the Lorentz structures present in eqs.~\eqref{eq:WLP}-\eqref{eq:WgenuineNLP}:
\begin{align}
    4(\gamma_T^\mu)_{\alpha\beta}(\gamma_T^\nu)_{\gamma\delta}
    =&
    -g_T^{\mu\nu}
    \Bigl[
    (\gamma^+)_{\alpha\delta} (\gamma^-)_{\gamma\beta}
    +(\gamma^-)_{\alpha\delta} (\gamma^+)_{\gamma\beta}
    \Bigr] \\
    \nn&+\img\epsilon_T^{\mu\nu}\Bigl[
    (\gamma^+)_{\alpha\delta} (\gamma^-\gamma^5)_{\gamma\beta}
    -(\gamma^-)_{\alpha\delta} (\gamma^+\gamma^5)_{\gamma\beta}\\
    \nn&
    \qquad\qquad+(\gamma^+\gamma^5)_{\alpha\delta} (\gamma^-)_{\gamma\beta}
    -(\gamma^-\gamma^5)_{\alpha\delta} (\gamma^+)_{\gamma\beta}\Bigr]+ \dots\,,\\[2ex]
    4 (\gamma_T^\mu)_{\alpha\beta} (\mathbb{1})_{\gamma\delta}
    &=
    \img\epsilon_T^{\mu\alpha}
    \Bigl[(\gamma^-)_{\alpha\delta}(\img\sigma^{\alpha+}\gamma^5)_{\gamma\beta}
    +(\img\sigma^{\alpha-} \gamma^5)_{\alpha\delta} (\gamma^+)_{\gamma\beta}
     \\
    \nn&\qquad\qquad-(\gamma^+)_{\alpha\delta}(\img\sigma^{\alpha-}\gamma^5)_{\gamma\beta}
    -(\img\sigma^{\alpha+} \gamma^5)_{\alpha\delta} (\gamma^-)_{\gamma\beta}\Bigr]\\
    \nn&\qquad+ (\gamma^+\gamma^5)_{\alpha\delta} (\img\sigma^{\mu-} \gamma^5)_{\gamma\beta} 
    +(\img\sigma^{\mu-} \gamma^5)_{\alpha\delta} (\gamma^+ \gamma^5)_{\gamma\beta}\\
    \nn&\qquad+ (\gamma^-\gamma^5)_{\alpha\delta} (\img\sigma^{\mu+} \gamma^5)_{\gamma\beta} 
    +(\img\sigma^{\mu+} \gamma^5)_{\alpha\delta} (\gamma^- \gamma^5)_{\gamma\beta}+\dots\,.
\end{align}
The dots denote additional structures arising from the Fierz decomposition, that do not contribute to the NLP cross section (see ref.~\cite{delCastillo:2023rng} for complete expressions). 

This process results in two twist-2 TMD jet functions and four twist-2 TMDPDFs without Lorentz indices:
\begin{align}
     \nn
    (\gamma^-)_{\gamma\beta}
    \blu{\bigl[\mathcal{D}_{q,11}(b_T)\bigr]_{\beta\gamma}}
    &=
    2N_c\, \blu{J^q_{11}(b_T)}\,,&
    (\gamma^-)_{\alpha\delta}
    \blu{\bigl[\mathcal{D}_{\bar{q},11}(b_T)\bigr]_{\delta\alpha}}
    &=
    2N_c\, \blu{J^{\bar{q}}_{11}(b_T)}\,,\\[1ex]
    \nn(\gamma^+)_{\alpha\delta}
    \grn{\bigl[\mathcal{F}_{q,11}(x,b_T)\bigr]_{\delta\alpha}}
    &=
    \, \grn{F^q_{11}(x,b_T)}\,, &
    (\gamma^+)_{\gamma\beta}
    \grn{\bigl[\mathcal{F}_{\bar{q},11}(x,b_T)\bigr]_{\beta\gamma}}
    &=
    \, \grn{F^{\bar{q}}_{11}(x,b_T)}\,,
    \\[1ex]
    (\gamma^+\gamma^5)_{\alpha\delta}
    \grn{\bigl[\mathcal{F}_{q,11}(x,b_T)\bigr]_{\delta\alpha}}
    &=
    \, \grn{G^q_{11}(x,b_T)}\,,&
    -(\gamma^+\gamma^5)_{\gamma\beta}
    \grn{\bigl[\mathcal{F}_{\bar{q},11}(x,b_T)\bigr]_{\beta\gamma}}
    &=
    \, \grn{G^{\bar{q}}_{11}(x,b_T)}\,.
    \label{eq:IndicesContractionTwist2Jets}
\end{align}
Here, we have included a minus sign for the anti-quark distribution $G^{\bar{q}}_{11}$ after inspecting its
$C$-conjugation property, see f.i.~refs.~\cite{Boer_1998,Boer_2003}. For the NLP case, after applying the Fierz identities, we obtain eight twist-3 TMD jet functions and four twist-3 TMDPDFs, each carrying one open Lorentz index:
\begin{align}
    -\epsilon_T^{\mu\alpha}(\sigma^{\alpha-} \gamma^5)_{\gamma\beta}
    \blu{\bigl[\mathcal{D}_{q/\bar{q},21}(\xi,b_T)\bigr]_{\beta\gamma}}
    &=
    2N_c\, \blu{J^{\mu,q/\bar{q}}_{21}(\xi,b_T)}\,,
    \label{eq:J8}
    \\[1ex]
     +\epsilon_T^{\mu\alpha}(\sigma^{\alpha-} \gamma^5)_{\gamma\beta}
    \blu{\bigl[\mathcal{D}_{q/\bar{q},12}(\xi,b_T)\bigr]_{\beta\gamma}}
    &=
    2N_c\, \blu{J^{\mu,q/\bar{q}}_{12}(\xi,b_T)}\,,
    \nonumber\\[1ex]
    -g_T^{\mu\alpha}(\sigma^{\alpha-} \gamma^5)_{\gamma\beta}
    \blu{\bigl[\mathcal{D}_{q/\bar{q},21}(\xi,b_T)\bigr]_{\beta\gamma}}
    &=
    2N_c\, \blu{\Tilde{J}^{\mu,q/\bar{q}}_{21}(\xi,b_T)}\,,
    \nonumber
    \\[1ex]
     +g_T^{\mu\alpha}(\sigma^{\alpha-} \gamma^5)_{\gamma\beta}
    \blu{\bigl[\mathcal{D}_{q/\bar{q},12}(\xi,b_T)\bigr]_{\beta\gamma}}
    &=
    2N_c\, \blu{\Tilde{J}^{\mu,q/\bar{q}}_{12}(\xi,b_T)}\,,
    \nonumber
    \\[1ex]
    +\epsilon_T^{\mu\alpha}(\sigma^{\alpha+} \gamma^5)_{\alpha\delta}
    \grn{\bigl[\mathcal{F}_{q/\bar{q},21}(x,\xi,b_T)\bigr]_{\delta\alpha}}
    &=
    \, \grn{F^{\mu,q/\bar{q}}_{21}(x,\xi,b_T)}\,,\nn
    \\[1ex]
   -\epsilon_T^{\mu\alpha}(\sigma^{\alpha+} \gamma^5)_{\alpha\delta}
    \grn{\bigl[\mathcal{F}_{q/\bar{q},12}(x,\xi,b_T)\bigr]_{\delta\alpha}}
    &=
    \, \grn{F^{\mu,q/\bar{q}}_{12}(x,\xi,b_T)}\,.\nn
\end{align}
The factor $2N_c$ accounts for spin and color averaging of the initial quark in jet distributions.

To fully factorize the hadronic tensor, we decompose distributions with open transverse Lorentz indices, including NLP functions and the kinematic power corrections of LP functions. In the case of the kinematic power corrections, we found that the subtraction procedure yields terms of the form shown in eq.~\eqref{eq:HadronicTensorSubtractedKinematicCorrections}, which are free of rapidity divergences. Accordingly, we define the derivatives of the jet functions (denoted by a prime subscript) as follows:
\begin{align} \label{eq:J_prime}
    &\biggl[\partial_\rho-\frac{1}{2}\bigl[\partial_\rho K\bigr]\ln\biggl(\frac{\zeta}{\bar\zeta}\biggr)\biggr] \blu{J^{q/\bar q}_{11}(b_T)}
    =\frac{b_\rho}{b^2} \blu{J^{q/\bar q\,\,\prime}_{11}(\mathbf{b}^2)}\,,
\end{align}

On the other hand, parity symmetry leads to the following relations for twist-3 jet distributions,
\begin{align}
    \label{eq:JetParametrization}&\blu{J^{\rho,q/\bar q}_{21/12}(\xi,b_T)}
    =\frac{b^\rho}{b^2} \blu{J^{q/\bar q}_{21/12}(\xi,\mathbf{b}^2)}\,,\\[1ex]
    &\nn\blu{\Tilde{J}^{\rho,q/\bar{q}}_{21/12}(\xi,b_T)}
    =\frac{\epsilon_T^{\rho\alpha}b_\alpha}{b^2} \blu{J^{q/\bar q}_{21/12}(\xi,\mathbf{b}^2)}\,,
\end{align}
reducing the number of independent twist-3 jet functions to four. For TMDPDFs, we use the conventional parameterization of twist-2 TMD distributions \cite{Ralston:1979ys,Tangerman:1994eh,Mulders:1995dh,Boer_1998,Bacchetta:2006tn,Boer:2011xd} and adopt the following parametrization for twist-3 distributions: 
\begin{align}
    \label{eq:Twist2TMDPDFsParametrization}\grn{F^{q/\bar q}_{11}(x,b_T)}
    =&\,\grn{f^{q/\bar q}_1(x,\mathbf{b}^2)}+\img\epsilon_T^{\alpha\beta}b_\alpha S_{T,\beta}M\grn{f_{1T}^{\perp,q/\bar q}(x,\mathbf{b}^2)}\,,\\[1ex]
    \nn\grn{G^{q/\bar q}_{11}(x,b_T)}
    =&\,S_L\,\grn{g_1^{q/\bar q}(x,\mathbf{b}^2)}+\img(b\cdot S_{T})M\grn{g_{1T}^{\perp,q/\bar q}(x,\mathbf{b}^2)}\,,\\[2ex]
    \label{eq:F21Parametrization}\grn{F^{\rho,q/\bar q}_{21/12}(x,\xi,b_T)}
    =&\,\frac{ib^\rho}{b^2} \grn{f^{\perp,q/\bar q}_{21/12}(x,\xi,\mathbf{b}^2)}+S_{L}\frac{i\epsilon_T^{\rho\alpha}b_\alpha}{b^2} \grn{g^{\perp,q/\bar q}_{21/12,L}(x,\xi,\mathbf{b}^2)}\\
    \nn&+\epsilon_T^{\rho\alpha}S_{T,\alpha}M\grn{f^{q/\bar q}_{21/12,T}(x,\xi,\mathbf{b}^2)}\\
    \nn&+\epsilon_T^{\rho\alpha}S_T^\beta\left(\frac{g_{T,\alpha\beta}}{2}-\frac{b_\alpha b_\beta}{b^2}\right)M\grn{f_{21/12,T}^{\perp,q/\bar q}(x,\xi,\mathbf{b}^2)}\,.
\end{align}
This increases the number of twist-3 TMDPDFs to sixteen. Appendix \ref{Appendix:Twist3Parametrization} relates our own twist-3 parametrization in eq.~\eqref{eq:F21Parametrization} to the one used in refs.~\cite{Rodini:2023plb,Rodini:2022wki}. Note that the transverse derivative $\partial/\partial b^\rho$ appearing in the kinematic part of the NLP contribution to the hadronic tensor must act on the right-hand side of eq.~\eqref{eq:Twist2TMDPDFsParametrization}. Based on the subtraction procedure discussed in the sec.~\ref{sec:OverlapSubtractionNLP}, we found that the correct definition of the subtracted derivative for the TMDPDFs takes a form analogous to the jet function case in eq.~\eqref{eq:J_prime}, with the difference that the order of the rapidity regulators $\zeta$ and $\bar \zeta$ in the logarithmic term is reversed. We therefore define:
\begin{equation}
    \label{eq:Fprime}
    \biggl[\partial_\rho+\frac{1}{2}\bigl[\partial_\rho K\bigr]\ln\biggl(\frac{\zeta}{\bar\zeta}\biggr)\biggr] \grn{F(\mathbf{b}^2)}
    =\frac{b_\rho}{b^2} \grn{F^{\prime}(\mathbf{b}^2)}\,
\end{equation}

Discrete $C$,$P$, and $T$ symmetries provide an additional reduction of independent functions, leading to the following relations for LP TMD and jet functions:
\begin{align}
    \blu{J_{11}(b_T)}&\equiv \blu{J^{\bar{q}}_{11}(b_T)}=
    \blu{J^{q}_{11}(b_T)}\,, &
    [\blu{J_{11}(b_T)}]^*&=\blu{J_{11}(-b_T)}\,,\\[0.8ex]
    \nn\big[\grn{f^{q/\bar{q}}_{1}\left(x,b_T\right)}\big]^*&=\grn{f^{q/\bar{q}}_{1}\left(x,-b_T\right)}\,,&
    \qquad\big[\grn{f^{\perp,q/\bar{q}}_{1T}\left(x,b_T\right)}\big]^*&=\grn{f^{\perp,q/\bar{q}}_{1T}\left(x,-b_T\right)}\,,\\[0.8ex]
   \nn \big[\grn{g^{q/\bar{q}}_{1}\left(x,b_T\right)}\big]^*&=\grn{g^{q/\bar{q}}_{1}\left(x,-b_T\right)}\,, &
   \big[\grn{g^{\perp,q/\bar{q}}_{1T}\left(x,b_T\right)}\big]^*&=\grn{g^{\perp,q/\bar{q}}_{1T}\left(x,-b_T\right)}\,\,,
\end{align}
which reduces the number of independent twist-2 jet functions to one. Similarly, for twist-3 TMDPDFs and jet functions, we find: \begin{align}
     \blu{J_{21}(\xi,b_T)}&\equiv \blu{J^{\bar{q}}_{21}(\xi,b_T)}=\blu{J^{q}_{21}(\xi,b_T)}\,,
    \\[1ex]
    \nn\blu{J_{12}(\xi,b_T)}&\equiv \blu{J^{\bar{q}}_{12}(\xi,b_T)}=\blu{J^{q}_{12}(\xi,b_T)}\,,\\[1ex]
    \nn\big[\blu{J_{12}(\xi,b_T)}\big]^*&=\blu{J_{21}(\xi,-b_T)}\,,\\[1ex]
   \nn\big[\grn{f^{\perp,q/\bar q}_{12}\left(x,\xi,b_T\right)}\big]^*&=-\grn{f^{\perp,q/\bar q}_{21}\left(x,\xi,-b_T\right)}\equiv -\grn{f^{\perp,q/\bar{q}}_{2}\left(x,\xi,b_T\right)}\,,\\[1ex]
   \nn\big[\grn{g^{\perp,q/\bar q}_{12,L}\left(x,\xi,b_T\right)}\big]^*&=-\grn{g^{\perp,q/\bar q}_{21,L}\left(x,\xi,-b_T\right)}\equiv-\grn{g^{\perp,q/\bar{q}}_{2L}\left(x,\xi,-b_T\right)}\,,\\[1ex]
   \nn\big[ \grn{f^{q/\bar{q}}_{12,T}\left(x,\xi,b_T\right)}\big]^*&=-\grn{f^{q/\bar{q}}_{21,T}\left(x,\xi,-b_T\right)}\equiv-\grn{f^{q/\bar{q}}_{2T}\left(x,\xi,-b_T\right)}\,,\\[1ex]
   \nn\big[ \grn{f_{12,T}^{\perp,q/\bar{q}}\left(x,\xi,b_T\right)}\big]^*&=-\grn{f_{21,T}^{\perp,q/\bar{q}}\left(x,\xi,-b_T\right)}\equiv-\grn{f_{2T}^{\perp,q/\bar{q}}\left(x,\xi,-b_T\right)}\,.
\end{align}
These restrictions decrease the number of independent twist-3 TMDPDFs from sixteen to eight and the number of independent twist-3 jet functions from four to one. Since quark and anti-quark jet functions are identical, we omit this label from now on.  

Lastly, since all functions depend only on $|\bm{b}|$, we integrate over the angular component using:
\begin{align}
    \int\frac{\df^2 b}{(2\pi)^2}\,e^{\img b\cdot q}\,f(\mathbf{b}^2)
    =&
    \int_0^\infty\df|\mathbf{b}|^2\,
    \frac{J_0(|\mathbf{b}||\mathbf{q}|)}{4\pi}\,
    f(\mathbf{b}^2)\,,
     \\[1ex]
    \nn\int\frac{\df^2 b}{(2\pi)^2}\,e^{\img b\cdot q}\,
    \frac{b^\rho}{b^2}\,f(\mathbf{b}^2)
    =&\,
    \img \,q_T^\rho\int_0^\infty\df|\mathbf{b}|^2\,
    \frac{J_1(|\mathbf{b}||\mathbf{q}|)}{4\pi|\mathbf{b}||\mathbf{q}|}\,
    f(\mathbf{b}^2)\\[1ex] 
    \nn\int\frac{\df^2 b}{(2\pi)^2}\,e^{\img b\cdot q}\,
    \frac{b^\rho b^\sigma}{b^2}\,f(\mathbf{b}^2)
    =&
    \,g_T^{\rho\sigma}\int_0^\infty\df|\mathbf{b}|^2\,
    \frac{J_1(|\mathbf{b}||\mathbf{q}|)}{4\pi|\mathbf{b}||\mathbf{q}|}\,
    f(\mathbf{b}^2)\\
    &+q_T^{\rho}q_T^{\sigma}\int_0^\infty\df|\mathbf{b}|^2\,
    \frac{J_2(|\mathbf{b}||\mathbf{q}|)}{4\pi|\mathbf{q}|^2}\,
    f(\mathbf{b}^2)\,,
\end{align}
Here, $J_n$ denotes the Bessel functions of the first kind. 

Applying all of these relations and definitions, we obtain a factorized form for the hadronic tensor. At leading power, this expression reads:
\begin{align}\label{eq:HadronicTensorLP}
        W_{\mathrm{LP}}^{\mu \nu}(q)=&\left(-g_T^{\mu \nu}\right) H_1\big(Q^2\big) \int \mathrm{d} \bm{b}^2 \frac{J_0(|\bm{b} \| \bm{q}|)}{8 \pi} \blu{J_{11}\big(\bm{b}^2\big) }\grn{f^{q\bar{q}}_{1}\big(x,\bm{b}^2\big)}\\
        \nn&+S_\parallel\left(\img\epsilon_T^{\mu \nu}\right) H_1\big(Q^2\big) \int \mathrm{d} \bm{b}^2 \frac{J_0(|\bm{b} \| \bm{q}|)}{8 \pi}\blu{J_{11}\big(\bm{b}^2\big)}\grn{g^{q\bar{q}}_{1}\big(x,\bm{b}^2\big)}\\
        \nn&-g_T^{\mu \nu}\epsilon_T^{\alpha\beta}q_{T,\alpha} S_{T,\beta}H_1\big(Q^2\big) \int \mathrm{d} \bm{b}^2\frac{M|\bm{b}|J_1(|\bm{b} \| \bm{q}|)}{8 \pi|\bm{q}|}\blu{J_{11}\big(\bm{b}^2\big)} \grn{f^{\perp,q\bar{q}}_{1T}\big(x,\bm{b}^2\big)}\\
        \nn&+\left(\img\epsilon_T^{\mu \nu}\right)(q_{T}\cdot S_{T})H_1\big(Q^2\big) \int \mathrm{d} \bm{b}^2 \frac{M|\bm{b}|J_1(|\bm{b} \| \bm{q}|)}{8 \pi|\bm{q}|}\blu{J_{11}\big(\bm{b}^2\big)} \grn{g^{\perp,q\bar{q}}_{1T}\big(x,\bm{b}^2\big)}\,,
\end{align}
Exploiting the quark-flavor independence of the jet functions, we find it convenient to introduce the superscript $q\bar q$ to denote the sum over quark and anti-quark TMD distributions, e.g.
\begin{equation}
    \grn{f_1^{q\bar{q}}\big(x,\bm{b}^2\big)}\equiv \sum_q\Big(\grn{f_1^{q}\big(x,\bm{b}^2\big)}+\grn{f_1^{\bar{q}}\big(x,\bm{b}^2\big)}\Big)\,,
\end{equation}

The twist-2 TMD jet functions and parton distributions appearing in eq.~\eqref{eq:HadronicTensorLP} are defined in terms of the matrix elements of the corresponding building-block operators and follow from their previous definitions in eqs.$\,$\eqref{eq:F_11_indices}-\eqref{eq:D_11_indices}, to which we apply the manipulations in eqs.~\eqref{eq:IndicesContractionTwist2Jets} and \eqref{eq:Twist2TMDPDFsParametrization}.  At next-to-leading power, the factorized expression for the kinematic part of the hadronic tensor becomes 
\begin{align}\label{eq:HadronicTensorkNLP}
\notag&W_{\mathrm{kNLP}}^{\mu \nu}(q)= -H_1\big(Q^2\big) \int_0^{\infty} \frac{\mathrm{d} \bm{b}^2}{8 \pi|\bm{b}| |\bm{q}|}\Bigg( J_1(|\bm{b} \| \bm{q}|)\,\bigg\lbrace \blu{J_{11}\big(\bm{b}^2\big)} \bigg[\left(\frac{n^\mu q_T^\nu}{q^{+}}+\frac{n^\nu q_T^\mu}{q^{+}}\right)  \grn{f^{q\bar{q}\,\,\prime}_{1}\big(x,\bm{b}^2\big)}\\
\notag&+S_\parallel\left(\frac{n^\mu(\img\epsilon_T^{\nu\rho} q_T^\rho)}{q^{+}}-\frac{n^\nu (\img\epsilon_T^{\mu\rho} q_T^\rho)}{q^{+}}\right) \grn{g^{q\bar{q}\,\,\prime}_{1}\big(x,\bm{b}^2\big)}\bigg]\\
\notag&+\blu{J_{11}^{\prime}\big(\bm{b}^2\big)}\bigg[\left(\frac{\bar{n}^\mu q_T^\nu}{q^{-}}+\frac{\bar{n}^\nu q_T^\mu}{q^{-}}\right)   \grn{f^{q\bar{q}}_{1}\big(x,\bm{b}^2\big)}+ S_\parallel\left(\frac{\bar{n}^\mu(\img\epsilon_T^{\nu\rho} q_T^\rho)}{q^{-}}-\frac{\bar{n}^\nu (\img\epsilon_T^{\mu\rho} q_T^\rho)}{q^{-}}\right)\grn{g^{q\bar{q}}_{1}\big(x,\bm{b}^2\big)}\bigg]\bigg\rbrace\\
\notag&+\blu{J_{11}\big(\bm{b}^2\big)} \bigg\lbrace\left(\frac{n^\mu(\epsilon_T^{\nu\rho} S_T^\rho)}{q^{+}}+\frac{n^\nu (\epsilon_T^{\mu\rho} S_T^\rho)}{q^{+}}\right)\bigg[M|\bm{b}||\bm{q}|J_0(|\bm{b} \| \bm{q}|)\grn{f^{\perp,q\bar{q}}_{1T}\big(x,\bm{b}^2\big)}\\
\nn&+MJ_1(|\bm{b} \| \bm{q}|)\grn{f^{\perp,q\bar{q}\,\,\prime}_{1T}\big(x,\bm{b}^2\big)}\bigg]+\left(\frac{n^\mu(\epsilon_T^{\nu\rho} S_T^\rho)}{q^{+}}-\frac{n^\nu (\epsilon_T^{\mu\rho} S_T^\rho)}{q^{+}}\right)\\
\nn&\times\bigg[M|\bm{b}||\bm{q}|J_0(|\bm{b} \| \bm{q}|)\grn{g^{\perp,q\bar{q}}_{1T}\big(x,\bm{b}^2\big)}+MJ_1(|\bm{b} \| \bm{q}|)\grn{g^{\perp,q\bar{q}\,\,\prime}_{1T}\big(x,\bm{b}^2\big)}\bigg]\bigg\rbrace\\
\notag&+MJ_1(|\bm{b} \| \bm{q}|)\blu{J^\prime_{11}\big(\bm{b}^2\big)} \bigg\lbrace\left(\frac{\bar{n}^\mu(\epsilon_T^{\nu\rho} S_T^\rho)}{q^{-}}+\frac{\bar{n}^\nu (\epsilon_T^{\mu\rho} S_T^\rho)}{q^{-}}\right)\grn{f^{\perp,q\bar{q}}_{1T}\big(x,\bm{b}^2\big)}\\
\notag&+\left(\frac{\bar n^\mu(\epsilon_T^{\nu\rho} S_T^\rho)}{q^{-}}-\frac{\bar n^\nu (\epsilon_T^{\mu\rho} S_T^\rho)}{q^{-}}\right)\grn{g^{\perp,q\bar{q}}_{1T}\big(x,\bm{b}^2\big)}\bigg\rbrace\\
\notag&+\frac{M|\bm{b}|J_2(|\bm{b} \| \bm{q}|)}{| \bm{q}|} \bigg\lbrace\epsilon_T^{\alpha\beta}q_{T}^{\alpha}S_{T}^{\beta} \bigg[\left(\frac{n^\mu q_T^\nu}{q^{+}}+\frac{n^\nu q_T^\mu}{q^{+}}\right) \blu{J_{11}\big(\bm{b}^2\big)}\grn{f^{\perp,q\bar{q}\,\,\prime}_{1T}\big(x,\bm{b}^2\big)}\\
\notag&+\left(\frac{\bar n^\mu q_T^\nu}{q^{-}}+\frac{\bar n^\nu q_T^\mu}{q^{-}}\right)  \blu{J^\prime_{11}\big(\bm{b}^2\big)}\grn{f^{\perp,q\bar{q}}_{1T}\big(x,\bm{b}^2\big)}\bigg]\\
\notag&+g_T^{\alpha\beta}q_{T}^{\alpha}S_{T}^{\beta}\bigg[\left(\frac{ n^\mu(\img\epsilon_T^{\nu\rho} q_T^\rho)}{q^{+}}-\frac{n^\nu (\img\epsilon_T^{\mu\rho} q_T^\rho)}{q^{+}}\right) \blu{J_{11}\big(\bm{b}^2\big)}\grn{g^{\perp,q\bar{q}\,\,\prime}_{1T}\big(x,\bm{b}^2\big)}\\
&+\left(\frac{\bar n^\mu(\img\epsilon_T^{\nu\rho} q_T^\rho)}{q^{-}}-\frac{\bar n^\nu (\img\epsilon_T^{\mu\rho} q_T^\rho)}{q^{-}}\right) \blu{J^\prime_{11}\big(\bm{b}^2\big)}\grn{g^{\perp,q\bar{q}}_{1T}\big(x,\bm{b}^2\big)}\bigg]\bigg\rbrace\Bigg)\,.
\end{align}
Finally, the factorized expression for the genuine twist-3 part reads
\begin{align}\label{eq:HadronicTensorgNLP} \nn&W_{\mathrm{gNLP}}^{\mu \nu}(q)=\int \mathrm{d} \xi \,H_2\big(x,\xi,Q^2\big)\int_0^{\infty} \frac{\mathrm{d} \bm{b}^2}{8 \pi|\bm{b} \| \bm{q}|}\left(\frac{\bar{n}^\mu }{q^{-}}-\frac{n^\mu }{q^{+}}\right)\Bigg\lbrace J_1(|\bm{b} \| \bm{q}|)\\
\notag&\times\bigg[q_T^\nu\Big(-\blu{J_{21}\big(\xi,\bm{b}^2\big)}\grn{f^{q\bar{q}}_{1}\big(x,\bm{b}^2\big)}+\img\blu{J_{11}\big(\bm{b}^2\big) }\grn{f^{\perp,q\bar{q}}_{2}\big(x,\xi,\bm{b}^2\big)}  \Big)\\
\notag& +S_\parallel(\img\epsilon_T^{\nu\rho} q_T^\rho)\Big( \blu{J_{11}\big(\bm{b}^2\big)}\grn{g^{\perp,q\bar{q}}_{2L}\big(x,\xi,\bm{b}^2\big)}+\blu{J_{21}\big(\xi,\bm{b}^2\big)}\grn{g^{q\bar{q}}_{1}\big(x,\bm{b}^2\big)}\Big)\bigg]\\
\notag& -M|\bm{b}||\bm{q}|J_0(|\bm{b}\|\bm{q}|)(\img\epsilon_T^{\nu\rho} S_T^\rho)\blu{J_{11}\big(\bm{b}^2\big)}  \grn{f^{q\bar{q}}_{2T}\big(x,\xi,\bm{b}^2\big)}\\
\notag& -MJ_1(|\bm{b}\|\bm{q}|)(\epsilon_T^{\nu\rho} S_T^\rho) \blu{J_{21}\big(\xi,\bm{b}^2\big) }\Big(\grn{f^{\perp,q\bar{q}}_{1T}\big(x,\bm{b}^2\big)}-\img\grn{g^{\perp,q\bar{q}}_{1T}\big(x,\bm{b}^2\big)}\Big)\\
\notag& +M|\bm{b}||\bm{q}|J_2(|\bm{b}\|\bm{q}|)(\img\epsilon_T^{\nu\alpha}S_T^\beta)\bigg(\frac{g^{\alpha\beta}_T}{2}+\frac{q_T^\alpha q_T^\beta}{|\bm{q}|^2}\bigg) \blu{J_{11}\big(\bm{b}^2\big)}  \grn{f^{\perp,q\bar{q}}_{2T}\big(x,\xi,\bm{b}^2\big)}\\
\notag& -\frac{M|\bm{b}|J_2(|\bm{b}\|\bm{q}|)}{|\bm{q}|}\blu{J_{21}\big(\xi,\bm{b}^2\big)}\bigg[q_T^\nu(\epsilon_T^{\alpha\beta} q_T^\alpha S_T^\beta)  \grn{f^{\perp,q\bar{q}}_{1T}\big(x,\bm{b}^2\big)}\\
&-(\img\epsilon_T^{\nu\rho}q_T^\rho)(g_T^{\alpha\beta} q_T^\alpha S_T^\beta)  \grn{g^{\perp,q\bar{q}}_{1T}\big(x,\bm{b}^2\big)}\bigg]\Bigg\rbrace+(\text{c.c}\,\,\text{and}\,\,\,\mu\leftrightarrow \nu)\,.
\end{align}
In this case, the twist-3 TMD and jet functions definitions in terms of matrix elements arise from eqs.~\eqref{eq:F_21_indices}-\eqref{eq:D_12_indices}. In the next section, we employ this factorized structure of the hadronic tensor to derive the expression for the factorized cross section.

\subsection{Cross section and form factors} \label{sec:Xsec&FF}

To determine the cross section for SIDIS with an observed jet, we must combine the hadronic and leptonic tensors. Specifically, this requires contracting the Lorentz structures found in eqs.~\eqref{eq:HadronicTensorLP}, \eqref{eq:HadronicTensorkNLP}, and \eqref{eq:HadronicTensorgNLP} with the leptonic tensor from eq.~\eqref{eq:LeptonicTensor}. These contractions, detailed in Appendix \ref{Appendix:LeptonicContractions}, allow us to express the cross section in a compact form that depends explicitly on the angles $\phi_J$ and $\phi_S$ in eqs.~\eqref{eq:AnglePhiJ} and \eqref{eq:AnglePhiS}, as well as a set of form factors.

Furthermore, by employing integration by parts and Bessel function identities, we can systematically transfer all derivatives acting on the TMDPDFs onto the jet functions. Specifically, we use the identity:
 \begin{align}
   &\int_0^{\infty}\mathrm{d}|\bm{b}|^2 \frac{J_n(|\bm{b} \| \bm{q}|)}{(|\bm{b}||\bm{q}|)^m}  \blu{J\big(\bm{b}^2\big)}\grn{F^{\prime}\big(x,\bm{b}^2\big)}=-\int_0^{\infty}\mathrm{d}|\bm{b}|^2 \frac{J_n(|\bm{b} \| \bm{q}|)}{(|\bm{b}||\bm{q}|)^m}  \blu{J^{\prime}\big(\bm{b}^2\big)}\grn{F\big(x,\bm{b}^2\big)}\\
   \nn&-\int_0^{\infty}\mathrm{d}|\bm{b}|^2\bigg(\frac{2-m+n}{2n}\frac{J_{n-1}(|\bm{b} \| \bm{q}|)}{(|\bm{b}||\bm{q}|)^{m-1}}+\frac{2-m-n}{2n}\frac{J_{n+1}(|\bm{b} \| \bm{q}|)}{(|\bm{b}||\bm{q}|)^{m-1}}\bigg)\blu{J\big(\bm{b}^2\big)}\grn{F\big(x,\bm{b}^2\big)}\,,
 \end{align}
 where $n>0$. Consequently, the cross section for SIDIS with a jet is:

\begin{align}\label{eq:FactorizedCrossSection}
 \frac{\text{d}\sigma}{\text{d}x\,\text{d}y\,\text{d}\phi_J\,\text{d}\phi_S\,\text{d}\bm{q}^2}=&\frac{\alpha_{\text{em}}^2y}{8Q^2}\Bigg\lbrace \frac{2-2y+y^2}{y^2}F_{UU,T}+ \frac{2(2-y)\sqrt{1-y}}{y^2}\cos\phi_J  F_{UU}^{\cos\phi_J}
  \\
  \nn&+\colorTwo{\lambda_e}
  \frac{2\sqrt{1-y}}{y} \sin\phi_J F_{LU}^{\sin\phi_J}+
  \colorThree{S_{\parallel}}\frac{2(2-y)\sqrt{1-y}}{y^2}\sin{\phi_J} F_{UL}^{\sin\phi_J}\\
  \nn &+\colorTwo{\lambda_e}\colorThree{S_{\parallel}}\left[\frac{2-y}{y}F_{LL}
  +\frac{2\sqrt{1-y}}{y} \cos{\phi_J} F_{LL}^{\cos\phi_J})\right]\\
  \nn&+\colorOne{|S_\perp|}\bigg[\frac{2-2y+y^2}{y^2}\sin{(\phi_J-\phi_S)}F^{\sin(\phi_J-\phi_S)}_{UT,T}\\
  &\nn+\frac{2(2-y)\sqrt{1-y}}{y^2}\Big(\sin{(\phi_S)}F^{\sin(\phi_S)}_{UT}+\sin{(2\phi_J-\phi_S)}F^{\sin(2\phi_J-\phi_S)}_{UT}\Big)\bigg]\\
  \nn&+\colorTwo{\lambda_e}\colorOne{|S_\perp|}\bigg[\frac{2-y}{y}\cos{(\phi_J-\phi_S)}F^{\cos(\phi_J-\phi_S)}_{LT}\\
  \nn&+\frac{2\sqrt{1-y}}{y}\Big(\cos{(\phi_S)}F^{\cos(\phi_S)}_{LT}+\cos{(2\phi_J-\phi_S)}F^{\cos(2\phi_J-\phi_S)}_{LT}\Big)\bigg]\Bigg\rbrace\,,
\end{align}
In this expression, we use color coding to highlight different dependencies: purple for the electron helicity $\colorTwo{\lambda_e}$, yellow for the hadron longitudinal polarization $\colorThree{S_\parallel}$, and red for its transverse polarization magnitude $\colorOne{|S_\perp|}$. The form factors are explicitly given by:
\begin{align}
\label{eq:FUUT}
    F_{UU,T}=&\,\mathcal{J}_{0,0}\big[\grn{f_{1}}\blu{ J_{11}}\big]\,,\\[1ex]
    F_{UU}^{\cos\phi_J}=&-\frac{2|\bm{q}|}{Q}\bigg\lbrace\mathcal{J}_{0,0}\big[\grn{f_{1}}\blu{ J_{11}}\big]+\mathcal{J}_{1,1}\big[\grn{f_{1}} \blu{J^{\prime}_{11}}\big]\\
    \nn&\qquad \qquad+\text{Re}\left(\mathcal{J}^{(2)}_{1,1}\big[\grn{f_{1}}\blu{J_{21}}\big]\right)+\text{Im}\left(\mathcal{J}^{(2)}_{1,1}\big[\grn{f^{\perp }_{2}}\blu{J_{11}}\big]\right)\bigg\rbrace\,,\\[1ex]
    F_{LU}^{\sin\phi_J}=&\,\frac{2|\bm{q}|}{Q}\bigg\lbrace\text{Im}\left(\mathcal{J}^{(2)}_{1,1}\big[\grn{f_{1}} \blu{J_{21}}\big]\right)-\text{Re}\left(\big[\grn{f^{\perp }_{2}}\blu{J_{11}}\big]\right)\bigg\rbrace\,,\\[1ex]
    F_{UL}^{\sin\phi_J}=&-\frac{2|\bm{q}|}{Q}\text{Im}\left(\mathcal{J}^{(2)}_{1,1}\big[\grn{g_{1}}\blu{J_{21}}+\grn{g^{\perp }_{2L}}\blu{J_{11}}\big]\right)\,,\\[1ex]
    F_{LL}=&\,\mathcal{J}_{0,0}\big[\grn{g_{1}}\blu{J_{11}}\big]\,,\\[1ex]
    F_{LL}^{\cos\phi_J}=&\,\frac{2|\bm{q}|}{Q}\bigg\lbrace\mathcal{J}_{0,0}\big[\grn{g_{1}}\blu{J_{11}}\big]+\mathcal{J}_{1,1}\big[\grn{g_{1}}\blu{J^{\prime}_{11}}\big]\\
    \nn&\qquad \quad+\text{Re}\left(\mathcal{J}^{(2)}_{1,1}\big[\grn{g_{1}}\blu{J_{21}}+\grn{g^{\perp }_{2L}}\blu{J_{11}}\big]\right)\bigg\rbrace\,,\\[1ex]
    F^{\sin(\phi_J-\phi_S)}_{UT,T}=&-\widetilde{\mathcal{J}}_{1,0}\big[\grn{f_{1T}^{\perp }}\blu{J_{11}}\big]\,,\\[1ex]
    F^{\sin(\phi_S)}_{UT}=&\frac{|\bm{q}|}{Q}\bigg\lbrace\widetilde{\mathcal{J}}_{0,1}\big[\grn{f_{1T}^{\perp }}\blu{J^{\prime}_{11}}\big]-\widetilde{\mathcal{J}}_{1,0}\big[\grn{f_{1T}^{\perp }}\blu{J_{11}}\big]\\
    &\nn\qquad+\text{Re}\left(\widetilde{\mathcal{J}}^{(2)}_{0,1}\big[\grn{f_{1T}^{\perp }}\blu{J_{21}}\big]\right)+\text{Im}\left(\widetilde{\mathcal{J}}^{(2)}_{0,1}\big[\grn{g_{1T}^{\perp }}\blu{J_{21}}-2\grn{f_{2T}}\blu{J_{11}}\big]\right)\bigg\rbrace\,,\\[1ex]
    F^{\sin(2\phi_J-\phi_S)}_{UT}=&\,\frac{|\bm{q}|}{Q}\bigg\lbrace\widetilde{\mathcal{J}}_{1,0}\big[\grn{f_{1T}^{\perp }}\blu{J_{11}}\big]+\widetilde{\mathcal{J}}_{2,1}\big[\grn{f_{1T}^{\perp }}\blu{J^{\prime}_{11}}\big]\\
    &\nn\qquad+\text{Re}\left(\widetilde{\mathcal{J}}^{(2)}_{2,1}\big[\grn{f_{1T}^{\perp }}\blu{J_{21}}\big]\right)-\text{Im}\left(\widetilde{\mathcal{J}}^{(2)}_{2,1}\big[\grn{g_{1T}^{\perp }}\blu{J_{21}}+\grn{f_{2T}^{\perp }}\blu{J_{11}}\big]\right)\bigg\rbrace\,,\\[1ex]
    F^{\cos(\phi_J-\phi_S)}_{LT}=&\,\widetilde{\mathcal{J}}_{1,0}\big[\grn{g_{1T}^{\perp }}\blu{J_{11}}\big]\,,\\[1ex]
    F^{\cos(\phi_S)}_{LT}=&\,\frac{|\bm{q}|}{Q}\bigg\lbrace\widetilde{\mathcal{J}}_{0,1}\big[\grn{g_{1T}^{\perp }}\blu{J^{\prime}_{11}}\big]-\widetilde{\mathcal{J}}_{1,0}\big[\grn{g_{1T}^{\perp }}\blu{J_{11}}\big]\\
    &\nn\qquad+\text{Im}\left(\widetilde{\mathcal{J}}^{(2)}_{0,1}\big[\grn{f_{1T}^{\perp }}\blu{J_{21}}\big]\right)-\text{Re}\left(\widetilde{\mathcal{J}}^{(2)}_{0,1}\big[\grn{g_{1T}^{\perp }}\blu{J_{21}}-2\grn{f_{2T}}\blu{J_{11}}\big]\right)\bigg\rbrace\,,\\[1ex]
    \label{eq:FLTc(2fj-fs)}
    F^{\cos(2\phi_J-\phi_S)}_{LT}=&-\frac{|\bm{q}|}{Q}\bigg\lbrace\widetilde{\mathcal{J}}_{1,0}\big[\grn{g_{1T}^{\perp }}\blu{J_{11}}\big]+\widetilde{\mathcal{J}}_{2,1}\big[\grn{g_{1T}^{\perp }}\blu{J^{\prime}_{11}}\big]\\
    &\nn\quad\qquad-\text{Im}\left(\widetilde{\mathcal{J}}^{(2)}_{2,1}\big[\grn{f_{1T}^{\perp }}\blu{J_{21}}\big]\right)-\text{Re}\left(\widetilde{\mathcal{J}}^{(2)}_{2,1}\big[\grn{g_{1T}^{\perp }}\blu{J_{21}}+\grn{f_{2T}^{\perp }}\blu{J_{11}}\big]\right)\bigg\rbrace\,.
\end{align}
Here we used the following shorthand notation to represent the Hankel-type transformation of the product of two distributions, which involves the hard factors and the $|\bm{b}|$-space integration:
\begin{align}\label{BesselIntegral}
    &\mathcal{J}_{n,m}\big[\grn{F}\blu{J}\big]=H_1(Q^2)\sum_{q}\int_0^{\infty} \mathrm{d}|\bm{b}|^2 \frac{J_n(|\bm{b} \| \bm{q}|)}{4 \pi(|\bm{b}||\bm{q}|)^m}  \blu{J\big(\bm{b}^2\big)}\grn{F^{q\bar q}\big(x,\bm{b}^2\big)}\,,\\[1ex]
    &\mathcal{J}_{n,m}^{(2)}\big[\grn{F}\blu{J}\big]=\sum_q\int d\xi\,H_2(x,\xi,Q^2)\int_0^{\infty} \mathrm{d}|\bm{b}|^2 \frac{J_n(|\bm{b} \| \bm{q}|)}{4 \pi(|\bm{b}||\bm{q}|)^m}  [\grn{F^{q\bar q}}\blu{J}](x,\xi,\bm{b}^2)\,.\label{BesselIntegral2}
\end{align}
Additionally, $\widetilde{\mathcal{J}}$ includes an extra factor of $M|\bm{b}|$ inside the integral, specifically:
\begin{equation}
    \widetilde{\mathcal{J}}\big[\grn{F}\blu{J}\big]=\mathcal{J}\big[M|\bm{b}|\grn{F}\blu{J}\big]\,.
\end{equation}
We emphasize that the TMD jet functions and parton distributions appearing in eq.~\eqref{eq:FactorizedCrossSection} are bare quantities, meaning they have not undergone UV renormalization. Consequently, when extracting phenomenological results in the next section, these bare functions must be replaced by their renormalized counterparts.

\subsection{Framework for phenomenological predictions}\label{sec:Phenomenology}

In order to present some phenomenological results, we set $\lambda_e=0$ and $|S_\perp|=0$ and we focus on the $\sin\phi_J$ asymmetry of the cross section for EIC encoded in the $F_{UL}^{\sin\phi_J}$ form factor of the cross section in eq.~(\ref{eq:FactorizedCrossSection}).

Additionally, we integrate the differential cross section over the inelasticity range $0.01 \leq y \leq 0.95$, as this interval is relevant for the EIC, and over the angles $\phi_J$ and $\phi_S$ such that the only dependence is given by the $F_{UL}^{\sin\phi_J}$ form factor. Therefore, the integrated cross section reads
\begin{equation}
    \text{d}\Sigma|_{\sin\left(\phi_J\right)}\equiv\int_0^{2\pi}\!\text{d}\phi_J\,\frac{\sin{\left(\phi_J\right)}}{\pi}
    \int_0^{2\pi}\! \text{d}\phi_S
    \int_{0.01}^{0.95} \!\text{d}y\, \left.\frac{\text{d}\sigma}{\text{d}x\,\text{d}y\,\text{d}\phi_J\,\text{d}\phi_S\,\text{d}\bm{q}^2}\right|_{\lambda_e,\,|S_\perp|=0}.
\end{equation}

Incorporating the results from the previous subsections, we define the following observable for our study:
\begin{align}
\label{eq:CrossSectionSine}
\left.\text{d}\Sigma\right|_{\sin\left(\phi_J\right)}&=\text{d}\Sigma_{g_{2L}^\perp}+\text{d}\Sigma_{J_{21}}\\
\nn&=-\frac{\pi\alpha_{\text{em}}^2R_yS_{\parallel}}{Q^2}\,\frac{|\bm{q}|}{Q}  \text{Im}\bigg(\sum_q\int \mathrm{d} \xi\, H_2\big(x,\xi,Q^2\big)\int_0^{\infty}\mathrm{d}|\bm{b}|^2\frac{J_1(|\bm{b}||\bm{q}|)}{4 \pi|\bm{b}||\bm{q}|}\\ \nn
&\qquad\times \Big\lbrace \blu{J^{\text{bare}}_{11}\big(\bm{b}^2;\bar\zeta\big)} \left[\grn{g^{\perp,\text{bare},q}_{2L}\big(x,\xi,\bm{b}^2;\zeta\big)}+\grn{g^{\perp,\text{bare},\bar{q}}_{2L}\big(x,\xi,\bm{b}^2;\zeta\big)}\right]\\
&\qquad\quad+\blu{J^{\text{bare}}_{21}\big(\xi,\bm{b}^2;\bar\zeta\big)}\left[\grn{g^{\text{bare},q}_{1}\big(x,\bm{b}^2;\zeta\big)}+\grn{g^{\text{bare},\bar{q}}_{1}\big(x,\bm{b}^2;\zeta\big)}\right]\Big\rbrace\bigg)\,,\nn
\end{align}
where we have introduced a factor coming from the integration over the inelasticity
\begin{equation}
    R_y=\int_{0.01}^{0.95}\frac{(2-y)\sqrt{1-y}}{y}\text{d}y\simeq7.33
\end{equation}
Note that $\left.\text{d}\Sigma\right|_{\sin\left(\phi_J\right)}$ depends only on the LP/NLP jet functions $J_{11}$ and $J_{21}$ and on the poorly known TMDPDFs $g_{1}$ and $g^\perp_{2L}$, making it a pure NLP prediction. Accordingly, we have separated the different NLP contributions in eq.~\eqref{eq:CrossSectionSine} as $\text{d}\Sigma_{g_{2L}^\perp}$ and $\text{d}\Sigma_{J_{21}}$.  To obtain predictions for these observables, we must address several aspects: the integration over $\xi$, the evolution, and the modeling of the relevant distributions. These points will be discussed in sections~\ref{sec:Treatment_g2} and~\ref{sec:Treatment_J21}.

\subsubsection{Treatment of \texorpdfstring{$g^\perp_{2L}$}{g2}}\label{sec:Treatment_g2}

First, we decompose the twist-3 TMD distributions into their T-even and T-odd components,
\begin{equation}\label{eq:OddAndEvenDistributions}
    \grn{g^{\perp,q/\bar{q}}_{2L}\left(x,\xi,\bm{b}^2\right)}=\grn{g^{\perp,q/\bar{q}}_{2L,\oplus}\left(x,\xi,\bm{b}^2\right)}-\img\,\grn{g^{\perp,q/\bar{q}}_{2L,\ominus}\left(x,\xi,\bm{b}^2\right)},
\end{equation}
which are real functions. Second, we must include the UV renormalization in eq.~\eqref{eq:CrossSectionSine}, 
\begin{align}
    \blu{J^{\text{bare}}_{11, \bar{n}}\big(\bm{b}^2;\bar\zeta\big)}&=Z_{J}(\mu,\bar\zeta)\blu{J_{11}\left(\bm{b}^2;\mu,\bar\zeta\right)},\\[1ex]
    \nn\grn{g^{\perp,\text{bare},q/\bar{q}}_{2L,\oplus}\big(x,\xi,\bm{b}^2;\zeta\big)}&=Z_{\oplus}(\xi;\mu,\zeta)\otimes \grn{g^{\perp,q/\bar{q}}_{2L,\oplus}\left(x,\xi,\bm{b}^2;\mu,\zeta\right)}\\[1ex]
    \nn\grn{g^{\perp,\text{bare},q/\bar{q}}_{2L,\ominus}\big(x,\xi,\bm{b}^2;\zeta\big)}&=Z_{\ominus}(\xi;\mu,\zeta)\otimes \grn{g^{\perp,q/\bar{q}}_{2L,\ominus}\left(x,\xi,\bm{b}^2;\mu,\zeta\right)}\,.
\end{align}
Here, $Z_{J}$, $Z_{\oplus}$, and $Z_{\ominus}$ represent the UV renormalization factors for the jet, as well as for the T-even and T-odd TMD distributions, respectively. The symbol $\otimes$ denotes the integral convolution in $\xi$ between $Z_{\oplus/\ominus}$ and the renormalized twist-3 TMDs; explicit expressions for these factors are provided in ref.~\cite{Rodini:2022wki}. The scales $\mu$ and $\zeta$ correspond to the renormalization of UV and rapidity divergences. 

Second, the TMD correlators $g^{\perp,\bar{q}/q}_{2L}\big(x,\xi,\bm{b}^2\big)$ vanish in specific regions of the variables $x$ and $\xi$. As shown in ref.~\cite{Rodini:2022wki}, the non-vanishing region is given by:
\begin{equation}\label{Eq:SupportTwist3}
     x_1\equiv-\frac{1-|x|}{|x|}<\xi<\frac{1}{|x|}\equiv x_2\quad(\text {for } -1<x<1).
\end{equation}
Consequently, the integration over $\xi$ for this distribution only needs to be carried out within these boundaries.

Following the derivation in \cite{Rodini:2023plb}, we decompose the hard function into its real and imaginary parts:
\begin{equation}
    H_2\big(x,\xi,Q^2\big)=H_2^R(x,\xi, Q^2)+\img\pi\, H_2^I(x,\xi, Q^2).
\end{equation}
The explicit expressions at leading order, adapted to our framework, read
\begin{align}
\label{eq:HardCoef}& H_2^R\left(x, \xi,Q^2\right)=1+\mathcal{O}(a_s)\,, \\[1ex]
& \nn H_2^I\left(x,\xi,Q^2\right)=\mathcal{O}(a_s),
\end{align}

 Using these decompositions, the twist-3 TMDPDF contribution to the cross section can be expressed as
\begin{equation}\label{eq:CrossSectionSineRenormalizedg2_Final}
    \begin{split}
\text{d}\Sigma_{g_{2L}^\perp}=&\frac{\pi\alpha_{\text{em}}^2R_yS_\parallel}{Q^2}\,\frac{|\bm{q}|}{Q}\sum_q\int_0^{\infty} \mathrm{d}|\bm{b}|^2 \frac{J_1(|\bm{b}||\bm{q}|)}{4 \pi|\bm{b}||\bm{q}|}R_J(\bm{b}^2;\mu,\zeta)Z_{J}(\mu,\zeta)\\
&\times \blu{J_{11}\left(\bm{b}^2;\mu,\zeta\right)}\bigg(\int_{x_1}^{x_2} \mathrm{d} \xi\, H_2^R\big(x,\xi,Q^2\big)R_2(\bm{b}^2;\mu,\zeta)\\
&\times Z_{\ominus}(\xi;\mu,\zeta)\otimes\left[\grn{g^{\perp,q}_{2L,\ominus}\left(x,\xi,\bm{b}^2;\mu,\zeta\right)}+\grn{g^{\perp,\bar{q}}_{2L,\ominus}\left(x,\xi,\bm{b}^2;\mu,\zeta\right)}\right]\\
&-\int_{x_1}^{x_2} \mathrm{d} \xi\, \pi H_2^I\big(x,\xi,Q^2\big)R_2(\bm{b}^2;\mu,\zeta)\\
&\times 
Z_{\oplus}(\xi;\mu,\zeta)\otimes\left[\grn{g^{\perp,q}_{2L,\oplus}\left(x,\xi,\bm{b}^2;\mu,\zeta\right)}+\grn{g^{\perp,\bar{q}}_{2L,\oplus}\left(x,\xi,\bm{b}^2;\mu,\zeta\right)}\right]\bigg).
\end{split}
\end{equation}
Analyzing the hard coefficient in eq.~(\ref{eq:HardCoef}), we observe that the imaginary part only contributes starting at next-to-leading order (NLO). As a result, the last two lines in eq.~(\ref{eq:CrossSectionSineRenormalizedg2_Final}) vanish at LO. On the other hand, the first three lines involve the real part of the coefficient, which at LO is simply one. Therefore, at this order, the only nontrivial task is to perform the $\xi$-integration over $g^{\perp,q/\bar{q}}_{2L,\oplus/\ominus}$. 

Since there are no experimental extractions of twist-3 TMD distributions, the T-odd function must be modeled. In ref.~\cite{Rodini:2023plb}, the size of some of these twist-3 distributions for $\xi=0$ is assumed to be comparable to the Boer-Mulders distribution in the small-$b$ limit. We assume a similar behavior for the odd part of $g_{2L}^\perp$. Furthermore, we incorporate a suppression of the distribution at large values of $\xi$, using a Gaussian damping factor for simplicity. Based on these considerations, we propose the following toy model for the twist-3 distribution:
\begin{equation}\label{eq:RelationBoerMulders}
    \grn{g^{\perp,q/\bar{q}}_{2L,\oplus/\ominus}\left(x,\xi,\bm{b}^2\right)}\simeq \frac{M^2|\bm{b}|^2}{\sqrt{2}\,\pi^{3/2}}\,\big[\Theta(\xi-x_1)-\Theta(\xi-x_2)\big]e^{-\xi^2/2}\grn{h_1^\perp\big(x,\bm{b}^2\big)}\,,
\end{equation}
where $h_1^\perp$ is the Boer-Mulders distribution, $\Theta(z)$ is the Heaviside step function defining the support for the distribution, and the additional factor in the numerator arises due to the different parametrization used in this work compared to ref.~\cite{Rodini:2023plb}, as explained in appendix~\ref{Appendix:Twist3Parametrization}.

Then, the integration for the T-odd part of the distribution yields:
\begin{align}
    \int_{-\infty}^{\infty}\text{d}\xi\,\grn{g_{2L,\ominus}\big(x,\xi,\bm{b}^2\big)}\simeq&\frac{M^2|\bm{b}|^2}{2\pi}\left(\text{erf}\left[\frac{1}{\sqrt{2}\,x}\right] - \text{erf}\left[-\frac{1 - x}{\sqrt{2}\,x}\right]\right)\grn{h_1^\perp\big(x,\bm{b}^2\big)}\,,
\end{align}
where $\text{erf}[z]$ is the Gauss error function. Recently, in ref.~\cite{Piloneta:2024aac}, the Boer-Mulders distribution was extracted using the following model:
\begin{equation}\label{BoerMuldersModel2}
\grn{h_{1}^\perp\big(x,\bm{b}^2\big)}= \frac{2^{\alpha+1}}{\Gamma(\alpha+1)}\, \frac{N}{\cosh{(\lambda}|\bm{b}|)}\,x\ln^\alpha(1/x)\,.
\end{equation}
The values for $\lambda$, $N$, and $\alpha$ can be found in ref.~\cite{Piloneta:2024aac}. Note that this model is constrained to $x<0.1$, as there is no current data beyond this point for extracting the Boer-Mulders distribution. 

For the twist-2 jet function, we adopt the LO expression, which is given by
\begin{equation}
    \blu{J_{11}\left(\bm{b}^2\right)}=1+\mathcal{O}(a_s)\,.
\end{equation}

Lastly, regarding the evolution of our distributions, we employ the framework used for the optimal TMD distribution in~\cite{Scimemi:2019cmh} for both the jet and the Boer-Mulders distribution
\begin{align}
    \grn{h_1^\perp\left(x,\bm{b}^2;Q,Q^2\right)}=\left(\frac{Q^2}{\zeta_Q(\bm{b}^2)}\right)^{-\frac{1}{2}K(\bm{b}^2,Q)}\grn{h_1^\perp\big(x,\bm{b}^2\big)}\,,\\[1ex]
     \nn \blu{J_{11}\left(\bm{b}^2;Q,Q^2\right)}=\left(\frac{Q^2}{\zeta_Q(\bm{b}^2)}\right)^{-\frac{1}{2}K(\bm{b}^2,Q)}\blu{J_{11}\left(\bm{b}^2\right)}\,,
\end{align}
where $K(\bm{b}^2,Q)$ is the Collins-Soper kernel, and $\zeta_Q(\bm{b}^2)$ defines the special equi-potential (or null-evolution) line. Explicit expressions for both are available in ref.~\cite{Scimemi:2019cmh}, and we implement these up to NLO accuracy. The integral over the running coupling in $K$ is computed analytically using the iterative solution derived in ref.~\cite{Deur_2016} at $\beta_1$ accuracy.

Based on all of these ingredients, we consider the following leading logarithm (LL) approximation:
\begin{equation}\label{eq:CrossSectionSineRenormalized_Plot2}
    \begin{split}
\text{d}\Sigma_{g_{2L}^\perp}^{\text{LL}}=&\frac{3\alpha_{\text{em}}^2M^2R_yS_\parallel}{2Q^2}\,\frac{|\bm{q}|}{Q} \int_0^{\infty} \mathrm{d}|\bm{b}|\,|\bm{b}|^2\frac{J_1(|\bm{b}||\bm{q}|)}{2 \pi|\bm{q}|}\left(\frac{Q^2}{\zeta_Q(\bm{b}^2)}\right)^{-K(\bm{b}^2,Q)}\\
&\times\left(\text{erf}\left[\frac{1}{\sqrt{2}\,x}\right] - \text{erf}\left[-\frac{1 - x}{\sqrt{2}\,x}\right]\right)\left(\grn{h_{1; q \leftarrow h}^\perp\big(x,\bm{b}^2\big)}+\grn{h_{1; \bar{q} \leftarrow h}^\perp\big(x,\bm{b}^2\big)}\right)\,.
\end{split}
\end{equation}
We refer to this approach as the Boer-Mulders model. In ref.~\cite{Piloneta:2024aac}, the quark and anti-quark distributions are parameterized using $|N|$ and $N$, respectively, with their analysis yielding $N<0$.  Since the cross section involves the sum of both quark and anti-quark contributions, this part of the cross section would cancel out. However, the absolute values of quark and anti-quark functions are unlikely to be identical. To account for this, we approximate the quark contribution as being ten times larger, noting that other choices will only affect the overall normalization. The parameters that we employ for this model are
\begin{equation}\label{eq:Boer-MuldersParameters}
   \lambda=0.2 \text{ GeV}\,,\quad N_q=0.854^{+0.378}_{-1.076}\,,\quad N_{\bar{q}}=-0.0854^{+0.1076}_{-0.0378}\,,\quad\alpha=9.4^{+5.4}_{-0.9},
\end{equation}
which ensures that $N_q N_{\bar{q}}=-0.0729$, consistent with the findings in ref.~\cite{Piloneta:2024aac}. 

Alternatively, as suggested in refs.~\cite{Barone:2009hw,Barone_2010,Barone_2015,Bastami_2019}, the Boer-Mulders function can be assumed to be proportional to the Sivers function, i.e., $h_{1; f \leftarrow h}^\perp=A_ff_{1T; f \leftarrow h}^\perp$.
Thus, another possible model for the Boer-Mulders function is given by:
\begin{equation}\label{BoerMuldersModel}
\grn{h_{1; f \leftarrow h}^{\perp}(x, \bm{b}^2)}=A_fN_f \frac{(1-x) x^{\beta_q}\left(1+\epsilon_q x\right)}{n(\beta_q, \epsilon_q)} \exp \left(-\frac{r_0+x\,r_1}{\sqrt{1+r_2\,x^2\,\bm{b}^2}} \bm{b}^2\right)\,,
\end{equation}
where $n(\beta_q, \epsilon_q)=(3+\beta_q+(1+\beta_q)\epsilon_q)\Gamma(\beta_q+1)/\Gamma(\beta_q+4)$.  
For the Sivers function, we adopt a parametrization similar to that in ref.~\cite{Bury:2021sue}, where the values of all the parameters above can be found. The proportionality constants $A_f$ are determined by imposing that, at $x=0.1$ and $|\bm{b}|=4\,\,\text{GeV}^{-1}$, both models in eqs.~\eqref{BoerMuldersModel2}~and~\eqref{BoerMuldersModel} agree for each flavor. This procedure leads to larger normalization factors for the strange and sea quarks; however, their contributions to the form factor remain negligible compared to those from the valence quarks (up and down).
For the reader's convenience, we present here the values taken from~\cite{Bury:2021sue}: 
\begin{equation}\label{eq:SiverParameters}
  r_0=0.54^{+0.60}_{-0.53}\text{ GeV}^{2}\,,\qquad r_1=5.22^{+1.18}_{-3.43}\text{ GeV}^{2}\,,\qquad r_2=203^{+71}_{-133}\text{ GeV}^{2}\,.
\end{equation}
The values that depend on the quark flavors are presented in Tab.\ref{tab:SiversModel}.

\begin{table}[t]
    \centering
    \begin{tabular}{c||c|c|c|c}
        \text{Parameter} & \text{up quark} & \text{down quark} & \text{strange quark} & \text{sea} \\
     \hhline{=||=|=|=|=}
         $A_f$ & -2.04 & 2.33 & 38.70 & 68.79\\
     \hhline{-||-|-|-|-}
        $N_f$ & $-0.017^{+0.011}_{-0.023}$ & $0.37^{+0.18}_{-0.17}$ & $0.76^{+0.89}_{-0.43}$ & $-0.47^{+0.21}_{-0.32}$\\
    \hhline{-||-|-|-|-}
        $\beta_f$ & $-0.36^{+0.09}_{-0.11}$ & $-0.70^{+0.77}_{-0.11}$ & $2.5^{+0.8}_{-0.7}$ & $2.5^{+0.8}_{-0.7}$\\
    \hhline{-||-|-|-|-}
        $\epsilon_f$ & $-3.9^{+0.6}_{-0.6}$ & $9.0^{+17.6}_{-9.1}$ & 0 & 0\\
    \end{tabular}
    \caption{Parameters for the Sivers model in \eqref{BoerMuldersModel}, obtained from ref.~\cite{Bury:2021sue}.}
    \label{tab:SiversModel}
\end{table}

Using these ingredients, we evaluate the contribution of $g_{2L}^\perp$ to the $\sin\phi_J$ asymmetry of the cross section at LL within the Sivers model:
\begin{equation}\label{eq:CrossSectionSineRenormalized_Plot3}
    \begin{split}
\text{d}\Sigma_{g_{2L}^\perp}^{\text{LL}}=&\frac{\alpha_{\text{em}}^2M^2R_yS_\parallel}{2Q^2}\,\frac{|\bm{q}|}{Q} \int_0^{\infty} \mathrm{d}|\bm{b}|\,|\bm{b}|^2 \frac{J_1(|\bm{b}||\bm{q}|)}{2\pi|\bm{q}|}\left(\frac{Q^2}{\zeta_Q(\bm{b}^2)}\right)^{-K(\bm{b}^2,Q)}\\
&\times\left(\text{erf}\left[\frac{1}{\sqrt{2}\,x}\right] - \text{erf}\left[-\frac{1 - x}{\sqrt{2}\,x}\right]\right)\bigg(A_u\,\grn{f_{1T; u \leftarrow h}^\perp\big(x,\bm{b}^2\big)}\\
&+A_d\,\grn{f_{1T; d \leftarrow h}^\perp\big(x,\bm{b}^2\big)}+A_s\,\grn{f_{1T; s \leftarrow h}^\perp\big(x,\bm{b}^2\big)}+A_{\text{sea}}\,\grn{f_{1T; \text{sea} \leftarrow h}^\perp\big(x,\bm{b}^2\big)}\bigg)\,.
\end{split}
\end{equation}

Since we lack a definitive method to determine which model more accurately represents the twist-3 distribution $g_{2L}^\perp$, we show results for both approaches. In the following subsection, we outline the framework necessary to account for the twist-3 jet contribution to the cross section.

\subsubsection{Treatment of \texorpdfstring{$J_{21}$}{J21}}\label{sec:Treatment_J21}
The expression for the bare (unsubtracted) twist-3 jet distribution was was derived in Ref.~\cite{delCastillo:2023rng} at LO:
\begin{align}\label{eq:NaiveJ21}
    \blu{J^{\text{naive}}_{21}(\xi,\bm{b}^2)}=-4a_sC_F\,\frac{\bar\xi}{\xi-\img\delta^-/q^-}\,\Theta(\xi)\Theta(\bar\xi)\frac{\xi^n+\bar{\xi}^n}{\bar\xi^{n-1}-\xi^{n-1}}+\mathcal{O}(a_s^2)\,.
\end{align}
Since  $J_{21}$ and $g_1$ are real functions, the imaginary part in eq.~\eqref{eq:CrossSectionSine} must arise solely from the hard coefficient. This implies that the $\xi$-integration in this part of the cross section begins at NLO:
\begin{align}
    \int_{-\infty}^{\infty}d\xi H_2^{I}(x,\xi,Q^2)\blu{J^{\text{naive}}_{21}\big(\xi,\bm{b}^2\big)}
    =\mathcal{O}(a_s^2)\,.
\end{align}
Therefore, when considering the physical distributions defined in eqs.~\eqref{eq:F_21_indices}–\eqref{eq:D_12_indices}, the LL contribution to the $\sin\phi_J$ asymmetry in the cross section arises solely from the twist-3 TMD distribution. That is,
\begin{align}
    \left.\text{d}\Sigma\right|^{\text{LL}}_{\sin\left(\phi_J\right)}=\text{d}\Sigma^{\text{LL}}_{g_{2L}^\perp}\,.
\end{align}

It is important to emphasize that this result holds only for the leading order hard function. At NLO, the imaginary part of the hard function contributes non-trivially. In that case, one must include the contributions from both the twist-3 jet function $J_{21}$ (after properly performing the overlap subtraction in eq.~\eqref{eq:NaiveJ21}$\,$) and the helicity TMD distribution $g_1$, using recent models such as those presented in refs.~\cite{Yang:2024wg,Bacchetta:2024yzl}. Nevertheless, based on the relative size of the jet and helicity terms, we expect the $g_1$ contribution to remain sub-leading compared to that of $g_{2L}^\perp$, even at higher orders.  With this formulation in place, we are now ready to present the phenomenological results for this observable in the next subsection.

\subsection{Phenomenological results for \texorpdfstring{$F_{UL}^{\sin\phi_J}$}{FUL}}

\begin{figure}[t]
    \centering
\includegraphics[width=0.8\textwidth]{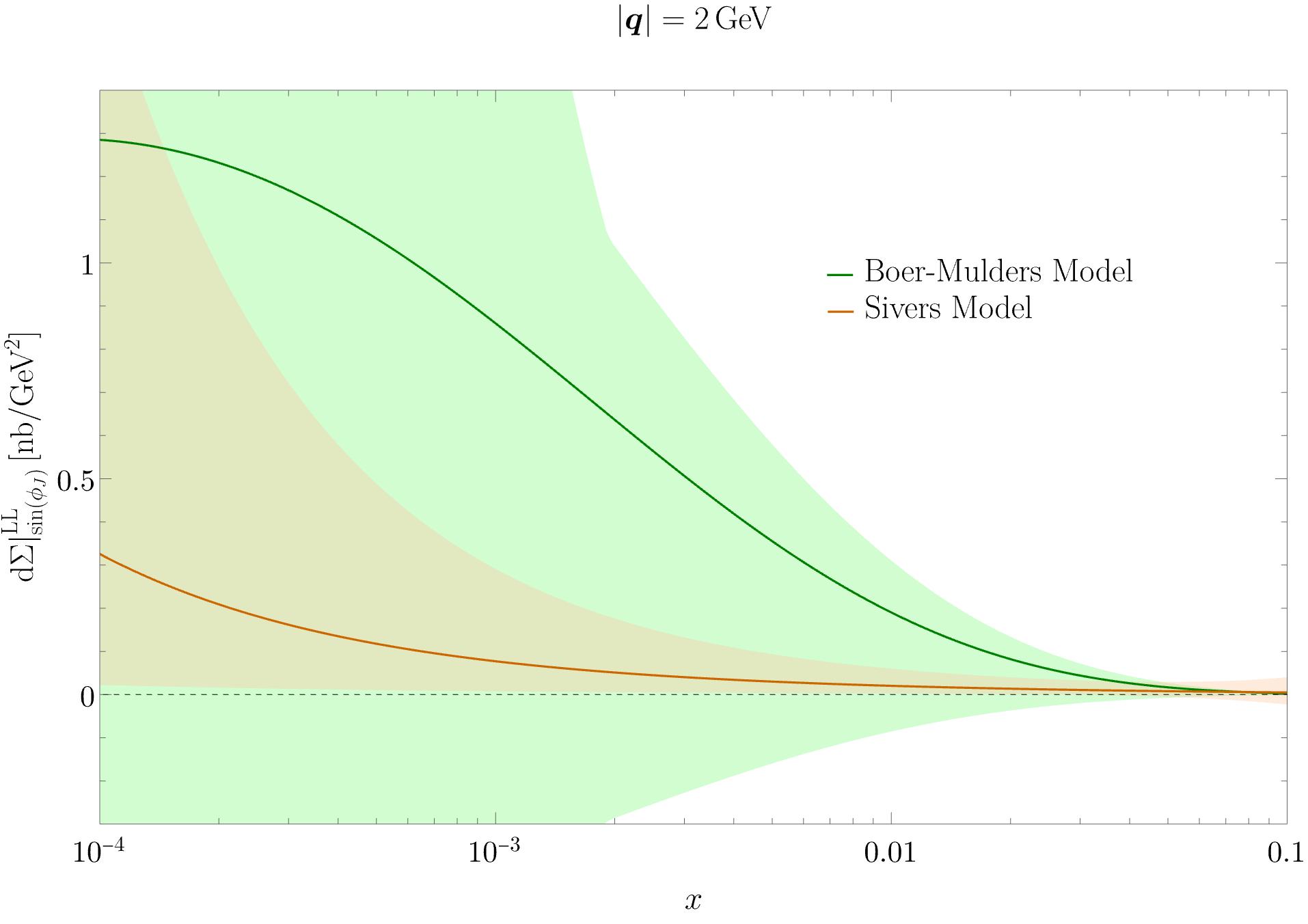}
    \caption{Comparison of the $\sin\phi_J$-asymmetry of the LL cross section for the Boer-Mulders and Sivers model in eqs.~\eqref{BoerMuldersModel2} and \eqref{BoerMuldersModel}, plotted as a function of $x$ for $|\bm{q}|=2$ GeV, and $Q=10$ GeV.}
\label{fig:SIDISwithJetCrossSectionNLPsineLogLinearPlotXBothModels2}
\end{figure}

\begin{figure}[t]
    \centering
\includegraphics[width=0.8\textwidth]{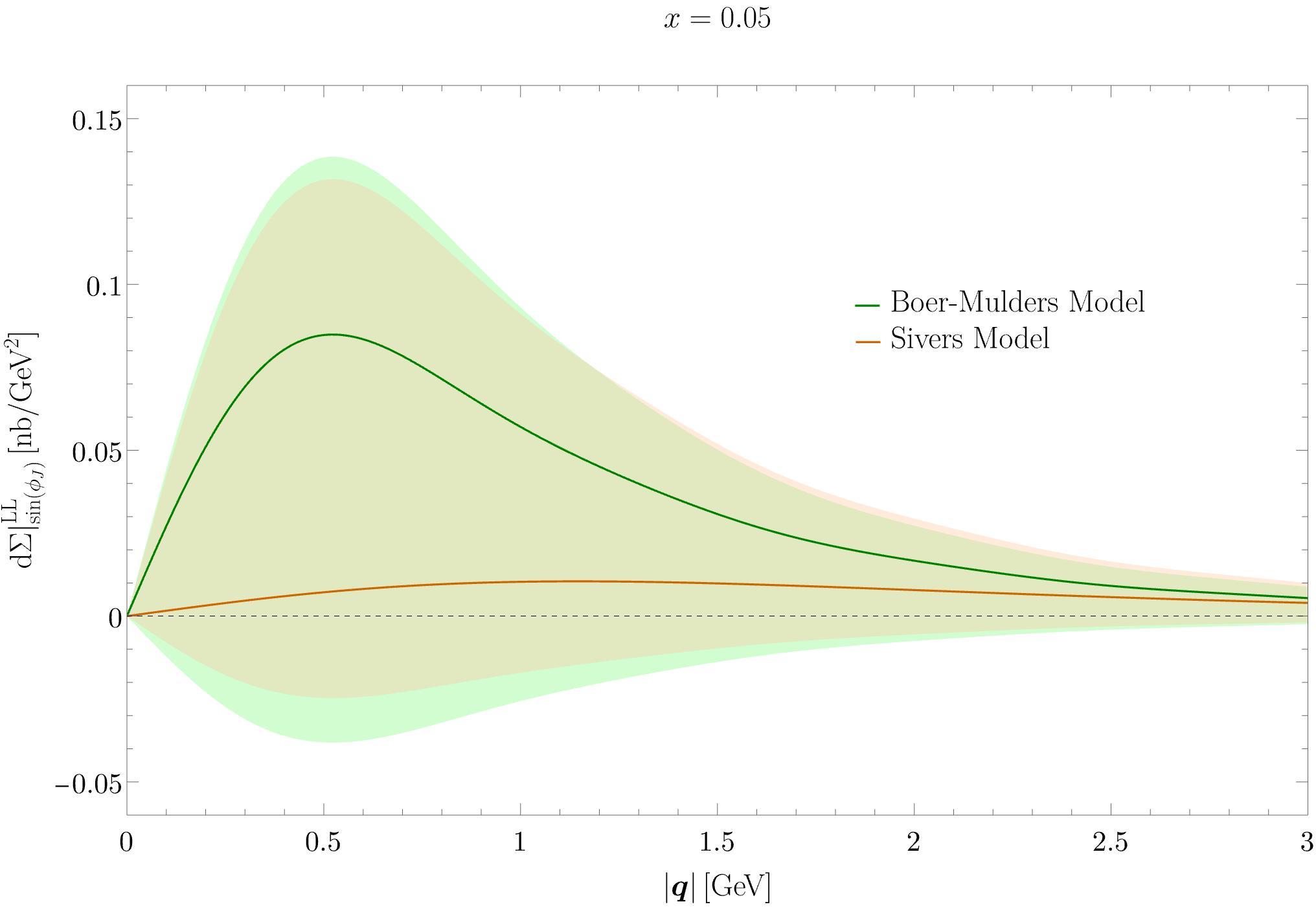}
    \caption{Comparison of the $\sin\phi_J$-asymmetry of the LL cross section for the Boer-Mulders and Sivers model in eqs.~\eqref{BoerMuldersModel2} and \eqref{BoerMuldersModel}, plotted as a function of $|\bm{q}|$ for $x=0.05$, and $Q=10$ GeV.}    \label{fig:SIDISwithJetCrossSectionNLPsinePlotQTBothModels}
\end{figure}

The  $\sin\phi_J$ asymmetry of the cross section at LL is plotted in terms of the momentum fraction $x$ and the transverse momentum $q_T$ in figs.~\ref{fig:SIDISwithJetCrossSectionNLPsineLogLinearPlotXBothModels2} and \ref{fig:SIDISwithJetCrossSectionNLPsinePlotQTBothModels}, respectively. These results are presented for  $Q=10$ GeV and $M=1$ GeV, and for the kinematic configuration that maximizes the asymmetry, i.e., $S_\parallel=1$.

Since two different models for the twist-3 $g_{2L}^\perp$ TMDPDF were considered, both are plotted, with each curve labeled according to the model used. While the models exhibit distinct behaviors, particularly at small $x$, their predictions remain consistent within the large error bands. These bands reflect the uncertainties in the nonperturbative TMD parameterizations, obtained by varying the model nonperturbative parameters within the uncertainties listed in eqs.~\eqref{eq:Boer-MuldersParameters}, \eqref{eq:SiverParameters}, and table~\ref{tab:SiversModel}.

For the Boer-Mulders model, the asymmetry rapidly decreases beyond $x = 0.1$, which is an artifact of the model due to the lack of data for $x>0.1\,$. Similarly, for the Sivers model, data for the large-$x$ region is currently limited to  $x\leq 0.5$. To avoid overinterpreting extrapolated behavior, we restrict our plots in fig.~\ref{fig:SIDISwithJetCrossSectionNLPsineLogLinearPlotXBothModels2} to the range $10^{-4}\leq x \leq 0.1$.
In the small-$x$ region, the Boer–Mulders model predicts a vanishing cross section as $x \to 0$, whereas the Sivers-based model diverges due to its $x^{\beta_q}$ behavior, with $\beta_u, \beta_d < 0$ as specified by the model. This divergence is inherited by the cross section.

Regarding the $|\bm{q}|$-dependence shown in fig.~\ref{fig:SIDISwithJetCrossSectionNLPsinePlotQTBothModels}, both models exhibit a Gaussian-like behavior. For the Sivers model, this behavior is less apparent in this plot in the central curve due to its smaller magnitude and higher $q_T$-value for the peak compared to the Boer–Mulders model. However, the error bands, obtained by varying the model parameters, show that these models are compatible within the current uncertainties.

\section{Conclusions}\label{sec:Conclusion}

In this work we derived a factorized expression for the 
transverse momentum distribution of jets in SIDIS, that holds up to next-to-leading power in $q_T/Q$. We compared against two results in the literature~\cite{Vladimirov:2021hdn,Ebert:2021jhy} that differ from each other in terms of the relevant degrees of freedom, and possibly disagree on the final result for the cross section, at higher orders in perturbation theory. To address this, we used a modified version of the background field method of ref.~\cite{Vladimirov:2021hdn},  including also background fields for soft modes. Our framework thus bridges the formulation of the BFM without soft modes~\cite{Vladimirov:2021hdn} with the SCET-II formulation of ref.~\cite{Ebert:2021jhy}. This unified treatment is potentially useful for other measurements that probe soft radiation beyond TMDs, e.g.~\cite{Beneke:2024cpq,Beneke:2025ufd}. Using our results, we make a phenomenological prediction for one of the angular-asymmetry form factors.

The first step towards the factorized cross section is the construction of an effective current operator. In our approach, the effective current is obtained by introducing background fields for collinear, soft, and anti-collinear modes and directly integrating out the remainder off-shell modes order-by-order in perturbation theory. To generalize this to higher orders, we matched onto gauge invariant operators using the building blocks of SCET, and constrained the Wilson coefficients of these operators using current conservation, reparameterization invariance, and discrete symmetries. From our expression for the effective current, we derived the hadronic tensor, finding that most of the sub-leading soft matrix elements vanished,
leaving only one soft gluon matrix element. To subtract the overlap (or zero-bin), we introduced a background field for the overlap region inspired by the approach of ref.~\cite{Idilbi:2007ff}.
By combining the NLP subtraction terms with the remaining soft gluon operators, we defined physical TMD distributions, yielding a cross section that is expressed in terms of these physical distributions and manifestly free of rapidity and endpoint divergences.

We have compared our method and results to two previous approaches, tracing a potential disagreement between them at higher orders in perturbation to a sub-leading soft contribution in ref.~\cite{Ebert:2021jhy}, for which it is not clear that it can be absorbed into the collinear and anti-collinear sectors. On the other hand, in ref.~\cite{Vladimirov:2021hdn} soft contributions arise only as the eikonal limits of the collinear and anti-collinear sectors and are thus always absorbed in the respective distributions. 
Our expression for the effective current operator differs from those of both refs.~\cite{Vladimirov:2021hdn,Ebert:2021jhy}. 
For ref.~\cite{Vladimirov:2021hdn}, this is simply due to the absence of soft background fields in their approach. 
Some differences with ref.~\cite{Ebert:2021jhy} stem from them using the label formalism of SCET rather than a position-space approach. A potential real difference in the effective current vanishes only if a non-trivial condition on the Wilson coefficients is satisfied— a condition that appears to hold so far but remains to be proven to all orders.

We applied our factorization formula to derive the factorized cross section for $e + h \rightarrow e + \text{jet} + X$ at small transverse momenta, presenting the general form factors for SIDIS with a jet measurement in terms of the physical twist-3 TMD distributions that we defined earlier. Additionally, we identified a particular angular asymmetry that depends only on NLP terms, which, interestingly, at LL receives contributions solely from a twist-3 TMD parton distribution. This observable thus presents a clean probe of sub-leading power dynamics and could serve as a useful input for future phenomenological studies. Higher-order corrections to this observable can be systematically incorporated through the perturbative computation of the relevant hard and jet functions, as outlined in this work. From a phenomenological perspective, the SIDIS form factors are especially relevant for future experimental analyses at the upcoming Electron-Ion Collider~\cite{AbdulKhalek:2021gbh}. 

While our framework successfully establishes a consistent factorization structure at NLP using the BFM with soft modes, a complete reconciliation with the SCET-based approach requires further investigation. 
Arriving at an agreement between all three approaches is essential for going beyond leading-logarithmic accuracy in making theoretical predictions and extracting twist-3 TMDs from experiment. 
Moreover, our method lays the groundwork for other sub-leading power studies, both to extend the analysis to other processes at NLP and to push the framework to higher orders in the power expansion, where soft contributions beyond the leading power soft function may contribute.

\acknowledgments
We thank Johannes Michel, Iain Stewart, and Alexey Vladimirov for discussions. This project is supported by the Spanish Ministerio de Ciencias y Innovaci\'on Grant No. PID2022-136510NB-C31 funded by MCIN/AEI/ 10.13039/501100011033. This project has received funding from the European Union Horizon programme under the Marie Skłodowska-Curie Actions Staff Exchanges with the reference number HORIZON-MSCA-2023-SE-01-101182937-HeI. O.$\,$dR.~is supported by the MIU (Ministerio de Universidades, Spain) fellowship FPU20/03110.

\appendix

\section{Different NLP parametrizations}\label{Appendix:Twist3Parametrization}

In this appendix, we discuss some differences in the parametrization of NLP distributions with refs.~\cite{Rodini:2023plb, Rodini:2022wki}, concerning the gluon field appearing in the twist-3 matrix elements, as well as the basis of twist-3 matrix elements.

An important difference between the twist-3 matrix elements in eqs.~\eqref{eq:F_21_indices}-\eqref{eq:D_12_indices} and the $\Phi_\rho$ functions found in these references, is that we use a SCET-like structure $\gamma_\rho \mathcal{A}^{\rho}_{T}$, whereas refs.~\cite{Rodini:2023plb, Rodini:2022wki} factor out the $\gamma_\rho$ along with the other Lorentz structures and employ the usual QCD field-strength $\mathcal{F}^{\rho\pm}$ inside the matrix element. The relation between the $\mathcal{A}$-fields and field strengths, for SIDIS, is given by
\begin{align}
    &\blu{\mathcal{A}_T^\mu(y)}
    =
    +g\int_0^{+\infty}\df z^+\,e^{-z^+ \delta^-}\,
    \blu{\mathcal{F}_T^{\mu-}(y + z^+ \bar{n})}\,,\\
    &
    \grn{\mathcal{A}_T^\mu(y)}
    =
    -g\int_{-\infty}^0\hspace{1.2ex}\df z^-\,e^{+z^- \delta^+}\,
    \grn{\mathcal{F}_T^{\mu+}(y + z^- n)}\,.
\end{align}
This substitution incorporates a $\delta$-regulator to control rapidity divergences. We can introduce the variable $\xi$, using the identities:
\begin{align}
    \int_0^\infty\df z^+\,e^{-z^+\delta^-}\,
    \blu{\mathcal{F}_T^{\mu-}(y + z^+ \bar{n})}
    &=
    \int_{-\infty}^{\infty}\df \xi\,
    \frac{-\img}{\pm\xi - \img\delta_E^-}
    \int\frac{\df z}{2\pi}\,e^{\pm \img \xi q^- z^+}\,
    \blu{\mathcal{F}_T^{\mu-}(y + z^+ \bar{n})}\,,
    \\
    \int_{-\infty}^0\df z^-\,e^{z^-\delta^+}\,
    \grn{\mathcal{F}_T^{\mu+}(y + z^- n)}
    &=
    \int_{-\infty}^{\infty}\df \xi\,
    \frac{-\img}{\pm\xi +\img\delta_E^+}
    \int\frac{\df z}{2\pi}\,e^{\pm \img \xi q^+ z^-}\,
   \grn{\mathcal{F}_T^{\mu+}(y + z^- n)}\,,
\end{align}
where $\delta_E^\pm=\delta^\pm/q^\pm$ is positive and ensures the integral is well-defined.
This makes the endpoint divergence in the variable $\xi$ explicit. In these identities, the sign of $\xi$ is arbitrary since it undergoes integration over all possible values. However, to match ref.~\cite{Rodini:2023plb}, we adopt the pole prescription $\xi - \img\delta_E^\pm$ for the $\mathcal{F}_{21}$ operators and $\xi + \img\delta_E^\pm$ for the $\mathcal{F}_{12}$ ones. 

The basis of TMDs used in this paper can be related to those in ref.~\cite{Rodini:2023plb}.
After a detailed comparison, taking into account the odd and even components as defined in eq.~\eqref{eq:OddAndEvenDistributions}, we arrive at the following relations for the bare distributions:
\begin{align}
    \label{eq:ConexionBetweenParametrizationsfperp}&f_{2,\oplus}^{\perp}=\frac{M^2|\bm{b}|^2}{\xi-\img\delta_E^+}\left(f_{\oplus}^{\perp,\text{\cite{Rodini:2023plb}}}+g_{\ominus}^{\perp,\text{\cite{Rodini:2023plb}}}\right)\,, &f_{2,\ominus}^{\perp}&=\frac{M^2|\bm{b}|^2}{\xi-\img\delta_E^+}\left(f_{\ominus}^{\perp,\text{\cite{Rodini:2023plb}}}-g_{\oplus}^{\perp,\text{\cite{Rodini:2023plb}}}\right)\,,\\
    &g_{2L,\oplus}^{\perp}=\frac{M^2|\bm{b}|^2}{\xi-\img\delta_E^+}\left(f_{\oplus L}^{\perp,\text{\cite{Rodini:2023plb}}}+g_{\ominus L}^{\perp,\text{\cite{Rodini:2023plb}}}\right)\,,&g_{2L,\ominus}^{\perp}&=\frac{M^2|\bm{b}|^2}{\xi-\img\delta_E^+}\left(f_{\ominus L}^{\perp,\text{\cite{Rodini:2023plb}}}-g_{\oplus L}^{\perp,\text{\cite{Rodini:2023plb}}}\right)\,,\\
    &f_{2T,\oplus}=\frac{1}{\xi-\img\delta_E^+}\left(f_{\oplus T}^{\text{\cite{Rodini:2023plb}}}+g_{\ominus T}^{\text{\cite{Rodini:2023plb}}}\right)\,, &f_{2T,\ominus}&=\frac{1}{\xi-\img\delta_E^+}\left(f_{\ominus T}^{\text{\cite{Rodini:2023plb}}}-g_{\oplus}^{\text{\cite{Rodini:2023plb}}}\right)\,,\\
    \label{eq:ConexionBetweenParametrizationsfTperp}&f_{2T,\oplus}^{\perp}=\frac{M^2|\bm{b}|^2}{\xi-\img\delta_E^+}\left(f_{\oplus T}^{\perp,\text{\cite{Rodini:2023plb}}}+g_{\ominus T}^{\perp,\text{\cite{Rodini:2023plb}}}\right)\,,&f_{2T,\ominus}^{\perp}&=\frac{M^2|\bm{b}|^2}{\xi-\img\delta_E^+}\left(f_{\ominus T}^{\perp,\text{\cite{Rodini:2023plb}}}-g_{\oplus}^{\perp,\text{\cite{Rodini:2023plb}}}\right)\,.
\end{align}
As is clear from the above equations, we don't (need to) isolate the $1/(\xi-\img\delta_E^+)$ poles. In our case these endpoint singularities are removed by the overlap subtraction.
Furthermore, while in refs.~\cite{Rodini:2023plb,Rodini:2022wki} the Dirac matrices that project the good components of quark spinors are $\gamma^+$ and $\gamma^+\gamma_5$ (corresponding to the unpolarized $f$ and polarized $g$ distributions, respectively), our approach employs the single projector $\epsilon_T^{\rho\alpha}\sigma^{\alpha+}\gamma_5$. This projector yields specific combinations of the distributions in~\cite{Rodini:2023plb}, as shown in eqs.~\eqref{eq:ConexionBetweenParametrizationsfperp}-\eqref{eq:ConexionBetweenParametrizationsfTperp}. This is not in conflict with ref.~\cite{Rodini:2023plb} as it was already noted there that the above combinations evolve together.

Finally, our definitions for the derivatives of the twist-2 distributions involve a different prefactor:
 \begin{equation}
     F^\prime=M^2|\bm{b}|^2\mathring{F}\,,
 \end{equation}
where the dotted function $\mathring{F}$ denotes the derivative of $F$ as defined in ref.~\cite{Rodini:2023plb}.

\section{Contractions with leptonic tensor}\label{Appendix:LeptonicContractions}

The contractions of the leptonic tensor given in eq.~\eqref{eq:LeptonicTensor} with the different Lorentz structures that appear for the hadronic tensor in eqs.~\eqref{eq:HadronicTensorLP}, \eqref{eq:HadronicTensorkNLP}, and \eqref{eq:HadronicTensorgNLP} are
\begin{align}
    (-g_T^{\mu\nu})L_{\mu\nu}
    &=
    +2Q^2 \frac{2-2y+y^2}{y^2}
    -4|\bm{q}|Q \cos(\phi_J) \frac{(2-y)\sqrt{1-y}}{y^2}+\mathcal{O}(|\bm{q}|^2)
    \,,\\
    (\img\epsilon_T^{\mu\nu})L_{\mu\nu}
    &=
    +2\colorTwo{\lambda_e} Q^2 \frac{2-y}{y}
    -4\colorTwo{\lambda_e} |\bm{q}|Q \cos(\phi_J) \frac{\sqrt{1-y}}{y}
    \,,\\
    \frac{n^\mu q_T^\nu}{q^{+}}L_{\mu\nu}
    &=
    -2|\bm{q}|Q\frac{\sqrt{1-y}}{y^2} \Big[(2-y)\cos(\phi_J) 
    +\img \colorTwo{\lambda_e} y \sin(\phi_J) \Big]+\mathcal{O}(|\bm{q}|^2)
    \,,\\
    \qquad\frac{n^\nu q_T^\mu}{q^{+}}L_{\mu\nu}
    &=
      -2|\bm{q}|Q\frac{\sqrt{1-y}}{y^2} \Big[(2-y)\cos(\phi_J) 
    -\img \colorTwo{\lambda_e} y \sin(\phi_J) \Big]+\mathcal{O}(|\bm{q}|^2)
    \,,\\
    \qquad\frac{\bar{n}^\mu q_T^\nu}{q^{-}}L_{\mu\nu}
    &= 2|\bm{q}|Q\frac{\sqrt{1-y}}{y^2} \Big[(2-y)\cos(\phi_J) 
    +\img \colorTwo{\lambda_e} y \sin(\phi_J) \Big]+\mathcal{O}(|\bm{q}|^2)
    \,,\\
    \qquad\frac{\bar{n}^\nu q_T^\mu}{q^{-}}L_{\mu\nu}
    &= 2|\bm{q}|Q\frac{\sqrt{1-y}}{y^2} \Big[(2-y)\cos(\phi_J) 
    -\img \colorTwo{\lambda_e} y \sin(\phi_J) \Big]+\mathcal{O}(|\bm{q}|^2)
    \,,\\
    \frac{n^\mu(\img\epsilon_T^{\nu\rho} q_T^\rho)}{q^{+}}L_{\mu\nu}
    &= -2|\bm{q}|Q\frac{\sqrt{1-y}}{y^2} \Big[
    \colorTwo{\lambda_e} y \cos(\phi_J) +(2-y)\img\sin(\phi_J) \Big]
    \,,\\
    \qquad\frac{n^\nu (\img\epsilon_T^{\mu\rho} q_T^\rho)}{q^{+}}L_{\mu\nu}
    &=
    2|\bm{q}|Q\frac{\sqrt{1-y}}{y^2} \Big[
    \colorTwo{\lambda_e} y \cos(\phi_J) -(2-y)\img\sin(\phi_J) \Big]
    \,,\\
    \qquad\frac{\bar{n}^\mu(\img\epsilon_T^{\nu\rho} q_T^\rho)}{q^{-}}L_{\mu\nu}
    &=
    2|\bm{q}|Q\frac{\sqrt{1-y}}{y^2} \Big[ 
    \colorTwo{\lambda_e} y \cos(\phi_J)+(2-y)\img\sin(\phi_J)\Big]
    \,,\\
    \qquad\frac{\bar{n}^\nu (\img\epsilon_T^{\mu\rho} q_T^\rho)}{q^{-}}L_{\mu\nu}
    &=
    -2|\bm{q}|Q\frac{\sqrt{1-y}}{y^2} \Big[ 
    \colorTwo{\lambda_e} y \cos(\phi_J)-(2-y)\img\sin(\phi_J)\Big]
  \,.
\end{align}
Note that we have neglected contributions with higher powers of $|\bm{q}|$ as they are sub-leading. These contractions can be reproduced by using the expressions of these tensors and vectors given in eqs.~\eqref{eq:n-vector}-\eqref{eq:Levi-Civita-Perp} and expressing them in terms of the relevant kinematic quantities. For the transverse spin structures, we also get other contractions that involve the angle with the spin of the incoming hadron $\phi_S$
\begin{align}
     &g_T^{\mu\nu}q_{T,\mu}S_{T,\nu}
    =|\bm{q}|\colorOne{|S_\perp|}\cos(\phi_J-\phi_S)
    \,,\\
    &\epsilon_T^{\mu\nu}q_{T,\mu}S_{T,\nu}
    =-|\bm{q}|\colorOne{|S_\perp|}\sin(\phi_J-\phi_S)
    \,,\\
    &\frac{n^\mu(\img\epsilon_T^{\nu\rho} S_T^\rho)}{q^{+}}L_{\mu\nu}
    =
    2 Q \colorOne{|S_\perp|}\frac{\sqrt{1-y}}{y^2} \Big[\colorTwo{\lambda_e} y \cos(\phi_S) 
    +(2-y) \img \sin(\phi_S) \Big]+\mathcal{O}(|\bm{q}|)
    ,\\ &
    \frac{n^\nu (\img\epsilon_T^{\mu\rho} S_T^\rho)}{q^{+}}L_{\mu\nu}
    = 
    -2 Q \colorOne{|S_\perp|}\frac{\sqrt{1-y}}{y^2} \Big[\colorTwo{\lambda_e} y \cos(\phi_S) 
    -(2-y) \img \sin(\phi_S) \Big]+\mathcal{O}(|\bm{q}|)
    \,,\\
    &\frac{\bar{n}^\mu(\img\epsilon_T^{\nu\rho} S_T^\rho)}{q^{-}}L_{\mu\nu}
    = 
    -2 Q \colorOne{|S_\perp|}\frac{\sqrt{1-y}}{y^2} \Big[\colorTwo{\lambda_e} y \cos(\phi_S) 
    +(2-y) \img \sin(\phi_S) \Big]+\mathcal{O}(|\bm{q}|)
    \,,\\
    &\frac{\bar{n}^\nu (\img\epsilon_T^{\mu\rho} S_T^\rho)}{q^{-}}L_{\mu\nu}
    = 
    2 Q \colorOne{|S_\perp|}\frac{\sqrt{1-y}}{y^2} \Big[\colorTwo{\lambda_e} y \cos(\phi_S) 
    -(2-y) \img \sin(\phi_S) \Big]+\mathcal{O}(|\bm{q}|)
    ,
\end{align}
\begin{align}
    \nn\frac{n^\mu(\img\epsilon_T^{\nu\alpha} S_T^\beta)}{q^{+}}\left(\frac{g_{T}^{\alpha\beta}}{2}+\frac{q_{T}^\alpha q_T^\beta}{|\bm{q}|^2}\right)L_{\mu\nu}
    =&- Q \colorOne{|S_\perp|}  \frac{\sqrt{1-y}}{y^2}
    \Big[\colorTwo{\lambda_e} y \cos(2\phi_J-\phi_S) 
    \\ &\quad+\img  (2-y)\sin(2\phi_J-\phi_S) \Big]+\mathcal{O}(|\bm{q}|)\,,&
    \\
    \nn\frac{n^\nu(\img\epsilon_T^{\mu\alpha} S_T^\beta)}{q^{+}}\left(\frac{g_{T}^{\alpha\beta}}{2}+\frac{q_{T}^\alpha q_T^\beta}{|\bm{q}|^2}\right)L_{\mu\nu}
    =&\,Q \colorOne{|S_\perp|}  \frac{\sqrt{1-y}}{y^2}
    \Big[\colorTwo{\lambda_e} y \cos(2\phi_J-\phi_S) 
    \\ &\quad-\img  (2-y)\sin(2\phi_J-\phi_S) \Big]+\mathcal{O}(|\bm{q}|)\,,
    \\
    \nn\frac{\bar n^\mu(\img\epsilon_T^{\nu\alpha} S_T^\beta)}{q^{+}}\left(\frac{g_{T}^{\alpha\beta}}{2}+\frac{q_{T}^\alpha q_T^\beta}{|\bm{q}|^2}\right)L_{\mu\nu}
    =&\,Q \colorOne{|S_\perp|}  \frac{\sqrt{1-y}}{y^2}
    \Big[\colorTwo{\lambda_e} y \cos(2\phi_J-\phi_S) 
    \\ &\quad+\img  (2-y)\sin(2\phi_J-\phi_S) \Big]+\mathcal{O}(|\bm{q}|)\,,
    \\
    \nn\frac{\bar n^\nu(\img\epsilon_T^{\mu\alpha} S_T^\beta)}{q^{+}}\left(\frac{g_{T}^{\alpha\beta}}{2}+\frac{q_{T}^\alpha q_T^\beta}{|\bm{q}|^2}\right)L_{\mu\nu}
    =&-Q \colorOne{|S_\perp|}  \frac{\sqrt{1-y}}{y^2}
    \Big[\colorTwo{\lambda_e} y \cos(2\phi_J-\phi_S) 
    \\ &\quad-\img  (2-y)\sin(2\phi_J-\phi_S) \Big]+\mathcal{O}(|\bm{q}|)
   \,.
\end{align}
Here, we have also employed eqs.~\eqref{eq:TransverseSpin} and \eqref{eq:PerpSpin} to perform the contractions. These results lead to the final expression of the cross section in \eqref{eq:FactorizedCrossSection}. 

\bibliography{Bibliography}

\end{document}